\documentclass[aps,prl,reprint,groupedaddress,amsmath,amssymb]{revtex4-1}

\usepackage{graphicx} 
\usepackage{bm} 
\usepackage{xfrac} 

\begin{document}

\title{Cell cycle heritability and localization phase transition in growing populations}

\author{Takashi Nozoe$^{1}$}
\author{Edo Kussell$^{1,2}$}
\affiliation{$^1$Department of Biology, New York University, 12 Waverly Place, New York, NY 10003. \\
$^2$Department of Physics, New York University, 726 Broadway, New York, NY 10003. }

\date{\today}

\begin{abstract}
The cell cycle duration is a variable cellular phenotype that underlies long-term population growth and age structures.  By analyzing the stationary solutions of a branching process with heritable cell division times, we demonstrate existence of a phase transition, which can be continuous or first-order, by which a non-zero fraction of the population becomes localized at a minimal division time. Just below the transition, we demonstrate coexistence of localized and delocalized age-structure phases, and power law decay of correlation functions. Above it, we observe self-synchronization of cell cycles, collective divisions, and slow `aging' of population growth rates.
\end{abstract}

\maketitle 

The duration of a cell-cycle, or inter-division time (IDT), is a fluctuating quantity
in cellular populations, and its statistical properties are thought to result from biological mechanisms that regulate cell growth and division \cite{Taheri-Araghi2015, Susman2018, Hashimoto2016}.
For single cells that grow in size exponentially, as in bacteria \cite{Taheri-Araghi2015} and yeast \cite{Nakaoka2017a},
models of cell-size regulation typically predict negative mother-daughter correlation of IDTs \cite{Amir2014a}.
Yet in a subset of bacterial experiments and in most observations on mammalian cells, positive IDT correlations have been measured
(see Table S1 
in \footnote{Supplemental Material at [URL will be inserted by publisher]}), indicating that the cell cycle duration is a heritable, fluctuating single-cell phenotype.
Such heritability enables selection to act on the distribution of IDTs in a population to increase the long-term population growth rate.  
Here we take up the question of how selection and cell cycle heritability interact to determine long-term population dynamics, a fundamental step toward understanding how evolution has shaped cell cycle control mechanisms.

To investigate the effect of IDT heritability on population growth, we considered a model with heritable IDTs that was first introduced in \cite{Lebowitz1974}, which we refer to as the Lebowitz-Rubinow model below.
We noticed the existence of a heritability threshold above which the population's distribution of IDTs would localize at the minimal IDT corresponding to the fastest possible single-cell growth rate.
We identify and characterize a localization phase transition in this model.  
We show the ancestral mean IDT provides an order parameter of the transition, and above the heritability threshold predict the emergence of arbitrarily long single cell lineages that maintain perfect inheritance of a minimal IDT. 
This prediction is confirmed in numerical simulations of finite populations.
We observe slow punctuated dynamics above the heritability threshold, characteristic of systems that exhibit aging, as well as cell-cycle synchronization in finite populations.
Our work provides fundamental connections between dynamics of age-structured populations \cite{Charlesworth1994}
and well-studied error-threshold phenomena of evolutionary theory \cite{Eigen1989, Hermisson2002},
both of which are described by the unifying framework of statistical physics of phase transitions.

\paragraph*{Model and stationary solutions. \label{par:Model-and-stationary}}
We model a proliferating population by a branching process in which $K\left(\tau,\tau^{\prime}\right)$ is the transition probability density from $\tau^{\prime}$ to $\tau$,
where $\tau^{\prime}$ and $\tau$ are the parent and offspring IDTs, respectively. To study the effect of heritability of IDTs, we focus
on the analytically tractable Lebowitz-Rubinow model \cite{Lebowitz1974}, which uses the transition kernel
\begin{equation}
K\left(\tau,\tau^{\prime}\right)=\beta\delta\left(\tau-\tau^{\prime}\right)+\left(1-\beta\right)k\left(\tau\right)\label{eq:trans-kernel-LRmodel} \ ,
\end{equation}
where $\delta\left(\cdot\right)$ is the Dirac delta function, $\beta$ represents the heritability of IDTs in the model  ($0\le\beta<1$), and $k\left(\tau\right)$ is a probability density function on the interval $\left[\tau_{0},\infty\right)$, with $0<\tau_{0}<\infty$.  A cell produces $\hat{z}$ offsprings at every division, where $\hat{z}$ is independently drawn from a fixed
probability distribution and its average is denoted by $z>0$. For
example, $z=2$ for binary division organisms and $z=1$ for the isolated
single cells. 

The dynamics of cell divisions in the population are governed by
\begin{equation}
n_{\textrm{div}}(\tau;t+\tau)=z\int_{0}^{\infty}K\left(\tau,\tau^{\prime}\right)n_{\textrm{div}}\left(\tau^{\prime};t\right)d\tau^{\prime}.\label{eq:gentimedist-time-evolution}
\end{equation}
where $n_{\textrm{div}}\left(\tau;t\right)d\tau dt$ is the expected
number of dividing cells (cells at the termination of cell-cycle)
between times $t$ and $t+dt$ with IDT between $\tau$ and $\tau+d\tau$ \footnote{IDTs of sibling cells may be correlated, which does not change Eq. (\ref{eq:gentimedist-time-evolution}), as long as the number of offspring is independent of the parent's IDT.}.
The expected number of divisions occurring between $t$ and $t+dt$ is given by $N_{\textrm{div}}\left(t\right)dt := dt\int_{0}^{\infty}n_{\textrm{div}}\left(\tau;t\right)d\tau$. 
The time-dependent population growth rate is defined by $\Lambda_{t}:=\sfrac{\left(z-1\right)N_{\text{div}}\left(t\right)}{N\left(t\right)}$, where  $N\left(t\right)$ is the expected population size at time $t$ (see \cite{Note1} for detailed derivations).

The steady-state, exponentially growing solution of \eqref{eq:gentimedist-time-evolution} can be found by substituting the form $n_{\textrm{div}}\left(\tau;t\right)=N_{\textrm{div}}\left(t\right)\cdot p_{\textrm{div}}\left(\tau\right)$, where $p_{\textrm{div}}\left(\tau\right)$ is the stationary probability density of IDTs of dividing cells, yielding
\begin{equation}
p_{\mathrm{div}}\left(\tau\right)=ze^{-\Lambda\tau}\int_{0}^{\infty}K\left(\tau,\tau^{\prime}\right)p_{\mathrm{div}}\left(\tau^{\prime}\right)d\tau^{\prime} \ ,
\label{eq:gentimedist-stationary}
\end{equation}
Using \eqref{eq:trans-kernel-LRmodel} above, and solving the equation one obtains
\begin{equation}
p_{\textrm{div}}\left(\tau\right)=q_{\beta,z,\Lambda}\left(\tau\right):=\frac{\left(1-\beta\right)k\left(\tau\right)ze^{-\Lambda\tau}}
{1-\beta ze^{-\Lambda\tau}} \label{eq:pdiv}
\end{equation}
where the steady-state growth rate $\Lambda$ is determined by the normalization condition
\begin{equation}
Q\left(\beta,z,\Lambda\right) :=\int_{0}^{\infty}q_{\beta,z,\Lambda}\left(\tau\right)d\tau=1. \label{eq:main-text-Q-function}
\end{equation}
If $z=1$, for example, then $\Lambda=0$ and $p_{\textrm{div}}\left(\tau\right)=k\left(\tau\right)$
for any $0\le\beta<1$. Namely, the stationary IDT distribution of isolated single cells is invariant of $\beta$.
We remark that $\beta$ is the correlation coefficient of IDTs between parent and offspring for  isolated, single cell growth at steady-state \cite{Note1}.

In the absence of IDT correlations ($\beta  = 0$), one recovers the well-known result $p_{\textrm{div}}\left(\tau\right)=ze^{-\Lambda\tau}k\left(\tau\right)$
where $\Lambda$ is the unique real root of the integral equation
$z\int_{0}^{\infty}e^{-\Lambda\tau}k\left(\tau\right)d\tau=1$ \cite{Fisher1930,Hashimoto2016}.
For $\beta > 0$, one additionally must have $1-\beta ze^{-\Lambda\tau}>0$ for all $\tau$ in the support of $k(\tau)$ to ensure $q_{\beta,z,\Lambda} \left(\tau\right) \geq 0$. If $k(\tau) > 0$ for all $\tau>\tau_0$, we find
\begin{equation}
\Lambda\ge\omega_{0}\left(\beta,z\right):=\sup_{\tau>\tau_{0}}\frac{\ln\left(\beta z\right)}{\tau}=\max\left(0,\frac{\ln\left(\beta z\right)}{\tau_{0}}\right); 
\end{equation}
(see \cite{Note1} for $k(\tau)$ with bounded support).
Since $Q\left(\beta,z,\Lambda\right)$ is a monotonically decreasing function of $\Lambda$ tending to zero as $\Lambda \rightarrow \infty$, Eq. \ref{eq:main-text-Q-function} has a unique root $\Lambda_{\beta,z}$ provided that $Q\left(\beta,z,\omega_{0}\left(\beta,z\right)\right) \geq 1$.
This condition holds for $\beta < \beta_c$, where $\beta_c$ is a critical heritability threshold defined by
\begin{equation}
Q\left(\beta_{c},z,\omega_{0}\left(\beta_{c},z\right)\right)=1 \ , \label{eq:beta_c}
\end{equation}
such that for $\beta > \beta_c$, $Q\left(\beta,z,\omega_{0}\left(\beta,z\right)\right) < 1$ and Eq. \ref{eq:main-text-Q-function} does not admit a real solution $\Lambda_{\beta,z}$.  For $\beta > \beta_c$, the solution $p_{\textrm{div}}\left(\tau\right)$ given in \eqref{eq:pdiv} is incomplete, as there is missing probability $1 - Q$.  For $z > 1$, the full solution is 
\begin{equation}
p_{\textrm{div}}\left(\tau\right)=q_{\beta,z,\Lambda_{\beta,z}}\left(\tau\right)+\left(1-Q\left(\beta,z,\Lambda_{\beta,z}\right)\right)\delta\left(\tau-\tau_{0}\right) \ , \label{eq:stationary-gentime-dist}
\end{equation}
and substitution into \eqref{eq:gentimedist-stationary} yields $\Lambda_{\beta,z}=\tau_{0}^{-1}\ln\left(\beta z\right)$ for the steady-state growth rate when $\beta > \beta_c$ \cite{Note1}.

\begin{figure}
\includegraphics[width=\linewidth]{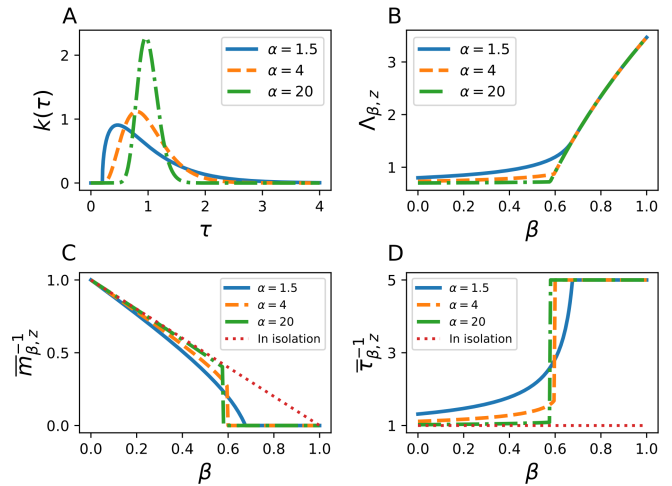}
\caption{Stationary solutions and their dependence on the heritability  parameter $\beta$.
(A) The IDT distribution, $k\left(\tau\right)$, is chosen to be 
a gamma distribution, shifted by $\tau_{0}=0.2$, with shape
parameters $\alpha=1.5$, $4$, and $20$; $k\left(\tau\right)=\Gamma\left(\alpha\right)^{-1}\theta^{-\alpha}\left(\tau-\tau_{0}\right)^{\alpha-1}e^{-\sfrac{\left(\tau-\tau_{0}\right)}{\theta}}$
for $\tau\ge\tau_{0}$.  For the mean IDT to be 1, the scale parameter $\theta$ is chosen as $\tau_{0}+\alpha\theta=1$.
(B-D) Dependence of key quantities of the stationary solutions on $\beta$ using  $z=2$; all other parameters are the same as in A.
$\beta_c = 0.68$, $0.60$ and $0.58$ respectively for $\alpha=1.5$, $4$, and $20$.
(B) Population growth rate, $\Lambda_{\beta,z}$.
(C) Reciprocal mean block size, $\overline{m}_{\beta,z}^{-1}$. (D) Reciprocal ancestral mean IDT, $\overline{\tau}_{\beta,z}^{-1}$. \label{fig:Fig1}}
\end{figure}

Further analysis of \eqref{eq:beta_c} shows that a heritability threshold $\beta_c < 1$ exists if and only if $\int_{0}^{\infty} d\tau k(\tau)/(\tau - \tau_0)$ converges; e.g. if $k(\tau) \sim (\tau - \tau_0)^{\gamma}$ near $\tau_0$ for some $\gamma > 0$.  The steady
state population growth rate $\Lambda_{\beta,z}$ qualitatively
changes as $\beta$ crosses the threshold: it depends on $k\left(\tau\right)$ for $\beta < \beta_c$, and becomes independent of it for $\beta > \beta_c$. The expected number of offspring having the same IDT as their parent
is $\beta z$, and only parents with $\tau' = \tau_{0}$ can generate offspring with $\tau = \tau_{0}$. Thus, the fraction of the population with IDT $\tau_{0}$ grows with rate $\tau_{0}^{-1}\ln\left(\beta z\right)$.
If we add a small fraction of $\tau_{0}$ cells to a population, they go extinct if $\beta z<1$, while if $\beta z>1$, they can constitute a giant cluster in the population's genealogy.
The subpopulation localized at $\tau = \tau_{0}$ will be outcompeted by the rest of the population for $z^{-1}<\beta<\beta_{c}$, with $\Lambda_{\beta,z}$ determined by \eqref{eq:main-text-Q-function}; or it will dominate the population for $\beta>\beta_{c}$, and thus dictate its growth rate to be $\Lambda_{\beta,z} = \tau_{0}^{-1}\ln\left(\beta z\right)$.
In Fig. \ref{fig:Fig1}A, we show a range of examples $k(\tau)$ which admit a threshold $\beta_c$. Increasing $\beta$ from 0 to 1, $\Lambda_{\beta,z}$  increases monotonically with a shallow slope, while past $\beta_{c}$, the slope of the growth rate changes markedly (Fig. \ref{fig:Fig1}B). 

\paragraph{Localization phase transition of population age structure.}
We now show that the threshold behavior identified above constitutes a phase transition in the strict sense.  We map the population to a statistical mechanical ensemble, as follows.  From the viewpoint of single cell lineages -- i.e. the history of an individual and all of its ancestors -- an age-structured population constitutes an ensemble of trajectories: the sequence of IDTs along a lineage $(\dots, \tau_{i-1}, \tau_{i}, \tau_{i+1}, \ldots)$ is analogous to a microscopic state of a large system (e.g. configuration of spins on a lattice, conformation of a polymer in space, etc.); while a single ancestral cell division $\tau_i$ specifies the state of a single component (e.g a spin, or a monomer). The population is an ensemble of lineages, and the steady-state population growth rate is its free energy \cite{Note1, Wakamoto2011}.

To analyze the structure of lineages observed above and below $\beta_c$, we consider the number of generations with which the same IDT is consecutively inherited, which we denote by $m$ and call the ``block size''.  The probability distribution of $m$ over lineages is analogous to a correlation function, and its mean measures the typical correlation length.  To compute these quantities, we let $\left(\tau^{\prime}, \tau\right)$ denote
a pair of IDTs, where $\tau^{\prime}$ and $\tau$ are parent and offspring IDTs, respectively. From (\ref{eq:gentimedist-stationary}) one can infer that
the stationary probability density to find $\left(\tau^{\prime}, \tau\right)$ is $ze^{-\Lambda\tau}K\left(\tau,\tau^{\prime}\right)p_{\textrm{div}}\left(\tau^{\prime}\right)$.
In particular, the probability of observing $(\tau, \tau)$ is $\beta ze^{-\Lambda\tau}p_{\textrm{div}}\left(\tau\right)$.
Repeatedly multiplying $\beta ze^{-\Lambda_{\beta,z}\tau}$, the probability for $\tau$ drawn from $p_{\textrm{div}}\left(\tau\right)$ to be
inherited at least $m-1$ times is $\left(\beta ze^{-\Lambda_{\beta,z}\tau}\right)^{m-1}$,
hence the joint probability distribution of block size $m$ and IDT $\tau$ is
\begin{equation}
p_{\textrm{lin}}\left(m,\tau\right) := \left(1-\beta ze^{-\Lambda_{\beta,z}\tau}\right)
\left(\beta ze^{-\Lambda_{\beta,z}\tau}\right)^{m-1}p_{\textrm{div}}\left(\tau\right) \ .
\label{eq:joint-dist-lin-m-tau}
\end{equation}
For $\beta<\beta_{c}$, the mean block size on lineage is
\begin{equation}
\overline{m}_{\beta,z} :=\sum_{m\ge1} \int_{0}^{\infty} m \cdot p_{\textrm{lin}}\left(m,\tau\right) d\tau<\infty \ ,
\end{equation}
and the probability distribution of IDT on lineage, also known as the ancestral distribution \cite{Hermisson2002}, is
\begin{equation}
p_{\textrm{lin}}\left(\tau\right) := \frac{\sum_{m\ge1} m \cdot p_{\textrm{lin}}\left(m,\tau\right)}{\overline{m}_{\beta,z}}\ , 
\label{eq:plin-of-tau}
\end{equation}
and represents the probability distribution of IDTs of ancestors in the
infinite past \cite{Note1}.
For $\beta>\beta_{c}$, $\overline{m}_{\beta,z}$ diverges and $p_{\textrm{lin}}\left(\tau\right)$ must be computed as an appropriate limit \cite{Note1}; we obtain 
\begin{equation}
p_{\textrm{lin}}\left(\tau\right) = \begin{cases}
\begin{array}{c}
\frac{\left(1-\beta\right)k\left(\tau\right)ze^{-\Lambda_{\beta,z}\tau}}{\overline{m}_{\beta,z}\left(1-\beta ze^{-\tau\Lambda_{\beta,z}}\right)^{2}},\\
\delta\left(\tau-\tau_{0}\right),
\end{array} & \begin{array}{c}
0\le\beta<\beta_{c}\\
\beta_{c}<\beta<1
\end{array}\end{cases} \ .
\label{eq:gentime-dist-lineage}
\end{equation}
The expression indicates that the IDT distribution on lineages in the population is localized entirely at $\tau = \tau_{0}$ above $\beta_{c}$, despite the fact that the IDT distribution of isolated lineages remains $k\left(\tau\right)$. 

The mean block size $\overline{m}_{\beta,z}$ serves as an
order parameter that diverges for $\beta>\beta_{c}$, and the continuity
of the phase transition can be characterized by its behavior near
$\beta_{c}$. If $\lim_{\beta\uparrow\beta_{c}}\overline{m}_{\beta,z}=\infty$,
the transition is a continuous phase transition, while if $\lim_{\beta\uparrow\beta_{c}}\overline{m}_{\beta,z} < \infty$ the transition is referred to as discontinuous, or `first-order'. Analogously, the lineage mean IDT, $\overline{\tau}_{\beta,z}:=\int_{0}^{\infty}\tau p_{\text{lin}}\left(\tau\right)d\tau$,
which is given by $\partial \Lambda_{\beta,z} / \partial \log z$ (i.e. a first-order derivative of the free energy \cite{Note1}),
exhibits the same type of phase transition. Examples of the $\beta$
dependence of $\overline{m}_{\beta,z}$ and $\overline{\tau}_{\beta,z}$
are shown in Fig. \ref{fig:Fig1}C and D. Assuming the law of large numbers, 
$\overline{\tau}_{\beta,z}^{-1}\simeq\sfrac{D}{t}$ and $\overline{m}^{-1}_{\beta,z}\simeq\sfrac{S}{D}$
hold where $D$ and $S$ denote the number of divisions and the number
of switches to different values of $\tau$ on lineage, respectively \cite{Note1}.

\begin{figure}
\includegraphics[width=\linewidth]{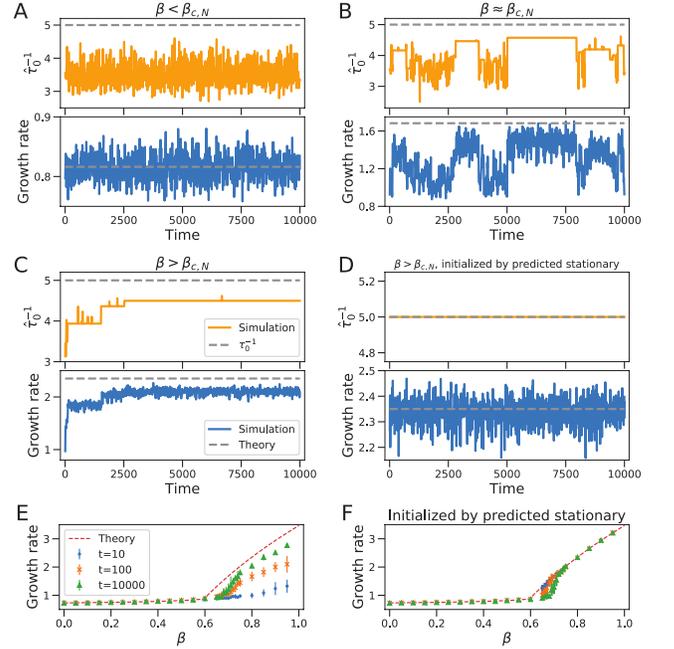}
\caption{Aging of growth rates in a finite population.
The parameters are as in Fig. \ref{fig:Fig1} using $\alpha = 4$, which exhibits a first-order localization transition.
In simulations, some of the cells in the population may be removed when newborn cells
appear in order to keep the population size $N = 100$ constant.
See \cite{Note1} for simulation details.
$\beta_{c,N}=0.73$.
(A-D) Time courses of reciprocal empirical minimal IDT $\hat\tau_0^{-1}$ (top, orange, solid) compared with the minimal IDT $\tau_0^{-1}$ (gray, dashed line) and growth rates (bottom, blue, solid) compared with predicted stationary value (gray, dashed line).
(A) is below the transition ($\beta=0.5$)
and the initial state is sampled from the delocalized state. (B)
shows coexistence of delocalized and localized states ($\beta=0.7$,
starting from the delocalized state). Time courses
of growth rates are compared to $\hat \tau_0^{-1}\log\left(\beta z\right)$
(orange dot). (C) Time courses above the transition showing
aging dynamics ($\beta=0.8$, starting from the delocalized state).
(D) For $\beta=0.8$, when the initial population
is sampled from predicted stationary solution, the growth rates fluctuate around the theoretically predicted stationary growth rate.
(E, F) Growth rate vs $\beta$. The initial condition was chosen
as delocalized state for (E) and predicted stationary for
(F). For each $\beta$ and stationary conditions, 10 simulations were
run, and the mean is plotted with error bars showing the standard
deviation. \label{fig:Fig2}}
\end{figure}

Using the form of $k\left(\tau\right)$ as in Fig. \ref{fig:Fig1}, where $k(\tau) \sim (\tau - \tau_0)^{\gamma}$ for $\tau$ near $\tau_0$, and computing $\overline{m}_{\beta,z}$ we find that the phase transition is continuous for $0<\gamma\le1$, and becomes discontinuous for $\gamma>1$.  For the latter type of transition, one expects to observe coexistence of two phases at the transition point, which is seen in numerical simulations of finite populations shown below (Fig. \ref{fig:Fig2}B).  We can also compute the marginal probability distribution $p_{\textrm{lin}}^{\textrm{M}}\left(m\right)$ of block size $m$, which decays exponentially below the transition, and follows power law statistics,
$p_{\textrm{lin}}^{\textrm{M}}\left(m\right)\sim m^{-\gamma-1}$ for
large $m$, in the limit $\beta\uparrow\beta_{c}$ as expected for correlation functions in the vicinity of a phase transition \cite{Note1}.
Such an appearance of long memory of IDT inheritance at the transition point is likewise observed in simulations (Fig. \ref{fig:Fig2}B).

\paragraph{Aging of long-term growth rate in finite populations.}

In finite sized populations, the stationary distribution \eqref{eq:stationary-gentime-dist} for $\beta>\beta_{c}$ is not achievable because any parent cell with IDT $\tau'>\tau_{0}$ has probability zero of generating offspring with IDT $\tau_{0}$.  Additionally, the distribution \eqref{eq:stationary-gentime-dist} cannot be maintained as a steady-state because cells with IDT $\tau_{0}$ can be lost from the population within finite time due to coalescence. 
To observe dynamics in finite populations, we conducted exact stochastic simulations in which cells are randomly removed to maintain a fixed population size \cite{Note1}.
We sampled the initial population with size $N=100$ independently from the stationary probability distribution with $\beta=0$, which we refer to as \emph{the delocalized state}, and simulated the population forward in time for a given value of $\beta>0$. 

In simulations with $\beta<\beta_{c}$, growth rates $\Lambda_t$ fluctuate around the expected steady-state growth rate (Fig. \ref{fig:Fig2}A). For $\beta$ slightly above $\beta_{c}$, however, $\Lambda_t$ exhibits sudden transitions between two distinct, long-lived states (Fig
\ref{fig:Fig2}B), which is expected for systems near a first-order phase transition. One of these states represents localization at the empirical
minimum IDT, denoted by $\hat \tau_0$, which is the minimum
IDT among all the dividing cells within each time bin \cite{Note1}.
In this state, the growth rate fluctuates around $\hat \tau_0^{-1}\ln\left(\beta z\right)$
over a sufficient period during which $\hat \tau_0$ is constant.
The other phase represents delocalization, where $\hat \tau_0$
exhibits large fluctuations. For higher values of $\beta$, the observed
growth rate increases stepwise and fluctuates around $\hat \tau_0^{-1}\ln\left(\beta z\right)$
(Fig. \ref{fig:Fig2}C). In this case, a fraction of the population localized at $\hat \tau_0$
is maintained over a significant time interval until a new value of $\hat\tau_0$
replaces the current empirical minimum IDT.  The time intervals between these replacement events become increasingly long as $\hat \tau_0$ approaches $\tau_0$, because IDTs that are shorter than the current minimum become increasingly rare. When the simulation starts with the population sampled from the predicted stationary probability distribution with the same
$\beta$ used in Fig. \ref{fig:Fig2}C, $\Lambda_t$ fluctuates around $\hat \tau_0^{-1}\ln\left(\beta z\right)$ and $\hat \tau_0 =\tau_{0}$ is preserved over the entire simulation time
(Fig. \ref{fig:Fig2}D). Values of $\Lambda_t$ at different simulation times are plotted as function of $\beta$ initialized at either the delocalized state (Fig. \ref{fig:Fig2}E) or the predicted stationary solution (Fig. \ref{fig:Fig2}F). For both initial conditions, the curve at
$t=10^{4}$ possesses an inflection point $\beta_{c,N}$ slightly
greater than $\beta_{c}$, which we refer to as the effective transition
point at fixed population size $N$ (see \cite{Note1} for detailed definition).
For both of these initial conditions, the population would reach the steady-state where $\hat \tau_0$
fluctuates around some specific $\tau_{\textrm{opt}}\left(>\tau_{0}\right)$
but it was not observed in reasonable simulation time due to the observed aging behavior of the dynamics.
Despite this challenge in numerically observing the true steady-state, due to the slow aging of the dynamics, our finite population simulations demonstrate the existence of the transition point above which the localized phase emerges in the dynamics, and that the dependence of the observed growth rate on $\beta$ is quantitatively predicted by theoretical analysis of the stationary solution. 

\begin{figure}
\includegraphics[scale=0.55]{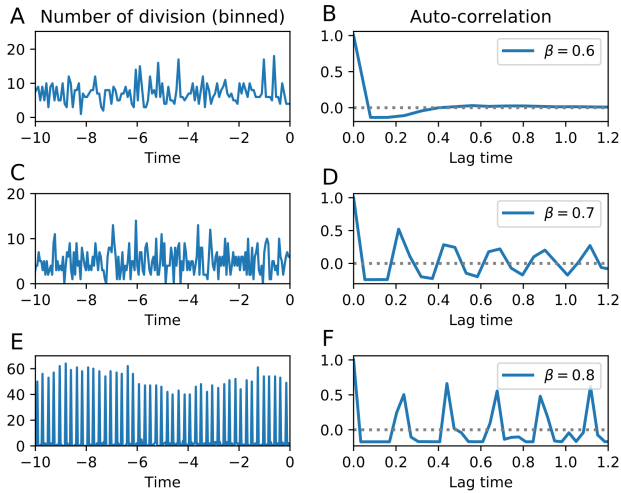}
\caption{Synchronization of cell-cycles above the transition.
The simulation is initialized by the delocalized state.
The parameters are as in Fig. \ref{fig:Fig2} using $\alpha = 4$ and $N=100$.
$\beta_{c,N}=0.73$.
Number of divisions binned over 0.1 doubling time (A, C, E) and their auto-correlations (B, D, F) are shown.
Simulations are run over $10^4$ time units and the last 10 time units of the series are shown for  A, C and E, where time point 0 indicates the simulation end.
Auto-correlations are computed over entire time series.
The correlation coefficient is normalized to equal 1 at zero lag. Dotted line indicates zero correlation.
(A,B) Below the transition ($\beta=0.6$)
timing of divisions are not synchronized. (C, D) Near the transition
point ($\beta=0.7$), timing of divisions are partially synchronized,
observed as decaying oscillation of the auto-correlation function. (E,
F) Above the transition ($\beta>\beta_{c,N}$), collective divisions
are observed indicating self-synchronization of cell cycles. \label{fig:Fig3}}
\end{figure}

\paragraph{Collective divisions and cell cycle synchronization.} In addition to aging dynamics, we observed collective divisions within the population above the transition (Fig. \ref{fig:Fig3}). The number of divisions that occurred, binned in short intervals, is plotted over time. Compared to Figs. \ref{fig:Fig3}A and C, Fig. \ref{fig:Fig3}E shows division events occurring collectively and periodically. Its auto-correlation
function does not oscillate below the transition (\ref{fig:Fig3}B), but exhibits decaying oscillations near the transition point (\ref{fig:Fig3}D) and sustained oscillations above the transition (\ref{fig:Fig3}F). The fact that the period of the autocorrelation function is close to
$\tau_{0}$ reflects localization at $\hat \tau_0$ close to $\tau_{0}$. We also tested the robustness of the transition properties to noise in the inheritance of IDTs by allowing small fluctuations of the offspring's IDT when it inherits its parent's IDT with probability $\beta$. To do so, we modified the transition kernel to be
\begin{equation}
K\left(\tau,\tau^{\prime}\right)=\beta p_{\text{norm}}\left(\tau-\tau^{\prime},\sigma\right)+\left(1-\beta\right)k\left(\tau\right)\label{eq:looser-inheritance-transitionprob}
\end{equation}
where $p_{\textrm{norm}}\left(x,\sigma\right)$ is a normal distribution
density function with standard deviation $\sigma$, truncated at $\tau_{0}$.
As a result, the population can reach equilibrium over a reasonable time
scale, that is, the long-term behaviors coincide between the two distinct initial conditions (Figs. S6 and S7 in \cite{Note1}).
Even without exact inheritance of parental IDTs, signatures of the phase transition are still observed (Figs. S8-S15 in \cite{Note1}).
Collective divisions are weaker but still detectable through the decaying oscillations of the autocorrelation function of divisions in population
(Figs. S16 and S17 in \cite{Note1}).
Division rate oscillations have also been predicted to arise in cell size control models,  due to negative IDT correlations \cite{Jafarpour2019}.

\paragraph{Discussion.}

We analyzed how the strength of the inter-division time heritability affects age-structured population dynamics.
In a model with heritable cell cycle durations first introduced in \cite{Lebowitz1974}, we demonstrated the existence of a localization phase transition.
While the existence of a heritability threshold $\beta_{c}$ was not explicitly suggested in \cite{Lebowitz1974},
the potential limit on the validity of the stationary solution was pointed out. A similar localization transition
in exponentially growing populations is known in mutation-selection models of theoretical population genetics
\cite{Eigen1971, Kingman1978, Hermisson2002}.
For example, Kingman's house-of-cards model \cite{Kingman1978} describes population dynamics on a specific type of fitness landscape, which exhibits localization at maximal fitness below a critical mutation rate. The localization phase transition as a stationary state in the house-of-cards model can be generalized to include cases where fitness is correlated between parent and offspring, and the existence and the uniqueness of the stationary solutions has been proven using the theory of positive linear operators
\cite{Burger1996}. 

We showed that the signature of localization phase transition can be observed even in finite populations as small as 100 cells, which is comparable to the capacity allowed by typical microfluidic experiments \cite{Wallden2011,Lambert2014,Hashimoto2016}, indicating that this transition could in principle be experimentally observed.  
Above the transition, population growth rates exhibit aging dynamics.
Similar phenomena are seen in evolution in random unbounded fitness
landscape with rare mutation rate in finite population, known as the ``diluted record process'' \cite{Park2008} or ``successional mutation regime'' \cite{Brotto2016}. While the distribution and auto-correlation of individual fitness are difficult to measure in
reality, measuring those of IDTs is highly feasible using \emph{in vivo} single cell tracking in microfluidics-based experiments.  The aging dynamics we observed in age-structured
populations are closely related to the emergence of self-synchronized cell-cycles (Fig. \ref{fig:Fig3}). Future work on the spectral structure of the linear semigroup of age-structured population dynamics \cite{Webb1986,Boulanouar2011} may further elucidate the emergence of synchronized growth above the transition.

While aging dynamics above the transition are specific to the noiseless
inheritance of IDT, we found that the signature of localization and self-synchronization
 can still be observable in ``noisier'' inheritance systems. This result indicates that our analytical results are applicable to real biological systems in which IDT heritability is noisy. The Lebowitz-Rubinow model does not include cell-cell interactions which
are thought to be a major mechanism for cell-cycle synchronization.  Remarkably, our analysis predicts strong but imperfect correlation may be enough to cause self-synchronization of cell cycles in finite populations in the absence of cell-cell
interactions. Our findings could provide a basis for the design of new types of synthetic biological oscillators which leverage population-level selective forces to establish robust cell cycle synchronization and to support sustained oscillations.

We thank Yuichi Wakamoto for discussions.  This work was supported by NIH grant R01-120231 to E.K.

\bibliography{ref_maintext}

\begin{thebibliography}{23}%
\makeatletter
\providecommand \@ifxundefined [1]{%
 \@ifx{#1\undefined}
}%
\providecommand \@ifnum [1]{%
 \ifnum #1\expandafter \@firstoftwo
 \else \expandafter \@secondoftwo
 \fi
}%
\providecommand \@ifx [1]{%
 \ifx #1\expandafter \@firstoftwo
 \else \expandafter \@secondoftwo
 \fi
}%
\providecommand \natexlab [1]{#1}%
\providecommand \enquote  [1]{``#1''}%
\providecommand \bibnamefont  [1]{#1}%
\providecommand \bibfnamefont [1]{#1}%
\providecommand \citenamefont [1]{#1}%
\providecommand \href@noop [0]{\@secondoftwo}%
\providecommand \href [0]{\begingroup \@sanitize@url \@href}%
\providecommand \@href[1]{\@@startlink{#1}\@@href}%
\providecommand \@@href[1]{\endgroup#1\@@endlink}%
\providecommand \@sanitize@url [0]{\catcode `\\12\catcode `\$12\catcode
  `\&12\catcode `\#12\catcode `\^12\catcode `\_12\catcode `\%12\relax}%
\providecommand \@@startlink[1]{}%
\providecommand \@@endlink[0]{}%
\providecommand \url  [0]{\begingroup\@sanitize@url \@url }%
\providecommand \@url [1]{\endgroup\@href {#1}{\urlprefix }}%
\providecommand \urlprefix  [0]{URL }%
\providecommand \Eprint [0]{\href }%
\providecommand \doibase [0]{https://doi.org/}%
\providecommand \selectlanguage [0]{\@gobble}%
\providecommand \bibinfo  [0]{\@secondoftwo}%
\providecommand \bibfield  [0]{\@secondoftwo}%
\providecommand \translation [1]{[#1]}%
\providecommand \BibitemOpen [0]{}%
\providecommand \bibitemStop [0]{}%
\providecommand \bibitemNoStop [0]{.\EOS\space}%
\providecommand \EOS [0]{\spacefactor3000\relax}%
\providecommand \BibitemShut  [1]{\csname bibitem#1\endcsname}%
\let\auto@bib@innerbib\@empty
\bibitem [{\citenamefont {Taheri-Araghi}\ \emph {et~al.}(2015)\citenamefont
  {Taheri-Araghi}, \citenamefont {Bradde}, \citenamefont {Sauls}, \citenamefont
  {Hill}, \citenamefont {Levin}, \citenamefont {Paulsson}, \citenamefont
  {Vergassola},\ and\ \citenamefont {Jun}}]{Taheri-Araghi2015}%
  \BibitemOpen
  \bibfield  {author} {\bibinfo {author} {\bibfnamefont {S.}~\bibnamefont
  {Taheri-Araghi}}, \bibinfo {author} {\bibfnamefont {S.}~\bibnamefont
  {Bradde}}, \bibinfo {author} {\bibfnamefont {J.~T.}\ \bibnamefont {Sauls}},
  \bibinfo {author} {\bibfnamefont {N.~S.}\ \bibnamefont {Hill}}, \bibinfo
  {author} {\bibfnamefont {P.~A.}\ \bibnamefont {Levin}}, \bibinfo {author}
  {\bibfnamefont {J.}~\bibnamefont {Paulsson}}, \bibinfo {author}
  {\bibfnamefont {M.}~\bibnamefont {Vergassola}},\ and\ \bibinfo {author}
  {\bibfnamefont {S.}~\bibnamefont {Jun}},\ }\href
  {https://doi.org/10.1016/j.cub.2014.12.009} {\bibfield  {journal} {\bibinfo
  {journal} {Curr. Biol.}\ }\textbf {\bibinfo {volume} {25}},\ \bibinfo {pages}
  {385} (\bibinfo {year} {2015})}\BibitemShut {NoStop}%
\bibitem [{\citenamefont {Susman}\ \emph {et~al.}(2018)\citenamefont {Susman},
  \citenamefont {Kohram}, \citenamefont {Vashistha}, \citenamefont {Nechleba},
  \citenamefont {Salman},\ and\ \citenamefont {Brenner}}]{Susman2018}%
  \BibitemOpen
  \bibfield  {author} {\bibinfo {author} {\bibfnamefont {L.}~\bibnamefont
  {Susman}}, \bibinfo {author} {\bibfnamefont {M.}~\bibnamefont {Kohram}},
  \bibinfo {author} {\bibfnamefont {H.}~\bibnamefont {Vashistha}}, \bibinfo
  {author} {\bibfnamefont {J.~T.}\ \bibnamefont {Nechleba}}, \bibinfo {author}
  {\bibfnamefont {H.}~\bibnamefont {Salman}},\ and\ \bibinfo {author}
  {\bibfnamefont {N.}~\bibnamefont {Brenner}},\ }\href
  {https://doi.org/10.1073/pnas.1615526115} {\bibfield  {journal} {\bibinfo
  {journal} {Proc. Natl. Acad. Sci. U. S. A.}\ }\textbf {\bibinfo {volume}
  {115}},\ \bibinfo {pages} {201615526} (\bibinfo {year} {2018})},\ \Eprint
  {https://arxiv.org/abs/1805.05058} {arXiv:1805.05058} \BibitemShut {NoStop}%
\bibitem [{\citenamefont {Hashimoto}\ \emph {et~al.}(2016)\citenamefont
  {Hashimoto}, \citenamefont {Nozoe}, \citenamefont {Nakaoka}, \citenamefont
  {Okura}, \citenamefont {Akiyoshi}, \citenamefont {Kaneko}, \citenamefont
  {Kussell},\ and\ \citenamefont {Wakamoto}}]{Hashimoto2016}%
  \BibitemOpen
  \bibfield  {author} {\bibinfo {author} {\bibfnamefont {M.}~\bibnamefont
  {Hashimoto}}, \bibinfo {author} {\bibfnamefont {T.}~\bibnamefont {Nozoe}},
  \bibinfo {author} {\bibfnamefont {H.}~\bibnamefont {Nakaoka}}, \bibinfo
  {author} {\bibfnamefont {R.}~\bibnamefont {Okura}}, \bibinfo {author}
  {\bibfnamefont {S.}~\bibnamefont {Akiyoshi}}, \bibinfo {author}
  {\bibfnamefont {K.}~\bibnamefont {Kaneko}}, \bibinfo {author} {\bibfnamefont
  {E.}~\bibnamefont {Kussell}},\ and\ \bibinfo {author} {\bibfnamefont
  {Y.}~\bibnamefont {Wakamoto}},\ }\href
  {https://doi.org/10.1073/pnas.1519412113} {\bibfield  {journal} {\bibinfo
  {journal} {Proc. Natl. Acad. Sci.}\ }\textbf {\bibinfo {volume} {113}},\
  \bibinfo {pages} {3251} (\bibinfo {year} {2016})}\BibitemShut {NoStop}%
\bibitem [{\citenamefont {Nakaoka}\ and\ \citenamefont
  {Wakamoto}(2017)}]{Nakaoka2017a}%
  \BibitemOpen
  \bibfield  {author} {\bibinfo {author} {\bibfnamefont {H.}~\bibnamefont
  {Nakaoka}}\ and\ \bibinfo {author} {\bibfnamefont {Y.}~\bibnamefont
  {Wakamoto}},\ }\href {https://doi.org/10.1371/journal.pbio.2001109}
  {\bibfield  {journal} {\bibinfo  {journal} {PLoS Biol.}\ }\textbf {\bibinfo
  {volume} {15}},\ \bibinfo {pages} {e2001109} (\bibinfo {year}
  {2017})}\BibitemShut {NoStop}%
\bibitem [{\citenamefont {Amir}(2014)}]{Amir2014a}%
  \BibitemOpen
  \bibfield  {author} {\bibinfo {author} {\bibfnamefont {A.}~\bibnamefont
  {Amir}},\ }\href {https://doi.org/10.1103/PhysRevLett.112.208102} {\bibfield
  {journal} {\bibinfo  {journal} {Phys. Rev. Lett.}\ }\textbf {\bibinfo
  {volume} {112}},\ \bibinfo {pages} {208102} (\bibinfo {year}
  {2014})}\BibitemShut {NoStop}%
\bibitem [{Note1()}]{Note1}%
  \BibitemOpen
  \bibinfo {note} {Supplemental Material at [URL will be inserted by
  publisher]}\BibitemShut {NoStop}%
\bibitem [{\citenamefont {Lebowitz}\ and\ \citenamefont
  {Rubinow}(1974)}]{Lebowitz1974}%
  \BibitemOpen
  \bibfield  {author} {\bibinfo {author} {\bibfnamefont {J.~L.}\ \bibnamefont
  {Lebowitz}}\ and\ \bibinfo {author} {\bibfnamefont {S.~I.}\ \bibnamefont
  {Rubinow}},\ }\href@noop {} {\bibfield  {journal} {\bibinfo  {journal} {J.
  Math. Biol.}\ }\textbf {\bibinfo {volume} {1}},\ \bibinfo {pages} {17}
  (\bibinfo {year} {1974})}\BibitemShut {NoStop}%
\bibitem [{\citenamefont {Charlesworth}(1994)}]{Charlesworth1994}%
  \BibitemOpen
  \bibfield  {author} {\bibinfo {author} {\bibfnamefont {B.}~\bibnamefont
  {Charlesworth}},\ }\href {https://doi.org/10.1017/CBO9780511525711} {\emph
  {\bibinfo {title} {{Evolution in Age-Structured Populations}}}},\ \bibinfo
  {edition} {2nd}\ ed.,\ Cambridge Studies in Mathematical Biology\ (\bibinfo
  {publisher} {Cambridge University Press},\ \bibinfo {year}
  {1994})\BibitemShut {NoStop}%
\bibitem [{\citenamefont {Eigen}\ \emph {et~al.}(1989)\citenamefont {Eigen},
  \citenamefont {McCaskill},\ and\ \citenamefont {Schuster}}]{Eigen1989}%
  \BibitemOpen
  \bibfield  {author} {\bibinfo {author} {\bibfnamefont {M.}~\bibnamefont
  {Eigen}}, \bibinfo {author} {\bibfnamefont {J.}~\bibnamefont {McCaskill}},\
  and\ \bibinfo {author} {\bibfnamefont {P.}~\bibnamefont {Schuster}},\
  }\href@noop {} {\bibfield  {journal} {\bibinfo  {journal} {Adv. Chem. Phys}\
  }\textbf {\bibinfo {volume} {75}},\ \bibinfo {pages} {149} (\bibinfo {year}
  {1989})}\BibitemShut {NoStop}%
\bibitem [{\citenamefont {Hermisson}\ \emph {et~al.}(2002)\citenamefont
  {Hermisson}, \citenamefont {Redner}, \citenamefont {Wagner},\ and\
  \citenamefont {Baake}}]{Hermisson2002}%
  \BibitemOpen
  \bibfield  {author} {\bibinfo {author} {\bibfnamefont {J.}~\bibnamefont
  {Hermisson}}, \bibinfo {author} {\bibfnamefont {O.}~\bibnamefont {Redner}},
  \bibinfo {author} {\bibfnamefont {H.}~\bibnamefont {Wagner}},\ and\ \bibinfo
  {author} {\bibfnamefont {E.}~\bibnamefont {Baake}},\ }\href
  {https://doi.org/10.1006/tpbi.2002.1582} {\bibfield  {journal} {\bibinfo
  {journal} {Theor. Popul. Biol.}\ }\textbf {\bibinfo {volume} {62}},\ \bibinfo
  {pages} {9} (\bibinfo {year} {2002})}\BibitemShut {NoStop}%
\bibitem [{Note2()}]{Note2}%
  \BibitemOpen
  \bibinfo {note} {IDTs of sibling cells may be correlated, which does not
  change Eq. (\ref {eq:gentimedist-time-evolution}), as long as the number of
  offspring is independent of the parent's IDT.}\BibitemShut {Stop}%
\bibitem [{\citenamefont {Fisher}(1930)}]{Fisher1930}%
  \BibitemOpen
  \bibfield  {author} {\bibinfo {author} {\bibfnamefont {R.}~\bibnamefont
  {Fisher}},\ }\href@noop {} {\emph {\bibinfo {title} {{The Genetical Theory of
  Natural Selection}}}}\ (\bibinfo  {publisher} {Clarendon},\ \bibinfo {year}
  {1930})\BibitemShut {NoStop}%
\bibitem [{\citenamefont {Wakamoto}\ \emph {et~al.}(2012)\citenamefont
  {Wakamoto}, \citenamefont {Grosberg},\ and\ \citenamefont
  {Kussell}}]{Wakamoto2011}%
  \BibitemOpen
  \bibfield  {author} {\bibinfo {author} {\bibfnamefont {Y.}~\bibnamefont
  {Wakamoto}}, \bibinfo {author} {\bibfnamefont {A.~Y.}\ \bibnamefont
  {Grosberg}},\ and\ \bibinfo {author} {\bibfnamefont {E.}~\bibnamefont
  {Kussell}},\ }\href {https://doi.org/10.1111/j.1558-5646.2011.01418.x}
  {\bibfield  {journal} {\bibinfo  {journal} {Evolution (N. Y).}\ }\textbf
  {\bibinfo {volume} {66}},\ \bibinfo {pages} {115} (\bibinfo {year}
  {2012})}\BibitemShut {NoStop}%
\bibitem [{\citenamefont {Jafarpour}(2019)}]{Jafarpour2019}%
  \BibitemOpen
  \bibfield  {author} {\bibinfo {author} {\bibfnamefont {F.}~\bibnamefont
  {Jafarpour}},\ }\href {https://doi.org/10.1103/PhysRevLett.122.118101}
  {\bibfield  {journal} {\bibinfo  {journal} {Phys. Rev. Lett.}\ }\textbf
  {\bibinfo {volume} {122}},\ \bibinfo {pages} {118101} (\bibinfo {year}
  {2019})},\ \Eprint {https://arxiv.org/abs/1809.10217} {arXiv:1809.10217}
  \BibitemShut {NoStop}%
\bibitem [{\citenamefont {Eigen}(1971)}]{Eigen1971}%
  \BibitemOpen
  \bibfield  {author} {\bibinfo {author} {\bibfnamefont {M.}~\bibnamefont
  {Eigen}},\ }\href {https://doi.org/10.1007/BF00623322} {\bibfield  {journal}
  {\bibinfo  {journal} {Naturwissenschaften}\ }\textbf {\bibinfo {volume}
  {58}},\ \bibinfo {pages} {465} (\bibinfo {year} {1971})}\BibitemShut
  {NoStop}%
\bibitem [{\citenamefont {Kingman}(1978)}]{Kingman1978}%
  \BibitemOpen
  \bibfield  {author} {\bibinfo {author} {\bibfnamefont {J.~F.~C.}\
  \bibnamefont {Kingman}},\ }\href {https://doi.org/10.1017/S0021900200105534}
  {\bibfield  {journal} {\bibinfo  {journal} {J. Appl. Probab.}\ }\textbf
  {\bibinfo {volume} {15}},\ \bibinfo {pages} {1} (\bibinfo {year}
  {1978})}\BibitemShut {NoStop}%
\bibitem [{\citenamefont {B{\"{u}}rger}\ and\ \citenamefont
  {Bomze}(1996)}]{Burger1996}%
  \BibitemOpen
  \bibfield  {author} {\bibinfo {author} {\bibfnamefont {R.}~\bibnamefont
  {B{\"{u}}rger}}\ and\ \bibinfo {author} {\bibfnamefont {I.~M.}\ \bibnamefont
  {Bomze}},\ }\href {https://doi.org/10.2307/1427919} {\bibfield  {journal}
  {\bibinfo  {journal} {Adv. Appl. Probab.}\ }\textbf {\bibinfo {volume}
  {28}},\ \bibinfo {pages} {227} (\bibinfo {year} {1996})}\BibitemShut
  {NoStop}%
\bibitem [{\citenamefont {Wallden}\ and\ \citenamefont
  {Elf}(2011)}]{Wallden2011}%
  \BibitemOpen
  \bibfield  {author} {\bibinfo {author} {\bibfnamefont {M.}~\bibnamefont
  {Wallden}}\ and\ \bibinfo {author} {\bibfnamefont {J.}~\bibnamefont {Elf}},\
  }\href {https://doi.org/10.1016/j.copbio.2010.10.004} {\bibfield  {journal}
  {\bibinfo  {journal} {Curr. Opin. Biotechnol.}\ }\textbf {\bibinfo {volume}
  {22}},\ \bibinfo {pages} {81} (\bibinfo {year} {2011})}\BibitemShut {NoStop}%
\bibitem [{\citenamefont {Lambert}\ and\ \citenamefont
  {Kussell}(2014)}]{Lambert2014}%
  \BibitemOpen
  \bibfield  {author} {\bibinfo {author} {\bibfnamefont {G.}~\bibnamefont
  {Lambert}}\ and\ \bibinfo {author} {\bibfnamefont {E.}~\bibnamefont
  {Kussell}},\ }\href {https://doi.org/10.1371/journal.pgen.1004556} {\bibfield
   {journal} {\bibinfo  {journal} {PLoS Genet.}\ }\textbf {\bibinfo {volume}
  {10}},\ \bibinfo {pages} {e1004556} (\bibinfo {year} {2014})}\BibitemShut
  {NoStop}%
\bibitem [{\citenamefont {Park}\ and\ \citenamefont {Krug}(2008)}]{Park2008}%
  \BibitemOpen
  \bibfield  {author} {\bibinfo {author} {\bibfnamefont {S.~C.}\ \bibnamefont
  {Park}}\ and\ \bibinfo {author} {\bibfnamefont {J.}~\bibnamefont {Krug}},\
  }\bibfield  {journal} {\bibinfo  {journal} {J. Stat. Mech. Theory Exp.}\
  }\textbf {\bibinfo {volume} {2008}},\ \href
  {https://doi.org/10.1088/1742-5468/2008/04/P04014}
  {10.1088/1742-5468/2008/04/P04014} (\bibinfo {year} {2008}),\ \Eprint
  {https://arxiv.org/abs/0711.1989} {arXiv:0711.1989} \BibitemShut {NoStop}%
\bibitem [{\citenamefont {Brotto}\ \emph {et~al.}(2016)\citenamefont {Brotto},
  \citenamefont {Bunin},\ and\ \citenamefont {Kurchan}}]{Brotto2016}%
  \BibitemOpen
  \bibfield  {author} {\bibinfo {author} {\bibfnamefont {T.}~\bibnamefont
  {Brotto}}, \bibinfo {author} {\bibfnamefont {G.}~\bibnamefont {Bunin}},\ and\
  \bibinfo {author} {\bibfnamefont {J.}~\bibnamefont {Kurchan}},\ }\bibfield
  {journal} {\bibinfo  {journal} {J. Stat. Mech. Theory Exp.}\ }\textbf
  {\bibinfo {volume} {2016}},\ \href
  {https://doi.org/10.1088/1742-5468/2016/03/033302}
  {10.1088/1742-5468/2016/03/033302} (\bibinfo {year} {2016}),\ \Eprint
  {https://arxiv.org/abs/1407.4669} {arXiv:1407.4669} \BibitemShut {NoStop}%
\bibitem [{\citenamefont {Webb}(1986)}]{Webb1986}%
  \BibitemOpen
  \bibfield  {author} {\bibinfo {author} {\bibfnamefont {G.~F.}\ \bibnamefont
  {Webb}},\ }\href {https://doi.org/10.1007/BF00276962} {\bibfield  {journal}
  {\bibinfo  {journal} {J. Math. Biol.}\ }\textbf {\bibinfo {volume} {23}},\
  \bibinfo {pages} {269} (\bibinfo {year} {1986})}\BibitemShut {NoStop}%
\bibitem [{\citenamefont {Boulanouar}(2011)}]{Boulanouar2011}%
  \BibitemOpen
  \bibfield  {author} {\bibinfo {author} {\bibfnamefont {M.}~\bibnamefont
  {Boulanouar}},\ }\href {https://doi.org/10.1007/s00028-011-0100-8} {\bibfield
   {journal} {\bibinfo  {journal} {J. Evol. Equations}\ }\textbf {\bibinfo
  {volume} {11}},\ \bibinfo {pages} {531} (\bibinfo {year} {2011})}\BibitemShut
  {NoStop}%
\end{thebibliography}%


\begin{thebibliography}{24}%
\makeatletter
\providecommand \@ifxundefined [1]{%
 \@ifx{#1\undefined}
}%
\providecommand \@ifnum [1]{%
 \ifnum #1\expandafter \@firstoftwo
 \else \expandafter \@secondoftwo
 \fi
}%
\providecommand \@ifx [1]{%
 \ifx #1\expandafter \@firstoftwo
 \else \expandafter \@secondoftwo
 \fi
}%
\providecommand \natexlab [1]{#1}%
\providecommand \enquote  [1]{``#1''}%
\providecommand \bibnamefont  [1]{#1}%
\providecommand \bibfnamefont [1]{#1}%
\providecommand \citenamefont [1]{#1}%
\providecommand \href@noop [0]{\@secondoftwo}%
\providecommand \href [0]{\begingroup \@sanitize@url \@href}%
\providecommand \@href[1]{\@@startlink{#1}\@@href}%
\providecommand \@@href[1]{\endgroup#1\@@endlink}%
\providecommand \@sanitize@url [0]{\catcode `\\12\catcode `\$12\catcode
  `\&12\catcode `\#12\catcode `\^12\catcode `\_12\catcode `\%12\relax}%
\providecommand \@@startlink[1]{}%
\providecommand \@@endlink[0]{}%
\providecommand \url  [0]{\begingroup\@sanitize@url \@url }%
\providecommand \@url [1]{\endgroup\@href {#1}{\urlprefix }}%
\providecommand \urlprefix  [0]{URL }%
\providecommand \Eprint [0]{\href }%
\providecommand \doibase [0]{https://doi.org/}%
\providecommand \selectlanguage [0]{\@gobble}%
\providecommand \bibinfo  [0]{\@secondoftwo}%
\providecommand \bibfield  [0]{\@secondoftwo}%
\providecommand \translation [1]{[#1]}%
\providecommand \BibitemOpen [0]{}%
\providecommand \bibitemStop [0]{}%
\providecommand \bibitemNoStop [0]{.\EOS\space}%
\providecommand \EOS [0]{\spacefactor3000\relax}%
\providecommand \BibitemShut  [1]{\csname bibitem#1\endcsname}%
\let\auto@bib@innerbib\@empty
\bibitem [{\citenamefont {B{\"{u}}rger}\ and\ \citenamefont
  {Bomze}(1996)}]{Burger1996}%
  \BibitemOpen
  \bibfield  {author} {\bibinfo {author} {\bibfnamefont {R.}~\bibnamefont
  {B{\"{u}}rger}}\ and\ \bibinfo {author} {\bibfnamefont {I.~M.}\ \bibnamefont
  {Bomze}},\ }\href {https://doi.org/10.2307/1427919} {\bibfield  {journal}
  {\bibinfo  {journal} {Adv. Appl. Probab.}\ }\textbf {\bibinfo {volume}
  {28}},\ \bibinfo {pages} {227} (\bibinfo {year} {1996})}\BibitemShut
  {NoStop}%
\bibitem [{\citenamefont {Hermisson}\ \emph {et~al.}(2002)\citenamefont
  {Hermisson}, \citenamefont {Redner}, \citenamefont {Wagner},\ and\
  \citenamefont {Baake}}]{Hermisson2002}%
  \BibitemOpen
  \bibfield  {author} {\bibinfo {author} {\bibfnamefont {J.}~\bibnamefont
  {Hermisson}}, \bibinfo {author} {\bibfnamefont {O.}~\bibnamefont {Redner}},
  \bibinfo {author} {\bibfnamefont {H.}~\bibnamefont {Wagner}},\ and\ \bibinfo
  {author} {\bibfnamefont {E.}~\bibnamefont {Baake}},\ }\href
  {https://doi.org/10.1006/tpbi.2002.1582} {\bibfield  {journal} {\bibinfo
  {journal} {Theor. Popul. Biol.}\ }\textbf {\bibinfo {volume} {62}},\ \bibinfo
  {pages} {9} (\bibinfo {year} {2002})}\BibitemShut {NoStop}%
\bibitem [{\citenamefont {Nozoe}\ \emph {et~al.}(2017)\citenamefont {Nozoe},
  \citenamefont {Kussell},\ and\ \citenamefont {Wakamoto}}]{Nozoe2017}%
  \BibitemOpen
  \bibfield  {author} {\bibinfo {author} {\bibfnamefont {T.}~\bibnamefont
  {Nozoe}}, \bibinfo {author} {\bibfnamefont {E.}~\bibnamefont {Kussell}},\
  and\ \bibinfo {author} {\bibfnamefont {Y.}~\bibnamefont {Wakamoto}},\ }\href
  {https://doi.org/10.1371/journal.pgen.1006653} {\bibfield  {journal}
  {\bibinfo  {journal} {PLOS Genet.}\ }\textbf {\bibinfo {volume} {13}},\
  \bibinfo {pages} {e1006653} (\bibinfo {year} {2017})}\BibitemShut {NoStop}%
\bibitem [{\citenamefont {Mosheiff}\ \emph {et~al.}(2018)\citenamefont
  {Mosheiff}, \citenamefont {Martins}, \citenamefont {Pearl-Mizrahi},
  \citenamefont {Gr{\"{u}}nberger}, \citenamefont {Helfrich}, \citenamefont
  {Mihalcescu}, \citenamefont {Kohlheyer}, \citenamefont {Locke}, \citenamefont
  {Glass},\ and\ \citenamefont {Balaban}}]{Mosheiff2018}%
  \BibitemOpen
  \bibfield  {author} {\bibinfo {author} {\bibfnamefont {N.}~\bibnamefont
  {Mosheiff}}, \bibinfo {author} {\bibfnamefont {B.~M.~C.}\ \bibnamefont
  {Martins}}, \bibinfo {author} {\bibfnamefont {S.}~\bibnamefont
  {Pearl-Mizrahi}}, \bibinfo {author} {\bibfnamefont {A.}~\bibnamefont
  {Gr{\"{u}}nberger}}, \bibinfo {author} {\bibfnamefont {S.}~\bibnamefont
  {Helfrich}}, \bibinfo {author} {\bibfnamefont {I.}~\bibnamefont
  {Mihalcescu}}, \bibinfo {author} {\bibfnamefont {D.}~\bibnamefont
  {Kohlheyer}}, \bibinfo {author} {\bibfnamefont {J.~C.~W.}\ \bibnamefont
  {Locke}}, \bibinfo {author} {\bibfnamefont {L.}~\bibnamefont {Glass}},\ and\
  \bibinfo {author} {\bibfnamefont {N.~Q.}\ \bibnamefont {Balaban}},\ }\href
  {https://doi.org/10.1103/PhysRevX.8.021035} {\bibfield  {journal} {\bibinfo
  {journal} {Phys. Rev. X}\ }\textbf {\bibinfo {volume} {8}},\ \bibinfo {pages}
  {021035} (\bibinfo {year} {2018})}\BibitemShut {NoStop}%
\bibitem [{\citenamefont {Kubitschek}(1962)}]{Kubitschek1962}%
  \BibitemOpen
  \bibfield  {author} {\bibinfo {author} {\bibfnamefont {H.~E.}\ \bibnamefont
  {Kubitschek}},\ }\href {https://doi.org/10.1016/0014-4827(62)90150-7}
  {\bibfield  {journal} {\bibinfo  {journal} {Exp. Cell Res.}\ }\textbf
  {\bibinfo {volume} {26}},\ \bibinfo {pages} {439} (\bibinfo {year}
  {1962})}\BibitemShut {NoStop}%
\bibitem [{\citenamefont {Hashimoto}\ \emph {et~al.}(2016)\citenamefont
  {Hashimoto}, \citenamefont {Nozoe}, \citenamefont {Nakaoka}, \citenamefont
  {Okura}, \citenamefont {Akiyoshi}, \citenamefont {Kaneko}, \citenamefont
  {Kussell},\ and\ \citenamefont {Wakamoto}}]{Hashimoto2016apdx}%
  \BibitemOpen
  \bibfield  {author} {\bibinfo {author} {\bibfnamefont {M.}~\bibnamefont
  {Hashimoto}}, \bibinfo {author} {\bibfnamefont {T.}~\bibnamefont {Nozoe}},
  \bibinfo {author} {\bibfnamefont {H.}~\bibnamefont {Nakaoka}}, \bibinfo
  {author} {\bibfnamefont {R.}~\bibnamefont {Okura}}, \bibinfo {author}
  {\bibfnamefont {S.}~\bibnamefont {Akiyoshi}}, \bibinfo {author}
  {\bibfnamefont {K.}~\bibnamefont {Kaneko}}, \bibinfo {author} {\bibfnamefont
  {E.}~\bibnamefont {Kussell}},\ and\ \bibinfo {author} {\bibfnamefont
  {Y.}~\bibnamefont {Wakamoto}},\ }\href
  {https://doi.org/10.1073/pnas.1519412113} {\bibfield  {journal} {\bibinfo
  {journal} {Proc. Natl. Acad. Sci.}\ }\textbf {\bibinfo {volume} {113}},\
  \bibinfo {pages} {3251} (\bibinfo {year} {2016})}\BibitemShut {NoStop}%
\bibitem [{\citenamefont {Yu}\ \emph {et~al.}(2017)\citenamefont {Yu},
  \citenamefont {Willis}, \citenamefont {Chau}, \citenamefont {Zambon},
  \citenamefont {Horowitz}, \citenamefont {Bhaya}, \citenamefont {Huang},\ and\
  \citenamefont {Quake}}]{Yu2017b}%
  \BibitemOpen
  \bibfield  {author} {\bibinfo {author} {\bibfnamefont {F.~B.}\ \bibnamefont
  {Yu}}, \bibinfo {author} {\bibfnamefont {L.}~\bibnamefont {Willis}}, \bibinfo
  {author} {\bibfnamefont {R.~M.~W.}\ \bibnamefont {Chau}}, \bibinfo {author}
  {\bibfnamefont {A.}~\bibnamefont {Zambon}}, \bibinfo {author} {\bibfnamefont
  {M.}~\bibnamefont {Horowitz}}, \bibinfo {author} {\bibfnamefont
  {D.}~\bibnamefont {Bhaya}}, \bibinfo {author} {\bibfnamefont {K.~C.}\
  \bibnamefont {Huang}},\ and\ \bibinfo {author} {\bibfnamefont {S.~R.}\
  \bibnamefont {Quake}},\ }\href {https://doi.org/10.1186/s12915-016-0344-4}
  {\bibfield  {journal} {\bibinfo  {journal} {BMC Biol.}\ }\textbf {\bibinfo
  {volume} {15}},\ \bibinfo {pages} {1} (\bibinfo {year} {2017})}\BibitemShut
  {NoStop}%
\bibitem [{\citenamefont {Hawkins}\ \emph {et~al.}(2009)\citenamefont
  {Hawkins}, \citenamefont {Markham}, \citenamefont {McGuinness},\ and\
  \citenamefont {Hodgkin}}]{Hawkins2009}%
  \BibitemOpen
  \bibfield  {author} {\bibinfo {author} {\bibfnamefont {E.~D.}\ \bibnamefont
  {Hawkins}}, \bibinfo {author} {\bibfnamefont {J.~F.}\ \bibnamefont
  {Markham}}, \bibinfo {author} {\bibfnamefont {L.~P.}\ \bibnamefont
  {McGuinness}},\ and\ \bibinfo {author} {\bibfnamefont {P.~D.}\ \bibnamefont
  {Hodgkin}},\ }\href {https://doi.org/10.1073/pnas.0905629106} {\bibfield
  {journal} {\bibinfo  {journal} {Proc. Natl. Acad. Sci.}\ }\textbf {\bibinfo
  {volume} {106}},\ \bibinfo {pages} {13457} (\bibinfo {year}
  {2009})}\BibitemShut {NoStop}%
\bibitem [{\citenamefont {Sasaki}\ \emph {et~al.}(1977)\citenamefont {Sasaki},
  \citenamefont {Yoshinaga},\ and\ \citenamefont {Kawano}}]{Sasaki1977}%
  \BibitemOpen
  \bibfield  {author} {\bibinfo {author} {\bibfnamefont {H.}~\bibnamefont
  {Sasaki}}, \bibinfo {author} {\bibfnamefont {H.}~\bibnamefont {Yoshinaga}},\
  and\ \bibinfo {author} {\bibfnamefont {K.}~\bibnamefont {Kawano}},\ }\href
  {https://doi.org/10.2307/3574706} {\bibfield  {journal} {\bibinfo  {journal}
  {Radiat. Res.}\ }\textbf {\bibinfo {volume} {72}},\ \bibinfo {pages} {364}
  (\bibinfo {year} {1977})}\BibitemShut {NoStop}%
\bibitem [{\citenamefont {Staudte}\ \emph {et~al.}(1984)\citenamefont
  {Staudte}, \citenamefont {Guiguet},\ and\ \citenamefont
  {D'Hooghe}}]{Staudte1984}%
  \BibitemOpen
  \bibfield  {author} {\bibinfo {author} {\bibfnamefont {R.~G.}\ \bibnamefont
  {Staudte}}, \bibinfo {author} {\bibfnamefont {M.}~\bibnamefont {Guiguet}},\
  and\ \bibinfo {author} {\bibfnamefont {M.~C.}\ \bibnamefont {D'Hooghe}},\
  }\href {https://doi.org/10.1016/S0022-5193(84)80115-0} {\bibfield  {journal}
  {\bibinfo  {journal} {J. Theor. Biol.}\ }\textbf {\bibinfo {volume} {109}},\
  \bibinfo {pages} {127} (\bibinfo {year} {1984})}\BibitemShut {NoStop}%
\bibitem [{\citenamefont {Hola}\ and\ \citenamefont {Riley}(1987)}]{Hola1987}%
  \BibitemOpen
  \bibfield  {author} {\bibinfo {author} {\bibfnamefont {M.}~\bibnamefont
  {Hola}}\ and\ \bibinfo {author} {\bibfnamefont {P.~A.}\ \bibnamefont
  {Riley}},\ }\href@noop {} {\bibfield  {journal} {\bibinfo  {journal} {J. Cell
  Sci.}\ }\textbf {\bibinfo {volume} {88 ( Pt 1)}},\ \bibinfo {pages} {73}
  (\bibinfo {year} {1987})}\BibitemShut {NoStop}%
\bibitem [{\citenamefont {Chakrabarti}\ \emph {et~al.}(2018)\citenamefont
  {Chakrabarti}, \citenamefont {Paek}, \citenamefont {Reyes}, \citenamefont
  {Lasick}, \citenamefont {Lahav},\ and\ \citenamefont
  {Michor}}]{Chakrabarti2018}%
  \BibitemOpen
  \bibfield  {author} {\bibinfo {author} {\bibfnamefont {S.}~\bibnamefont
  {Chakrabarti}}, \bibinfo {author} {\bibfnamefont {A.~L.}\ \bibnamefont
  {Paek}}, \bibinfo {author} {\bibfnamefont {J.}~\bibnamefont {Reyes}},
  \bibinfo {author} {\bibfnamefont {K.~A.}\ \bibnamefont {Lasick}}, \bibinfo
  {author} {\bibfnamefont {G.}~\bibnamefont {Lahav}},\ and\ \bibinfo {author}
  {\bibfnamefont {F.}~\bibnamefont {Michor}},\ }\href
  {https://doi.org/10.1038/s41467-018-07788-5} {\bibfield  {journal} {\bibinfo
  {journal} {Nat. Commun.}\ }\textbf {\bibinfo {volume} {9}},\ \bibinfo {pages}
  {1} (\bibinfo {year} {2018})}\BibitemShut {NoStop}%
\bibitem [{\citenamefont {Froese}(1964)}]{Froese1964}%
  \BibitemOpen
  \bibfield  {author} {\bibinfo {author} {\bibfnamefont {G.}~\bibnamefont
  {Froese}},\ }\href {https://doi.org/10.1016/0014-4827(64)90108-9} {\bibfield
  {journal} {\bibinfo  {journal} {Exp. Cell Res.}\ }\textbf {\bibinfo {volume}
  {35}},\ \bibinfo {pages} {415} (\bibinfo {year} {1964})}\BibitemShut
  {NoStop}%
\bibitem [{\citenamefont {{Van Wijk}}\ and\ \citenamefont {{Van De
  Poll}}(1979)}]{VanWijk1979}%
  \BibitemOpen
  \bibfield  {author} {\bibinfo {author} {\bibfnamefont {R.}~\bibnamefont {{Van
  Wijk}}}\ and\ \bibinfo {author} {\bibfnamefont {K.~W.}\ \bibnamefont {{Van De
  Poll}}},\ }\href {https://doi.org/10.1111/j.1365-2184.1979.tb00184.x}
  {\bibfield  {journal} {\bibinfo  {journal} {Cell Prolif.}\ }\textbf {\bibinfo
  {volume} {12}},\ \bibinfo {pages} {659} (\bibinfo {year} {1979})}\BibitemShut
  {NoStop}%
\bibitem [{\citenamefont {Absher}\ \emph {et~al.}(1983)\citenamefont {Absher},
  \citenamefont {Sylwester},\ and\ \citenamefont {Hart}}]{Absher1983}%
  \BibitemOpen
  \bibfield  {author} {\bibinfo {author} {\bibfnamefont {M.}~\bibnamefont
  {Absher}}, \bibinfo {author} {\bibfnamefont {D.}~\bibnamefont {Sylwester}},\
  and\ \bibinfo {author} {\bibfnamefont {B.~A.}\ \bibnamefont {Hart}},\ }\href
  {https://doi.org/https://doi.org/10.1016/0013-9351(83)90163-9} {\bibfield
  {journal} {\bibinfo  {journal} {Environ. Res.}\ }\textbf {\bibinfo {volume}
  {30}},\ \bibinfo {pages} {34} (\bibinfo {year} {1983})}\BibitemShut {NoStop}%
\bibitem [{\citenamefont {Absher}\ and\ \citenamefont
  {Cristofalo}(1984)}]{Absher1984}%
  \BibitemOpen
  \bibfield  {author} {\bibinfo {author} {\bibfnamefont {M.}~\bibnamefont
  {Absher}}\ and\ \bibinfo {author} {\bibfnamefont {V.~J.}\ \bibnamefont
  {Cristofalo}},\ }\href {https://doi.org/10.1002/jcp.1041190309} {\bibfield
  {journal} {\bibinfo  {journal} {J. Cell. Physiol.}\ }\textbf {\bibinfo
  {volume} {119}},\ \bibinfo {pages} {315} (\bibinfo {year}
  {1984})}\BibitemShut {NoStop}%
\bibitem [{\citenamefont {Sandler}\ \emph {et~al.}(2015)\citenamefont
  {Sandler}, \citenamefont {Mizrahi}, \citenamefont {Weiss}, \citenamefont
  {Agam}, \citenamefont {Simon},\ and\ \citenamefont {Balaban}}]{Sandler2015}%
  \BibitemOpen
  \bibfield  {author} {\bibinfo {author} {\bibfnamefont {O.}~\bibnamefont
  {Sandler}}, \bibinfo {author} {\bibfnamefont {S.~P.}\ \bibnamefont
  {Mizrahi}}, \bibinfo {author} {\bibfnamefont {N.}~\bibnamefont {Weiss}},
  \bibinfo {author} {\bibfnamefont {O.}~\bibnamefont {Agam}}, \bibinfo {author}
  {\bibfnamefont {I.}~\bibnamefont {Simon}},\ and\ \bibinfo {author}
  {\bibfnamefont {N.~Q.}\ \bibnamefont {Balaban}},\ }\href
  {https://doi.org/10.1038/nature14318} {\bibfield  {journal} {\bibinfo
  {journal} {Nature}\ }\textbf {\bibinfo {volume} {519}},\ \bibinfo {pages}
  {468} (\bibinfo {year} {2015})}\BibitemShut {NoStop}%
\bibitem [{\citenamefont {Miyamoto}\ \emph {et~al.}(1973)\citenamefont
  {Miyamoto}, \citenamefont {Zeuthen},\ and\ \citenamefont
  {Rasmussen}}]{Miyamoto1973}%
  \BibitemOpen
  \bibfield  {author} {\bibinfo {author} {\bibfnamefont {H.}~\bibnamefont
  {Miyamoto}}, \bibinfo {author} {\bibfnamefont {E.}~\bibnamefont {Zeuthen}},\
  and\ \bibinfo {author} {\bibfnamefont {L.}~\bibnamefont {Rasmussen}},\
  }\href@noop {} {\bibfield  {journal} {\bibinfo  {journal} {J. Cell Sci.}\
  }\textbf {\bibinfo {volume} {13}},\ \bibinfo {pages} {879} (\bibinfo {year}
  {1973})}\BibitemShut {NoStop}%
\bibitem [{\citenamefont {van Meeteren}\ \emph {et~al.}(1984)\citenamefont {van
  Meeteren}, \citenamefont {van Wijk}, \citenamefont {Stap},\ and\
  \citenamefont {Deys}}]{VanMeeteren1984}%
  \BibitemOpen
  \bibfield  {author} {\bibinfo {author} {\bibfnamefont {A.}~\bibnamefont {van
  Meeteren}}, \bibinfo {author} {\bibfnamefont {R.}~\bibnamefont {van Wijk}},
  \bibinfo {author} {\bibfnamefont {J.}~\bibnamefont {Stap}},\ and\ \bibinfo
  {author} {\bibfnamefont {B.~F.}\ \bibnamefont {Deys}},\ }\href
  {https://doi.org/10.1111/j.1365-2184.1984.tb00573.x} {\bibfield  {journal}
  {\bibinfo  {journal} {Cell Prolif.}\ }\textbf {\bibinfo {volume} {17}},\
  \bibinfo {pages} {105} (\bibinfo {year} {1984})}\BibitemShut {NoStop}%
\bibitem [{\citenamefont {Grasman}(1990)}]{Grasman1990}%
  \BibitemOpen
  \bibfield  {author} {\bibinfo {author} {\bibfnamefont {J.}~\bibnamefont
  {Grasman}},\ }\href {https://doi.org/10.1007/BF02462266} {\bibfield
  {journal} {\bibinfo  {journal} {Bull. Math. Biol.}\ }\textbf {\bibinfo
  {volume} {52}},\ \bibinfo {pages} {535} (\bibinfo {year} {1990})}\BibitemShut
  {NoStop}%
\bibitem [{\citenamefont {Shields}\ and\ \citenamefont
  {Smith}(1977)}]{Shields1977}%
  \BibitemOpen
  \bibfield  {author} {\bibinfo {author} {\bibfnamefont {R.}~\bibnamefont
  {Shields}}\ and\ \bibinfo {author} {\bibfnamefont {J.~A.}\ \bibnamefont
  {Smith}},\ }\href {https://doi.org/10.1002/jcp.1040910304} {\bibfield
  {journal} {\bibinfo  {journal} {J. Cell. Physiol.}\ }\textbf {\bibinfo
  {volume} {91}},\ \bibinfo {pages} {345} (\bibinfo {year} {1977})}\BibitemShut
  {NoStop}%
\bibitem [{\citenamefont {Staudte}\ \emph {et~al.}(1997)\citenamefont
  {Staudte}, \citenamefont {Huggins}, \citenamefont {Zhang}, \citenamefont
  {Axelrod},\ and\ \citenamefont {Kimmel}}]{Staudte1997}%
  \BibitemOpen
  \bibfield  {author} {\bibinfo {author} {\bibfnamefont {R.~G.}\ \bibnamefont
  {Staudte}}, \bibinfo {author} {\bibfnamefont {R.~M.}\ \bibnamefont
  {Huggins}}, \bibinfo {author} {\bibfnamefont {J.}~\bibnamefont {Zhang}},
  \bibinfo {author} {\bibfnamefont {D.~E.}\ \bibnamefont {Axelrod}},\ and\
  \bibinfo {author} {\bibfnamefont {M.}~\bibnamefont {Kimmel}},\ }\href
  {https://doi.org/10.1016/S0025-5564(97)00006-0} {\bibfield  {journal}
  {\bibinfo  {journal} {Math. Biosci.}\ }\textbf {\bibinfo {volume} {143}},\
  \bibinfo {pages} {103} (\bibinfo {year} {1997})}\BibitemShut {NoStop}%
\bibitem [{\citenamefont {Sennerstam}\ and\ \citenamefont
  {Str{\"{o}}mberg}(1988)}]{Sennerstam1988}%
  \BibitemOpen
  \bibfield  {author} {\bibinfo {author} {\bibfnamefont {R.}~\bibnamefont
  {Sennerstam}}\ and\ \bibinfo {author} {\bibfnamefont {J.~O.}\ \bibnamefont
  {Str{\"{o}}mberg}},\ }\href {https://doi.org/10.1016/S0022-5193(88)80232-7}
  {\bibfield  {journal} {\bibinfo  {journal} {J. Theor. Biol.}\ }\textbf
  {\bibinfo {volume} {131}},\ \bibinfo {pages} {151} (\bibinfo {year}
  {1988})}\BibitemShut {NoStop}%
\bibitem [{\citenamefont {Dawson}\ \emph {et~al.}(1965)\citenamefont {Dawson},
  \citenamefont {Madoc-Jones},\ and\ \citenamefont {Field}}]{Dawson1965}%
  \BibitemOpen
  \bibfield  {author} {\bibinfo {author} {\bibfnamefont {K.~B.}\ \bibnamefont
  {Dawson}}, \bibinfo {author} {\bibfnamefont {H.}~\bibnamefont
  {Madoc-Jones}},\ and\ \bibinfo {author} {\bibfnamefont {E.~O.}\ \bibnamefont
  {Field}},\ }\href
  {https://doi.org/https://doi.org/10.1016/0014-4827(65)90429-5} {\bibfield
  {journal} {\bibinfo  {journal} {Exp. Cell Res.}\ }\textbf {\bibinfo {volume}
  {38}},\ \bibinfo {pages} {75} (\bibinfo {year} {1965})}\BibitemShut {NoStop}%
\end{thebibliography}%

\end{document}


\title{Supplemental Material: Cell cycle heritability and localization phase transition in growing populations}

\author{Takashi Nozoe$^{1}$}
\author{Edo Kussell$^{1,2}$}
\affiliation{$^1$Department of Biology, New York University, 12 Waverly Place, New York, NY 10003. \\
$^2$Department of Physics, New York University, 726 Broadway, New York, NY 10003. }

\date{\today}

\maketitle
\tableofcontents 
\listoftables
\listoffigures

\newpage

\section{General stationary solutions \label{app:age-idt-model}}

Let $n\left(a,\tau;t\right)da \, d\tau$ be the expected number of cells with age $a$ and IDT $\tau$ at time $t$. $a$ is the time since division, which is reset to $0$ just after division ($0\le a<\tau$).
In particular, $n_{\mathrm{div}}\left(\tau;t\right)d\tau dt:=\lim_{a\rightarrow\tau}n\left(a,\tau;t\right)d\tau dt$  is the expected number of dividing cells at age $\tau$ during the infinitesimal interval $[t,t+dt]$.
For fixed $\tau$, $n\left(a,\tau;t\right)$ satisfies the partial
differential equation:
\begin{equation}
\frac{\partial n\left(a,\tau;t\right)}{\partial a}+\frac{\partial n\left(a,\tau;t\right)}{\partial t}=0\label{eq:PDE-age-idt-model}
\end{equation}
We denote the initial distribution of $a$ and $\tau$ by $n\left(a,\tau;0\right)=n_{0}\left(a,\tau\right)$.
Then the solution of Eq. \eqref{eq:PDE-age-idt-model} satisfies
\begin{equation}
n\left(a,\tau;t\right)=\begin{cases}
n\left(0,\tau ; t-a\right) & t > a\\
n_{0}\left(a-t,\tau\right) & t \le a
\end{cases}\label{eq:PDE-age-idt-model-solution}
\end{equation}
Let $K\left(\tau,\tau^{\prime}\right)$ denote the transition probability
density from $\tau^{\prime}$ to $\tau$. The boundary condition is
given by
\begin{equation}
n\left(0,\tau;t\right)=z\int_{0}^{\infty}K\left(\tau,\tau^{\prime}\right)n_{\mathrm{div}}\left(\tau^{\prime};t\right)d\tau^{\prime}\label{eq:PDE-age-idt-model-BC}
\end{equation}
Let 
\begin{equation}
N_{t}:=\int_{0}^{\infty} \int_{0}^{\tau}n\left(a,\tau;t\right)da \, d\tau
\end{equation}
denote the expected population size at time $t$ (relative to the initial population
size) and 
\begin{equation}
p\left(a,\tau;t\right):=n\left(a,\tau;t\right)\cdot N_{t}^{-1}
\end{equation}
denote the probability density of $(a,\tau)$ at time $t$. The marginal
distribution of age is denoted by 
\begin{equation}
p_{\text{age}}\left(a;t\right):=\int_{a}^{\infty}p\left(a,\tau;t\right)d\tau \ .
\end{equation}
By integration of Eq. \eqref{eq:PDE-age-idt-model-BC}, we have 
\begin{equation}
\int_{0}^{\infty}n\left(0,\tau;t\right)d\tau=z\int_{0}^{\infty}n_{\mathrm{div}}\left(\tau;t\right)d\tau\label{eq:balance-nb-division}
\end{equation}
This equation represents the balance between numbers of newly formed
cells and dividing cells. Thus the exponential growth rate at time
$t$, denoted by $\Lambda_{t}$, is
\begin{align}
\Lambda_{t} & :=\frac{d}{dt}\ln N_{t}=N_{t}^{-1}\int_{0}^{\infty}\left(\int_{0}^{\tau}\frac{\partial n\left(a,\tau;t\right)}{\partial t}da\right)d\tau\nonumber \\
 & =-N_{t}^{-1}\int_{0}^{\infty}\left(\int_{0}^{\tau}\frac{\partial n\left(a,\tau;t\right)}{\partial a}da\right)d\tau\nonumber \\
 & =N_{t}^{-1}\int_{0}^{\infty}\left(n\left(0,\tau;t\right)-n_{\mathrm{div}}\left(\tau;t\right)\right)d\tau\nonumber \\
 & =\left(1-z^{-1}\right)p_{\text{age}}\left(0;t\right)
\end{align}
This definition of time-dependent growth rate is equivalent to its
definition in Main Text ($\Lambda_{t}:=\sfrac{\left(z-1\right)N_{\text{div}}\left(t\right)}{N\left(t\right)}$)
because $p_{\text{age}}\left(0;t\right)=\sfrac{zN_{\text{div}}\left(t\right)}{N\left(t\right)}$.

Three interelated distributions of IDTs can be defined, as follows:  (i) the offspring's IDT distribution, 
\begin{equation}
p_{\mathrm{birth}}\left(\tau;t\right):=\frac{p\left(0,\tau;t\right)}{p_{\text{age}}\left(0;t\right)}=\frac{1-z^{-1}}{\Lambda_{t}}p\left(0,\tau;t\right) \ ,
\end{equation}
which is the probability distribution of $\tau$ conditioned on $a=0$; (ii) the parent's IDT distribution, 
\begin{align}
p_{\mathrm{div}}\left(\tau;t\right) & :=\frac{n_{\mathrm{div}}\left(\tau;t\right)}{\int_{0}^{\infty}n_{\mathrm{div}}\left(\tau;t\right)d\tau}=\frac{z\cdot n_{\mathrm{div}}\left(\tau;t\right)}{N_{t}p_{\text{age}}\left(0;t\right)}\nonumber \\
 & =\begin{cases}
z\frac{\Lambda_{t-\tau}}{\Lambda_{t}}e^{-\int_{t-\tau}^{t}\Lambda_{s}ds}p_{\mathrm{birth}}\left(\tau;t-\tau\right) & \tau<t\\
z\frac{\Lambda_{0}}{\Lambda_{t}}e^{-\int_{0}^{t}\Lambda_{s}ds} p\left(\tau-t,\tau;0\right) & \tau\ge t
\end{cases} \ ,
\end{align}
which is the probability distribution of age among dividing cells at
time $t$; and (iii) the population's IDT distribution, 
\begin{equation}
p_{\text{idt}}\left(\tau;t\right):=\int_{0}^{\tau}p\left(a,\tau;t\right)da \ ,
\end{equation}
i.e. the marginal of $p\left(a,\tau;t\right)$ with respect to $\tau$, which is 
the probability distribution of $\tau$ at time $t$.

The joint distribution of $\tau$ and $a$ at time $t$
can be expressed as
\begin{equation}
p\left(a,\tau;t\right)=\begin{cases}
\frac{\Lambda_{t}e^{-\int_{t-a}^{t}\Lambda_{s}ds}}{1-z^{-1}}p_{\mathrm{birth}}\left(\tau;t-a\right) & a<t\\
e^{-\int_{0}^{t}\Lambda_{s}ds}p\left(a-t,\tau;0\right) & a\ge t
\end{cases}
\end{equation}
Stationary distributions are found by substituting $p\left(a,\tau;t\right)=p\left(a,\tau\right)$,
which is independent of $t$. Accordingly, we denote $p_{\text{age}}\left(a\right):=\int_{a}^{\infty}p\left(a,\tau\right)d\tau$,
$p_{\text{idt}}\left(\tau\right):=\int_{0}^{\tau}p\left(a,\tau\right)da$,
$p_{\mathrm{birth}}\left(\tau\right):=p\left(0,\tau\right)/p_{\text{age}}\left(0\right)$
and $p_{\mathrm{div}}\left(\tau\right):=p\left(\tau,\tau\right)/\int_{0}^{\infty}p\left(\tau^{\prime},\tau^{\prime}\right)d\tau^{\prime}$.
Let $\Lambda:=\left(1-z^{-1}\right)p_{\text{age}}\left(0\right)$
denote the steady-state population growth rate. Then we formally obtain
\begin{align}
p\left(a,\tau\right) & =e^{-\Lambda a}p\left(0,\tau\right)\nonumber \\
 & =\frac{\Lambda}{1-z^{-1}}e^{-\Lambda a}p_{\text{birth}}\left(\tau\right)\label{eq:pst_pbirth_relation}
\end{align}
\begin{equation}
p_{\text{age}}\left(a\right)=\frac{\Lambda}{1-z^{-1}}e^{-\Lambda a}\int_{a}^{\infty}p_{\text{birth}}\left(\tau\right)d\tau
\end{equation}
\begin{equation}
p_{\text{div}}\left(\tau\right)=ze^{-\Lambda\tau}p_{\text{birth}}\left(\tau\right)\label{eq:pdiv_pbirth_relation}
\end{equation}
and
\begin{equation}
p_{\text{idt}}\left(\tau\right) =\frac{1-e^{-\Lambda\tau}}{1-z^{-1}}p_{\text{birth}}\left(\tau\right) \ ,
\end{equation}
where $\Lambda$ and $p_{\text{birth}}\left(\tau\right)$ are determined as the solutions of
\begin{equation}
p_{\text{birth}}\left(\tau\right)=z\int_{0}^{\infty}K\left(\tau,\tau^{\prime}\right)e^{-\Lambda\tau^{\prime}}p_{\text{birth}}\left(\tau^{\prime}\right)d\tau^{\prime} \ ,
\end{equation}
or equivalently, 
\begin{equation}
p_{\mathrm{div}}\left(\tau\right)=ze^{-\Lambda\tau}\int_{0}^{\infty}K\left(\tau,\tau^{\prime}\right)p_{\mathrm{div}}\left(\tau^{\prime}\right)d\tau^{\prime} \ .
\end{equation}
Using the above, we obtain the relation
\begin{equation}
p_{\text{birth}}\left(\tau\right)  =\frac{z-1}{z}p_{\text{idt}}\left(\tau\right) + \frac{1}{z}p_{\text{div}}\left(\tau\right) \ .
\end{equation}
Eq. \eqref{eq:PDE-age-idt-model-BC} becomes at stationary state
\begin{equation}
p_{\text{birth}}\left(\tau\right)=\int_{0}^{\infty}K\left(\tau,\tau^{\prime}\right)p_{\text{div}}\left(\tau^{\prime}\right)d\tau^{\prime} \ .
\end{equation}

\section{$\beta$ as correlation coefficient\label{app:beta-as-correlation}}

The covariance between mother and daughter IDTs of
isolated single cells at steady state is given by
\begin{align}
\int_{0}^{\infty}d\tau\int_{\tau_{0}}^{\infty}d\tau^{\prime}\tau\tau^{\prime}K\left(\tau,\tau^{\prime}\right)k\left(\tau^{\prime}\right)-\left(\int_{0}^{\infty}\tau k\left(\tau\right)d\tau\right)^{2}\nonumber \\
= \beta\left(\int_{0}^{\infty}\tau^{2}k\left(\tau\right)d\tau-\left(\int_{0}^{\infty}\tau k\left(\tau\right)d\tau\right)^{2}\right)
\end{align}
Thus, we find that $\beta$ is the Pearson correlation coefficient.

\section{Stationary solution in the Lebowitz-Rubinow model \label{app:Derivations-Stationary-LR}}

In this section, we derive the stationary distribution for $p_{\textrm{div}}$ given in the main text
Eq. (8), 
and determine the conditions in which a heritability threshold exists.

Let $\left[\tau_{0},\tau_{1}\right]$ be the support of the IDT distribution $k(\tau)$ 
\begin{equation}
\supp(k) := \overline{\left\{\tau\in\R\vert k(\tau)\neq0\right\}} = \left[\tau_{0},\tau_{1}\right].
\end{equation}
Here, we consider the general case $0\le\tau_{0}<\tau_{1}\le\infty$, while in the main text and throughout the other sections of the Supplement, we present results for $\tau_0 > 0$ and $\tau_{1} = \infty$. We recall that the function $k\left(\tau\right)$ is non-negative and normalized such that $\int_{\tau_{0}}^{\tau_{1}} k\left(\tau\right)d\tau=1$; 
the parameter $z$ denotes the average number of offspring ($0<z<\infty$); and  $\beta$ denotes the IDT heritability ($0\le\beta<1$).

The stationary distribution $p_{\textrm{div}}$ is a probability measure on the interval $\left[\tau_{0},\tau_{1}\right]$ (see Sec. \ref{subsec:stat-sol-above-betac} for details on the measure-theoretic formulation).  We wish to determine the range of $\beta$ for which $p_{\textrm{div}}$ can be expressed by a probability density function $p_{\textrm{div}}(\tau)$; in such cases 
substituting 
Eq. (1) in Eq. (3), we obtain
\begin{equation}
p_{\textrm{div}}\left(\tau\right) 
=\frac{\left(1-\beta\right)zk\left(\tau\right)}{e^{\Lambda\tau}-\beta z},
\label{smeq:stat-dist-below}
\end{equation}
and $\Lambda$ is determined by the normalization condition
\begin{equation}
\int_{\tau_{0}}^{\tau_{1}}p_{\textrm{div}} \left(\tau\right)d\tau = 
\int_{\tau_{0}}^{\tau_{1}} \frac{\left(1-\beta\right)zk\left(\tau\right)}{e^{\Lambda\tau}-\beta z}d\tau= 1 \ .
\end{equation}
If the above normalization condition does not admit a solution $\Lambda < \infty$ such that the integrand is a non-negative function of $\tau$, the measure $p_{\textrm{div}}$ is not equivalent to a density function (formally, it is not absolutely continuous with respect to Lebesgue measure), and, as we will show below, the measure contains a singular portion corresponding to a delta function (i.e. a Dirac measure).  

To determine the conditions such that $p_{\textrm{div}}$ is given by a density, we define 
\begin{equation}
q_{\beta,z,\lambda}\left(\tau\right):= \frac{\left(1-\beta\right)zk\left(\tau\right)}{e^{\lambda\tau}-\beta z},
\label{smeq:q-density}
\end{equation}
and its integral on $\left[\tau_{0},\tau_{1}\right]$,
\begin{equation}
Q\left(\beta,z,\lambda\right):=\int_{\tau_{0}}^{\tau_{1}}q_{\beta,z,\lambda}\left(\tau\right)d\tau.
\label{smeq:Q-function}
\end{equation}
For fixed $\beta$, $z$, and $k(\tau)$, $p_{\textrm{div}}$ is given by a density if and only if there exists $\lambda < \infty$ for which the following conditions both hold:
\begin{condition}
$q_{\beta,z, \lambda}\left(\tau\right) \ge 0$ for any $\tau\in[\tau_{0}, \tau_{1}]$.
\label{cond:cond1}
\end{condition}
\begin{condition}
$Q\left(\beta,z, \lambda\right)=1$.
\label{cond:cond2}
\end{condition}
As we will see below, condition \ref{cond:cond2}  may not be satisfied even though condition \ref{cond:cond1} holds for a range of $\lambda$.

We characterize the range of $\beta$ for which $p_{\textrm{div}}$ has a density by the value $\beta_{c}$, which is defined as
the supremum over the set of $\beta$ for which both conditions can be satisfied for some $\lambda < \infty$, given $z$ and $k(\tau)$ are fixed; that is, 
\begin{align}
\beta_{c} :=  \sup \{ \beta : 0 < &\beta < 1 \mathrm{\ and\ } \exists \ \lambda < \infty \label{eq:supp_beta_c} \\ 
& \mathrm{\ s. t.\ conditions\ 1 \ and\ 2\ both\ hold} \} , \nonumber
\end{align}
or $\beta_{c} := 0$ if the above set is empty.

When $\beta_{c}=1$, $p_{\textrm{div}}$ is given by a density for all values of $\beta$. If $\beta_c < 1$, then the system undergoes a phase transition as $\beta$ increases past $\beta_c$, such that for $\beta > \beta_c$, $p_{\textrm{div}}$ is not given by a density, and contains a singular portion that is localized at a single point.  We refer to a system for which $\beta_c < 1$ as one that exhibits a \emph{heritability threshold}; and refer to $p_{\textrm{div}}$ as being \emph{localized} if $\beta > \beta_c$ or \emph{delocalized} if $\beta < \beta_c$.  Below, we show that the behavior of $p_{\textrm{div}}$ at precisely $\beta = \beta_c$ depends on additional details including the form of $k(\tau)$. 
  
In Sec. \ref{subsec:Classification-of-stationary}, we determine the general conditions in which there exists a heritability threshold, i.e. when $\beta_c < 1$.
In Sec. \ref{subsec:stat-sol-above-betac}, we derive the form of $p_{\textrm{div}}$ in the localized state. In Sec. \ref{subsec:small_variation} we provide an explicit calculation of $\beta_c$ in a special case, and in Sec. \ref{subsec:heritab_threshold_beta0} we discuss the case $\beta = 0$.

\subsection{Heritability threshold\label{subsec:Classification-of-stationary} }

Throughout this subsection, we assume $0<\beta<1$ unless otherwise noted.  We consider the definition of $\beta_c$ given in \eqref{eq:supp_beta_c}, and observe that condition \ref{cond:cond1} is satisfied if and only if
\begin{equation}
\omega_{0}\left(\beta,z\right) \leq \lambda < \infty \label{eq:cond1_range}
\end{equation}
where
\begin{align}
\omega_{0}\left(\beta,z\right) & := 
\begin{cases}
\sfrac{\ln\left(\beta z\right)}{\tau_{1}}, & 0<\beta z<1\\
0, & \beta z = 1 \\
\sfrac{\ln\left(\beta z\right)}{\tau_{0}}, & \beta z > 1 \ .
\end{cases}
\end{align}
If $\tau_{0} >0$, then $\omega_0$ is finite and there exists a non-empty range of $\lambda$ satisfying \eqref{eq:cond1_range}.  If $\tau_0 = 0$, then the range \eqref{eq:cond1_range} is non-empty if and only if $\beta z \leq 1$ (since for $\beta z > 1$ we have $\omega_0 = \infty$).  

To additionally satisfy condition \ref{cond:cond2}, we first identify the values of $\lambda$ such that $Q\left(\beta,z,\lambda\right)<\infty$.
For any $\lambda$ in the range 
\begin{equation}
\omega_0(\beta,z) < \lambda < \infty \ , \label{eq:lambda_range}
\end{equation}
we can bound the integral  in \eqref{smeq:Q-function} yielding
\begin{align}
Q\left(\beta,z,\lambda\right) & \le\max\left(\frac{\left(1-\beta\right)z}{e^{\lambda\tau_{0}}-\beta z}, \frac{\left(1-\beta\right)z}{e^{\lambda\tau_{1}}-\beta z}\right) \label{eq:Q_bound}
\end{align}
and the right-hand side is finite if and only if $\beta z \neq 1$ or $\tau_0 \neq 0$.  Thus, we have $Q\left(\beta,z,\lambda\right)<\infty$ for all $\lambda$ satisfying \eqref{eq:lambda_range} except in the special case in which both $\beta z = 1$ and $\tau_0=0$.  We will treat that case separately when needed below.  

Given that $Q\left(\beta,z,\lambda\right)<\infty$, we have that $Q\left(\beta,z,\lambda\right)$ monotonically decreases as $\lambda$ increases because
\begin{equation}
\frac{\partial}{\partial\lambda}Q\left(\beta,z,\lambda\right)=-\int_{\tau_{0}}^{\tau_{1}}\frac{\left(1-\beta\right)z\tau e^{\lambda\tau}k\left(\tau\right)}{\left(e^{\lambda\tau}-\beta z\right)^{2}}d\tau\label{eq:Q-beta_z_lamba_dellambda} < 0 \ .
\end{equation}
For sufficiently large $\lambda$, 
\begin{equation}
Q(\beta, z, \lambda) \le \frac{(1-\beta)z}{e^{\lambda\tau_{0}}-\beta z} \ ,
\end{equation}
and thus $\lim_{\lambda\rightarrow\infty}Q\left(\beta,z,\lambda\right)=0$ if $\tau_{0}>0$.
If $\tau_{0}=0$ (and thus $\beta z \leq 1$ by condition 1), $\lim_{\lambda\rightarrow\infty}Q\left(\beta,z,\lambda\right)=0$ likewise holds because for any $\epsilon\in(0,\tau_1)$, 
\begin{equation}
Q(\beta, z, \lambda) \le  \frac{(1-\beta)z}{1-\beta z} \int_{0}^{\epsilon}k(\tau)d\tau + \frac{(1-\beta)z}{e^{\lambda\epsilon}-\beta z} \int_{\epsilon}^{\tau_1}k(\tau)d\tau \ .
\label{eq:Qlim0_tau0eq0}
\end{equation}
Therefore, over the range of $\lambda$ satisfying condition 1, we have shown that $Q\left(\beta,z,\lambda\right)$ monotonically decreases to $0$ with increasing $\lambda$.  We conclude that
condition \ref{cond:cond2} is satisfied if and only if 
\begin{equation}
Q_{0}\left(\beta,z\right) := \lim_{\lambda\downarrow\omega_{0}\left(\beta,z\right)}Q\left(\beta,z,\lambda\right) \geq 1 \ . \label{eq:Q0rel}
\end{equation}
In this case, $p_\mathrm{div}$ is delocalized and the long-term growth rate $\Lambda_{\beta,z}$ is given by the unique solution of $Q(\beta, z, \Lambda_{\beta,z}) = 1$, where $\omega_0(\beta, z) \leq \Lambda_{\beta,z} < \infty$.

To determine when \eqref{eq:Q0rel} is satisfied, we analyze separately the cases $z>1$, $z=1$ and $0<z<1$ below. We treat the case $\beta = 0$ separately in Sec.\ref{subsec:heritab_threshold_beta0}, while for the cases below we will always have $0 < \beta < 1$.
\\

%
%
\subsubsection*{Case I: $z>1$}
Given that $0 < \beta < 1$, we consider three sub-cases below, consisting of $\beta z < 1$,  $\beta z > 1$,  and $\beta z = 1$. \\ 
%
%
\paragraph*{Sub-case I-1: $\beta z < 1$.}
In this case, we have $\omega_0(\beta,z) = \ln (\beta z) / \tau_1 \leq 0$.  Since $Q(\beta,z, \lambda)$ increases for decreasing $\lambda$, we have by the definition in \eqref{eq:Q0rel} that $Q_0(\beta, z) \geq  Q(\beta, z, 0)$.  Since $Q(\beta, z, 0) = \sfrac{\left(1-\beta\right)z}{\left(1-\beta z\right)}$, and given that we have assumed $z > 1$, we have $Q(\beta, z, 0)> 1$.  Hence, $Q_0(\beta, z) > 1$ and condition 2 holds for $\beta < z^{-1}$.  We conclude that $\beta_c \geq z^{-1}$. 
%
%
\paragraph*{Sub-case I-2: $\beta z > 1$.}
We have already seen that for $\beta z > 1$ and $\tau_0 = 0$, condition 1 cannot be satisfied implying  that $\beta_c \leq z^{-1}$.  Since sub-case I-1 above yielded $\beta_c \geq z^{-1}$, which holds in general, we conclude that if $\tau_0 = 0$ then $\beta_c = z^{-1}$.

We now consider $\tau_0 > 0$.  Below, we show that for $\beta z > 1$, $Q_{0}\left(\beta,z\right) < \infty$ if and only if 
\begin{equation}
\int_{\tau_{0}}^{\tau_{1}} \frac{k\left(\tau\right) d\tau}{\tau-\tau_{0}}<\infty \ . \label{eq:Q0finite_condition}
\end{equation}
When $Q_0(\beta,z) = \infty$, Eq. \ref{eq:Q0rel} is satisfied, hence condition 2 is satisfied for all $\beta > z^{-1}$; this implies that $\beta_c = 1$. We thus turn to cases when $Q_0(\beta,z) < \infty$. We have $\omega_0(\beta,z) = \ln(\beta z) / \tau_0$, from which 
\begin{equation}
Q_{0}\left(\beta,z\right)=\frac{1-\beta}{\beta}\int_{\tau_{0}}^{\tau_{1}}\frac{k\left(\tau\right)}{\left(\beta z\right)^{\sfrac{\tau}{\tau_{0}-1}}-1}d\tau \label{eq:Q0bz_1}
\end{equation}
and
\begin{widetext}
\begin{equation}
\frac{\partial}{\partial\beta}Q_{0}\left(\beta,z\right)=-\frac{1}{\beta^{2}}\int_{\tau_{0}}^{\tau_{1}}\frac{\left(\left(1+\left(1-\beta\right)\left(\sfrac{\tau}{\tau_{0}-1}\right)\right)\left(\beta z\right)^{\sfrac{\tau}{\tau_{0}-1}}-1\right)k\left(\tau\right)}{\left(\left(\beta z\right)^{\sfrac{\tau}{\tau_{0}-1}}-1\right)^{2}}d\tau
<0,
\end{equation}
\end{widetext}
since $\left(1+\left(1-\beta\right)\left(\sfrac{\tau}{\tau_{0}-1}\right)\right)\left(\beta z\right)^{\sfrac{\tau}{\tau_{0}-1}}\ge1$
holds for $\tau\in[\tau_{0},\tau_{1}]$ with equality only when $\tau=\tau_{0}$.
That is, $Q_{0}\left(\beta,z\right)$ is monotonically decreasing as $\beta$ increases over the range $z^{-1}<\beta<1$.  Since relation \eqref{eq:Q0finite_condition} holds, Eq. \ref{eq:Q0bz_1} yields $\lim_{\beta \rightarrow z^{-1}} Q_{0}\left(\beta,z\right) = \infty$ and $\lim_{\beta \rightarrow 1} Q_{0}(\beta,z) = 0$; therefore, there exists $\beta_c$ such that $Q_{0}\left(\beta_{c},z\right)=1$ with $z^{-1} < \beta_c < 1$, i.e. a heritability threshold exists.  
 
 In summary, if $\tau_0 > 0$ and $z > 1$, we conclude that $z^{-1} < \beta_c < 1$ if $\int_{\tau_{0}}^{\tau_{1}}d\tau\sfrac{k\left(\tau\right)}{\left(\tau-\tau_{0}\right)} < \infty$, and $\beta_c = 1$ otherwise.  We know that $p_{\mathrm{div}}$ is localized (delocalized) if $\beta > \beta_c$ ($\beta < \beta_c$).  Moreover, since $Q_0(\beta_c, z) = 1$ satisfies \eqref{eq:Q0rel}, we have that at $\beta = \beta_c$, $p_{\mathrm{div}}$ is delocalized. If $\tau_0 = 0$, we have $\beta_c = z^{-1}$; the behavior of $p_{\mathrm{div}}$ at $\beta_c$ corresponds to the unique special case in which the bound given in \eqref{eq:Q_bound} is not finite, and $Q(\beta, z, \lambda) < \infty$ may or may not hold.  This special case is analyzed next. 
 %
%
\paragraph*{Sub-case I-3: $\beta=z^{-1}$.}  We only need to analyze the case $\tau_0 = 0$, since for $\tau_0 > 0$, we showed above that $\beta_c > z^{-1}$, hence we know that $p_{\mathrm{div}}$ is delocalized for $\beta = z^{-1}$.  For $\tau_0 = 0$, we have
\begin{equation}
Q\left(\beta,z,\lambda\right)  = Q\left(z^{-1},z,\lambda\right) = \int_{0}^{\tau_{1}}\frac{ \left(z-1\right) k\left(\tau\right)}{e^{\lambda\tau}-1}d\tau \ ,
\end{equation}
where the integral may diverge or converge for $\lambda>\omega_0(z^{-1}, z) = 0$.  If $\int_{0}^{\tau_{1}}\tau^{-1}k\left(\tau\right)d\tau=\infty$, $Q\left(z^{-1},z,\lambda\right)=\infty$ for any $\lambda>0$ and condition 2 cannot be satisfied.  If $\int_{0}^{\tau_{1}}\tau^{-1}k\left(\tau\right)d\tau<\infty$, $Q\left(z^{-1},z,\lambda\right) < \infty$ for all $\lambda > 0$, and the argument leading to relation \eqref{eq:Q0rel} applies.  

In summary, for $\tau_0 = 0$ and $z > 1$, the system exhibits a heritability threshold, given by $\beta_c = z^{-1}$.  Convergence or divergence of $\int_{0}^{\tau_{1}}\tau^{-1}k\left(\tau\right)d\tau$ affects the uniqueness of the stationary solution at $\beta = \beta_c$ (see Sec. \ref{subsec:stat-sol-above-betac}).
 \\

In conclusion, for $z > 1$ we have shown that
\begin{equation}
\beta_{c}
\begin{cases}
= z^{-1}, & \text{if }\tau_{0} = 0,\\
> z^{-1}, & \text{if }\tau_{0} > 0 \text{ and } \int_{\tau_{0}}^{\tau_{1}}\sfrac{k(\tau)}{(\tau-\tau_{0})}<\infty,\\
= 1, & \text{if }\tau_{0} > 0 \text{ and } \int_{\tau_{0}}^{\tau_{1}}\sfrac{k(\tau)}{(\tau-\tau_{0})}=\infty \ .
\end{cases}
\end{equation}

%
%
\subsubsection*{Case I\hspace{-.1em}I: $z=1$}

Since $Q\left(\beta,1,0\right)=1$, $\Lambda=0$ is the unique solution of $Q\left(\beta,1,\Lambda\right)=1$.
Thus $\beta_{c}=1$ for $z=1$.

%
%
\subsubsection*{Case I\hspace{-.1em}I\hspace{-.1em}I: $0<z<1$}
For $z < 1$, we have $0<\beta z<1$, $\omega_0(\beta,z) = \ln(\beta z) / \tau_1 \leq 0$, and \begin{equation}
Q_{0}\left(\beta,z\right)=\frac{1-\beta}{\beta}\int_{\tau_{0}}^{\tau_{1}}\frac{k\left(\tau\right)}{\left(\beta z\right)^{\sfrac{\tau}{\tau_{1}-1}}-1}d\tau \ . \label{eq:Q0_bz_lt1}
\end{equation}

If $\tau_{1}=\infty$, $Q_{0}\left(\beta,z\right)=\sfrac{\left(1-\beta\right)z}{\left(1-\beta z\right)}<1$ and thus by relation \eqref{eq:Q0rel} condition 2 cannot be satisfied for any value of $\beta$, implying $\beta_{c}=0$.

If $\tau_1<\infty$, we show below that for $\beta z < 1$,
$Q_0(\beta,z) < \infty$ if and only if 
\begin{equation}
\int_{\tau_{0}}^{\tau_{1}}\frac{k\left(\tau\right) d\tau}{\tau_{1}-\tau}<\infty \ . \label{eq:Q0finite_condition2}
\end{equation}
If $Q_{0}\left(\beta,z\right)=\infty$, then relation \eqref{eq:Q0rel} implies that $\beta_{c}=1$.
For $Q_{0}\left(\beta,z\right)<\infty$, we can show that
\begin{widetext}
\begin{align}
\frac{\partial}{\partial\beta}Q_{0}\left(\beta,z\right)
 & =-\frac{1}{\beta^{2}}\int_{\tau_{0}}^{\tau_{1}}\frac{\left(\left(1+\left(1-\beta\right)\left(\sfrac{\tau}{\tau_{1}-1}\right)\right)\left(\beta z\right)^{\sfrac{\tau}{\tau_{1}-1}}-1\right)k\left(\tau\right)}{\left(\left(\beta z\right)^{\sfrac{\tau}{\tau_{1}-1}}-1\right)^{2}}d\tau
 < 0 \ .
\end{align}
\end{widetext}
To prove this inequality, it suffices to prove that $\left(\left(1-\beta\right)x-1\right)\left(\beta z\right)^{-x}+1>0$ for $x\in\left(0,1\right)$; since $(\beta z)^x < \beta^x$ for $x>0$, the result will follow by showing that $\beta^x < 1-(1-\beta)x$ for $x\in\left(0,1\right)$.  To see this, we define $f(x):=1-(1-\beta)x-\beta^x$, which has a unique maximum at $x^*:=\ln[\sfrac{(\beta-1)}{\ln\beta}] / \ln\beta$.
$0<1-\beta<-\ln\beta$ implies $x^*>0$ and $\beta^{-1}-1> -\ln\beta$ implies $x^*<1$.
Together with $f(0)=f(1)=0$ and continuity of $f(x)$ and $f^\prime(x)$, $f(x) > 0$ is true for $x\in(0,1)$.

Next, we show that $\lim_{\beta\uparrow1}Q_{0}\left(\beta,z\right)=0$ and $\lim_{\beta\downarrow0}Q_{0}\left(\beta,z\right)=\infty$. We have
\begin{equation}
\lim_{\beta\uparrow1}Q_{0}\left(\beta,z\right)=0\cdot\int_{\tau_{0}}^{\tau_{1}}\frac{k\left(\tau\right)}{z^{\sfrac{\tau}{\tau_{1}-1}}-1}d\tau=0,
\end{equation}
since the integral converges by relation \eqref{eq:Q0finite_condition2} which holds. For $\tau_0>0$, we can bound the integral in \eqref{eq:Q0_bz_lt1} to obtain
\begin{align}
Q_{0}\left(\beta,z\right)
 & \ge\frac{\left(1-\beta\right)z}{\left(\beta z\right)^{\sfrac{\tau_{0}}{\tau_{1}}}-\beta z} , 
\end{align}
while for $\tau_{0}=0$, given $\epsilon\in\left(0,\tau_{1}\right)$ we have
\begin{equation}
Q_{0}\left(\beta,z\right)\ge\left(1-\beta\right)z\left(\left(\beta z\right)^{\sfrac{\epsilon}{\tau_{1}}}-\beta z\right)^{-1}\int_{\epsilon}^{\tau_{1}}k\left(\tau\right)d\tau ; 
\end{equation}
hence $\lim_{\beta\downarrow0}Q_{0}\left(\beta,z\right)=\infty$.
Therefore, there exists $0<\beta_{c}<1$ such that $Q_{0}\left(\beta_{c},z\right)=1$.

In conclusion, for $0 < z < 1$ we have shown that
\begin{equation}
\beta_{c}
\begin{cases}
= 0, & \text{if }\tau_{1} = \infty,\\
< 1, & \text{if }\tau_{1} < \infty \text{ and } \int_{\tau_{0}}^{\tau_{1}}\sfrac{k(\tau)}{(\tau_1-\tau)}<\infty,\\
= 1, & \text{if }\tau_{1} < \infty \text{ and } \int_{\tau_{0}}^{\tau_{1}}\sfrac{k(\tau)}{(\tau_1-\tau)}=\infty \ .
\end{cases}
\end{equation}
Table \ref{tab:LR-stationary-class} provides a summary of the results obtained in this subsection.

\subsubsection*{Necessary and sufficient condition for $Q_{0}\left(\beta,z\right)<\infty$ \label{subsec:iff_betac}}

It suffices to prove that $\int_{\tau_{0}}^{\tau_{1}}\sfrac{k(\tau) d\tau}{(\tau-\tau_{0})}<\infty$ is equivalent to $Q_{0}\left(\beta,z\right)<\infty$ when $\tau_{0}>0$, $z>1$ and $\beta z > 1$.
In the same way one can show that for $\tau_{1}<\infty$, $0<z<1$, and $\beta z <1$, $\int_{\tau_{0}}^{\tau_{1}}\sfrac{k(\tau)}{(\tau_{1}-\tau)}<\infty$ is the equivalent condition.

We write
\begin{align}
Q_{0}&\left(\beta,z\right) \cdot\beta/\left(1-\beta\right) = \int_{\tau_0}^{\tau_{1}}\frac{k\left(\tau\right)}{\left(\beta z\right)^{\tau/\tau_{0}-1}-1}d\tau \\
&= \int_{\tau_{0}}^{\tau_{0}+\epsilon}\frac{k\left(\tau\right)}{\left(\beta z\right)^{\tau/\tau_{0}-1}-1}d\tau\nonumber +\int_{\tau_{0}+\epsilon}^{\tau_{1}}\frac{k\left(\tau\right)}{\left(\beta z\right)^{\tau/\tau_{0}-1}-1}d\tau
\end{align}
where $0<\epsilon \ll 1$. The second term of RHS is finite, while the integrand of the first term can be expanded with respect to $\tau-\tau_{0}$, yielding
\begin{align}
\int_{\tau_{0}}^{\tau_{0}+\epsilon} \frac{k\left(\tau\right)}{\omega_{0}}&\left(\frac{1}{\tau-\tau_{0}}+\mathcal{O}\left(1\right)\right)d\tau\nonumber \\
= & \frac{1}{\omega_{0}}\int_{\tau_{0}}^{\tau_{0}+\epsilon}\frac{k\left(\tau\right)}{\tau-\tau_{0}}d\tau+\mathcal{O}\left(\epsilon\right)
\end{align}
where $\omega_{0}=\tau_{0}^{-1}\ln\left(\beta z\right)$. Thus, for
any $\beta$ satisfying $z^{-1}<\beta<1$, $Q_{0}\left(\beta,z\right)$
is finite if and only if $\int_{\tau_{0}}^{\tau_{0}+\epsilon}\sfrac{k\left(\tau\right) d\tau}{(\tau-\tau_{0})}$
is finite for small $\epsilon>0$. This is equivalent to $\int_{\tau_{0}}^{\tau_{1}}\sfrac{k\left(\tau\right) d\tau}{(\tau-\tau_{0})}<\infty$.

For example, if $k\left(\tau\right)$ behaves power of $\tau-\tau_{0}$
near $\tau_{0}$, that is, there exists a real number $\gamma$ and
$\lim_{\tau\downarrow\tau_{0}}\frac{k\left(\tau\right)}{\left(\tau-\tau_{0}\right)^{\gamma}}=A$
where $A$ is a positive finite constant, then the dominant term of $\int_{\tau_{0}}^{\tau_{0}+\epsilon}\frac{k\left(\tau\right)}{\tau-\tau_{0}}d\tau$
will be $A\int_{0}^{\epsilon}x^{\gamma-1}dx$. Thus $\int_{\tau_{0}}^{\tau_{0}+\epsilon}\frac{k\left(\tau\right)}{\tau-\tau_{0}}d\tau < \infty$
if and only if $\gamma>0$.

\subsection{Stationary solution for $\beta > \beta_c$ \label{subsec:stat-sol-above-betac}}
For $\beta>\beta_{c}$, the measure $p_{\textrm{div}}$ does not have a density relative to Lebesgue measure $\mu$ on the interval $\left[\tau_{0},\tau_{1}\right]$. We find that in this regime $p_{\textrm{div}}$ includes a Dirac delta function, and we obtain the exact form of the solution, which we will show below to be unique. 

We seek a solution of the form
\begin{equation}
p_{\mathrm{div}}\left(\tau\right)=\widetilde{p}\left(\tau\right)+C_{1}\delta\left(\tau-\tau_{*}\right) \label{eq:stationary_solution}
\end{equation}
where $\tau_{*}\in\left[\tau_{0},\tau_{1}\right]$, $C_{1}>0$, $\delta\left(\cdot\right)$ denotes the Dirac delta function, and $\widetilde{p}\left(\tau\right) \geq 0$ is
a Lebesgue integrable function on $\left[\tau_{0},\tau_{1}\right]$, i.e. $\widetilde p \in L^1(\mu)$.
The normalization is given by $\int_{\tau_{0}}^{\tau_{1}}p_{\mathrm{div}}\left(\tau\right)d\tau=1$.
Substituting Eq. \eqref{eq:stationary_solution} in Eq. (3)
we have
\begin{subequations}
\begin{align}
\left(1-\beta ze^{-\Lambda\tau}\right)\widetilde{p}\left(\tau\right) &= \left(1-\beta\right)ze^{-\Lambda\tau}k\left(\tau\right),\\
\left(1-\beta ze^{-\Lambda\tau_{*}}\right)C_{1}&=0.
\end{align}
\end{subequations}
Since $\widetilde{p}\left(\tau\right)\ge0$ for $\tau\in\left[\tau_{0},\tau_{1}\right]$, $\Lambda\ge\omega_{0}\left(\beta,z\right)$; and $C_{1}>0$ implies $\Lambda=\ln\left(\beta z\right) / \tau_{*} \le \omega_{0}\left(\beta,z\right)$ for $\tau_{*}\in\left[\tau_{0},\tau_{1}\right]$.
Thus
\begin{equation}
\Lambda = \omega_{0}\left(\beta,z\right) = \ln\left(\beta z\right) / \tau_{*} \ . \label{eq:lambda_above_transition}
\end{equation}
Using $\beta > \beta_c \geq z^{-1}$ for $z>1$, we have $\beta z > 1$ hence $\omega_0 = \ln (\beta z) / \tau_0$; while for $z < 1$, $\beta z < 1$ yields $\omega_0 = \ln (\beta z) / \tau_1$.  These relations together with \eqref{eq:lambda_above_transition} imply
\begin{equation}
\tau_{*}=\begin{cases}
\tau_{0}, & z>1\\
\tau_{1}, & z<1 \ .
\end{cases}
\label{smeq:taustar}
\end{equation}
Due to the normalization $\int_{\tau_{0}}^{\tau_{1}}p_{\mathrm{div}}\left(\tau\right)d\tau=1$,
\begin{align}
C_{1} & =1-Q\left(\beta,z,\omega_{0}\left(\beta,z\right)\right)\nonumber \\
 & =1-\int_{\tau_{0}}^{\tau_{1}}\frac{\left(1-\beta\right)zk\left(\tau^{\prime}\right)}{\left(\beta z\right)^{\sfrac{\tau^{\prime}}{\tau_{*}-1}}-\beta z}d\tau^{\prime} \ ,
\end{align}
and $0<C_{1} \leq 1$ holds because $Q\left(\beta,z,\omega_{0}\left(\beta,z\right)\right) = Q_{0}(\beta,z)<1$ for $\beta>\beta_{c}$.

To summarize, for $\tau_{0}>0$, $z>1$ and $\beta>\beta_{c}$, 
\begin{subequations}
\begin{align}
\Lambda&=\frac{\ln\left(\beta z\right)}{\tau_{0}},\\
p_{\textrm{div}}\left(\tau\right)&=q_{\beta,z,\Lambda}\left(\tau\right)+\left(1-Q\left(\beta,z,\Lambda\right)\right)\delta\left(\tau-\tau_{0}\right),
\label{smeq:stat_above_betac_zge1_2}
\end{align}
\end{subequations}
and for $\tau_{1}<\infty$, $z<1$ and $\beta>\beta_{c}$, 
\begin{subequations}
\begin{align}
\Lambda&=\frac{\ln\left(\beta z\right)}{\tau_{1}},\\
p_{\textrm{div}}\left(\tau\right)&=q_{\beta,z,\Lambda}\left(\tau\right)+\left(1-Q\left(\beta,z,\Lambda\right)\right)\delta\left(\tau-\tau_{1}\right) \ .
\label{smeq:stat_above_betac_zle1_2}
\end{align}
\label{smeq:stat_above_betac_zle1}
\end{subequations}

We note that in the limits $\tau_0 \rightarrow 0$ and $\tau_1 \rightarrow \infty$, the limiting forms of the solution above are likewise valid.  The case in which $z>1$ and $\tau_0 \rightarrow 0$ is unique in that we have $\Lambda = \infty$ and $Q(\beta, z, \Lambda) = 0$, hence $p_{\mathrm{div}} = \delta(\tau - \tau_0)$ for all $\beta > \beta_c = z^{-1}$, and the entire population is localized at $\tau_0$ above the transition.  Moreover, in this special case, uniqueness of the solution at $\beta = \beta_c$ is dependent on whether or not $\int_0^{\tau_1} \tau^{-1} k(\tau) d\tau$ converges, as discussed below.

For cases in which $z<1$, a finite population will eventually go extinct, and in practice the stationary solution Eq. \eqref{smeq:stat_above_betac_zle1} may or may not be realized before extinction occurs.
Lastly, we remark that Eqs. \eqref{smeq:stat_above_betac_zge1_2} and \eqref{smeq:stat_above_betac_zle1_2} also hold for $\beta < \beta_{c}$ as $Q(\beta,z,\Lambda)=1$. \\

\paragraph*{Uniqueness of the solution.}
We establish existence and uniqueness of the stationary solution by the same approach used in \cite{Burger1996} to study Kingman's ``house of cards'' model in evolutionary dynamics. 

We set $X = \left[\tau_{0},\tau_{1}\right]$, and let $\mathcal{B}(X)$ be the collection of Borel sets in $X$,  let $\mathcal{P}$ denote the set of all Borel probability measures on $X$, and let $\mu$ denote the Lebesgue measure on $X$.
Eq. (3) 
is rewritten for a stationary probability measure $P \in\mathcal{P}$ as
\begin{equation}
P \left(A\right) = \int_{A}\beta ze^{-\Lambda\tau} P \left(d\tau\right) + \int_{A}\left(1-\beta\right)ze^{-\Lambda\tau}k\left(\tau\right)\mu\left(d\tau\right)
\label{smeq:LR-stationary-equation-probmeas}
\end{equation}
for any set $A\subset \mathcal{B}(X)$, and where $P \left(X\right) = 1$.  

Given that $P\in\mathcal{P}$ of Eq. \eqref{smeq:LR-stationary-equation-probmeas} exists,
then by the Lebesgue decomposition theorem there exist unique $P_{a}, P_{s} \in \mathcal{P}$ and a number $C_{1}\in\left[0,1\right]$ such that
\begin{equation}
P = \left(1-C_{1}\right)P_{a} + C_{1} P_{s}
\end{equation}
where $P_{a}$ is absolutely continuous with respect to $\mu$ and $P_{s}$ is singular with respect to $\mu$;
i.e. there exists $f \geq 0 \in L^1(\mu)$ such that $P_{a}(A) = \int_{A}f(\tau) \mu(d \tau)$ and $P_{s}$ is supported on a $\mu$-null set.
Decomposing Eq. \eqref{smeq:LR-stationary-equation-probmeas} into absolutely continuous and singular parts with respect to $\mu$, we have
\begin{subequations}
\begin{align}
\left(1-C_{1}\right)\int_{A}(1-\beta z e^{-\Lambda\tau})P_{a}(d\tau) \nonumber\\
= \int_{A}\left(1-\beta\right)ze^{-\Lambda\tau}k\left(\tau\right)\mu\left(d\tau\right)\label{smeq:stat-Pa-eq}\\
C_{1}\int_{A}(1-\beta z e^{-\Lambda\tau})P_{s}(d\tau) & = 0 \label{smeq:stat-Ps-eq}
\end{align}
\end{subequations}
From Eq. \eqref{smeq:stat-Pa-eq},
\begin{equation}
\widetilde{p}\left(\tau\right) = \frac{\left(1-\beta\right)ze^{-\Lambda\tau}k\left(\tau\right)}{1-\beta z e^{-\Lambda\tau}} \text{, }\mu\text{-a.s. on }\left[\tau_{0}, \tau_{1}\right]
\end{equation}
where $\widetilde{p}\left(\tau\right)$ is a non-negative Lebesgue integrable function such that
$(1-C_{1})P_{a}(A) = \int_{A}\widetilde{p}\left(\tau\right)\mu\left(d\tau\right)$ for any measurable set $A\in\left[\tau_{0},\tau_{1}\right]$,
denoted by $(1-C_{1})P_{a} = \widetilde{p} \, \mu$ in short.
$\widetilde{p}\left(\tau\right)\ge 0 \text{, }\mu\text{-a.s. on }\left[\tau_{0}, \tau_{1}\right]$ implies $\Lambda\ge\omega_{0}\left(\beta,z\right)$.
By the normalization $P_{a}\left(\left[\tau_{0}, \tau_{1}\right]\right) = 1$,  we have
\begin{equation}
C_{1} = 1 - \int_{\tau_{0}}^{\tau_{1}}\widetilde{p}\left(\tau\right)\mu\left(d\tau\right) = 1 - Q\left(\beta,z,\Lambda\right) \ ,
\end{equation}
for $Q(\beta, z, \Lambda) \leq 1$. From Eq. \eqref{smeq:stat-Ps-eq},
$C_{1} = 0$ or
\begin{equation}
e^{\Lambda \tau} = \beta z  \text{, }P_{s}\text{-a.s. on }\left[\tau_{0}, \tau_{1}\right].
\end{equation}
Therefore, if $C_1 > 0$, $P_{s}$ is the Dirac measure concentrated on $\tau_{*}$, denoted by $\delta_{\tau_{*}}$, with $\tau_{*} = \sfrac{\ln\left(\beta z\right)}{\Lambda}\in\left[\tau_{0}, \tau_{1}\right]$.
This implies that if $C_1 > 0$ then $\Lambda \le \omega_{0}\left(\beta,z\right)$ and thus $\Lambda = \omega_{0}\left(\beta,z\right)$.  

From the discussion in Sec. \ref{subsec:Classification-of-stationary},
for $\tau_0 > 0$ and $0<\beta \le \beta_{c}$, there exists $\Lambda \ge \omega_{0}\left(\beta,z\right)$ such that $Q\left(\beta,z,\Lambda\right) = 1$ and $P = \widetilde{p} \, \mu$.
For $\beta > \beta_{c}$, on the other hand, $\Lambda = \omega_{0}\left(\beta,z\right)$ and
$P = \widetilde{p} \, \mu + \left(1-Q\left(\beta,z, \omega_{0}\left(\beta,z\right)\right)\right)\delta_{\tau_{*}}$
where $\tau_{*}$ is given by Eq. \eqref{smeq:taustar}. 

In the special case of $\tau_0 = 0$, for $\beta > \beta_c = z^{-1}$, $\Lambda = \ln (\beta z) / \tau_0 = \infty$, which yields $\tilde p(\tau) = 0$, $Q(\beta, z, \Lambda) = 0$, and $C_1 = 1$; hence the stationary solution $P$ is given by Dirac measure $\delta_{\tau_0}$ concentrated at $\tau_0 = 0$. At $\beta = \beta_c = z^{-1}$, if $\int_0^{\tau_1} \tau^{-1} k(\tau) d\tau = \infty$, we have $Q(\beta,z,\Lambda) = \infty$, and it is not possible to normalize $\tilde p(\tau)$ for any $\Lambda < \infty$; in this case, the only possible solution of Eq. \ref{smeq:stat-Pa-eq} can be understood as the limiting case $\beta z \downarrow 1$, which yields $C_1 = 1$,  $\Lambda = \infty$, and $P = \delta_{\tau_0}$.  If on the other hand $\int_0^{\tau_1} \tau^{-1} k(\tau) d\tau < \infty$, then for all $0 \leq C_1 \leq 1$, there exists $\Lambda(C_1)$ such that $Q(z^{-1}, z, \Lambda(C_1)) = 1-C_1$, or 
\begin{equation}
1 - C_1 = \int_0^{\tau_1} \frac{(z - 1)  k(\tau)}{e^{\Lambda(C_1) \tau} - 1} \mu(d\tau) \ ,
\end{equation}
and we obtain solutions at all growth rates $\Lambda \geq \Lambda(0)$. 

\subsection{Small variation limit of IDT distribution \label{subsec:small_variation}}

We consider the limit in which $k(\tau)$ has vanishingly small variation.  We fix $\bar{\tau}\in\left[\tau_{0},\tau_{1}\right]$, and taking the limit $k\left(\tau\right)\rightarrow\delta\left(\tau-\bar{\tau}\right)$, 
\begin{equation}
Q\left(\beta,z,\lambda\right)=\frac{\left(1-\beta\right)z}{e^{\lambda \bar{\tau}}-\beta z}
\end{equation}
and 
\begin{equation}
Q_{0}\left(\beta,z\right)=\begin{cases}
\frac{\left(1-\beta\right)z}{\left(\beta z\right)^{\sfrac{\bar{\tau}}{\tau_{0}}}-\beta z}, & 0<\beta z<1\\
\frac{\left(1-\beta\right)z}{\left(\beta z\right)^{\sfrac{\bar{\tau}}{\tau_{1}}}-\beta z}, & \beta z > 1 \ .
\end{cases}
\end{equation}
For $z > 1$, we have $\beta_c z > 1$, and $Q_{0}\left(\beta_c,z\right)=1$ implies $\beta_{c}=z^{\sfrac{\tau_{0}}{\bar{\tau}}-1}$; while for $z<1$, we find $\beta_{c}=z^{\sfrac{\tau_{1}}{\bar{\tau}}-1}$.
Therefore
\begin{equation}
\beta_{c}=\begin{cases}
z^{\sfrac{\tau_{1}}{\bar{\tau}}-1}, & 0<z<1\\
1, & z=1\\
z^{\sfrac{\tau_{0}}{\bar{\tau}}-1}, & z>1
\end{cases}
\end{equation}
We also observe $\beta_{c}\downarrow0$ as $\tau_{1}\uparrow\infty$
for $z<1$ and $\beta_{c}\downarrow\dot{z^{-1}}$ as $\tau_{0}\downarrow0$
for $z>1$.
This limiting case is also included in Table \ref{tab:LR-stationary-class}.

\subsection{Independent divisions: $\beta=0$\label{subsec:heritab_threshold_beta0}}
For $\beta = 0$, condition \ref{cond:cond1} is automatically satisfied for any $\lambda$, since $q_{0, z, \lambda}(\tau) = z e^{-\lambda \tau} k(\tau)$; here we determine when condition \ref{cond:cond2} is satisfied.

For $z>1$, since $Q(0,z,0)=z>1$, it suffices to show that  $Q(0,z,\lambda)$ is monotonically decreasing (which follows from \eqref{eq:Q-beta_z_lamba_dellambda}) to $0$ for $\lambda\ge0$.
If $\tau_{0}>0$, $Q\left(0,z,\lambda\right)$ is finite and $\lim_{\lambda\rightarrow\infty}Q\left(0,z,\lambda\right)=0$
because $0\le Q\left(0,z,\lambda\right)\le ze^{-\lambda\tau_{0}}\rightarrow 0$ as $\lambda\rightarrow\infty$.
If $\tau_{0}=0$, $\lim_{\lambda\rightarrow\infty}Q\left(0,z,\lambda\right)=0$ holds since the argument given at Eq. \eqref{eq:Qlim0_tau0eq0} holds for $\beta=0$. We thus have a unique $\lambda>0$ satisfying conditions \ref{cond:cond1} and \ref{cond:cond2}.

For $z=1$, $Q(0,1,\lambda)$ is monotonically decreasing to $0$ for $\lambda\ge 0$.
Since $Q(0,1,\lambda)>1$ for any $\lambda<0$, $\lambda=0$ is the unique root of $Q(0,1,\lambda)=1$.

Finally, let $0<z<1$.
If $\tau_{1}<\infty$, then $Q(0,z,\lambda)\le ze^{-\lambda\tau_{1}}<\infty$ for any $\lambda\le0$; in particular, $Q(\beta, z, 0) < 1$.  
For some $\epsilon\in(0,\tau_{1})$,
\begin{equation}
Q\left(0,z,\lambda\right)\ge z\int_{\epsilon}^{\tau_{1}}e^{-\lambda\tau}k\left(\tau\right)\ge ze^{-\lambda\epsilon}\int_{\epsilon}^{\tau_{1}}k\left(\tau\right)
\end{equation}
Thus $\lim_{\lambda\rightarrow-\infty}Q\left(0,z,\lambda\right)=\infty$.
Since $Q\left(0,z,\lambda\right)$ decreases monotonically with respect to $\lambda$, there is a unique $\lambda<0$ satisfying conditions \ref{cond:cond1} and \ref{cond:cond2}.
If $\tau_{1}=\infty$, $\lambda$ may be bounded below by $\lambda_{c}\in(-\infty, 0)$ for $Q\left(0,z,\lambda\right)$
to be finite. In this case, a unique $\lambda<0$ satisfying conditions \ref{cond:cond1} and \ref{cond:cond2} exists
as long as $\lim_{\lambda\downarrow\lambda_{c}}Q\left(0,z,\lambda\right)>1$ holds.
For example, consider $k\left(\tau\right) = \sfrac{\tau^{-2}e^{-\tau}}{\int_{1}^{\infty}\tau^{-2}e^{-\tau}d\tau}$ for $\tau\ge1$ ($\tau_{0}=1, \tau_{1}=\infty$).
Then $Q\left(0,z,\lambda\right)<\infty$ for $\lambda\ge-1$ and $Q\left(0,z,\lambda\right)=\infty$ for $\lambda<-1$.
In particular, $Q\left(0,z,-1\right) = z/\int_{1}^{\infty}\tau^{-2}e^{-\tau}d\tau$.
Thus if $z<\int_{1}^{\infty}\tau^{-2}e^{-\tau}d\tau (<1)$, there is no $\lambda$ such that $Q(0,z,\lambda)=1$.

\section{Statistics of Lineages \label{app:Cluster-size-distributiion}}

\subsection{Mean block size for $\beta > \beta_{c}$ in finite time \label{subapp:Interval-compliments}}
We consider the case in which a heritability threshold $\beta_{c}$ exists, and analyze  $\beta > \beta_{c}$.
Then $\Lambda_{\beta,z}=\omega_{0}=\tau_{0}^{-1}\ln\left(\beta z\right)$.
We choose $t>0$ large enough so that $\int_{t}^{\infty}k\left(\tau\right)d\tau\ll1$ and
let $m_t = m_t(m, \tau)$ denote the block size with cutoff $\left\lfloor \sfrac{t}{\tau}\right\rfloor$; we define $m_t = m $ if $m<\left\lfloor \sfrac{t}{\tau}\right\rfloor$
and otherwise $m_t = \left\lfloor \sfrac{t}{\tau}\right\rfloor$, where $\left\lfloor x\right\rfloor $ denotes the integer part of a real $x$.
Let $\rho(\tau) := \beta z e^{-\omega_{0}\tau}$.
Note that $\rho(\tau) < 1$ for $\tau>\tau_0$ and  $\rho(\tau) =1$ for $\tau=\tau_0$.
Using Eq. (9), 
the joint probability distribution of $m_t$ and $\tau$ is
\begin{equation}
\widetilde{p}_{\textrm{lin}}\left(m_{t},\tau\right) =
\widetilde{p}_{\textrm{lin}}\left(m_{t}\vert\tau\right) p_{\textrm{div}}(\tau),
\label{eq:ptilde-lin}
\end{equation}
where
\begin{equation}
\widetilde{p}_{\textrm{lin}}\left(m_{t}\vert\tau\right) = \begin{cases}
(1-\rho(\tau))\rho(\tau)^{m_t-1}, & 1\le m_{t}<\left\lfloor \sfrac{t}{\tau}\right\rfloor \\
\rho(\tau)^{m_t-1}, & m_{t}=\left\lfloor \sfrac{t}{\tau}\right\rfloor
\end{cases}
\end{equation}
Noting 
\begin{align}
\sum_{m_t=1}^{\left\lfloor \sfrac{t}{\tau}\right\rfloor} m_t\cdot\widetilde{p}_{\textrm{lin}}\left(m_{t}\vert\tau\right) =
& \sum_{m_t=1}^{\left\lfloor \sfrac{t}{\tau}\right\rfloor-1}m_t(1-\rho(\tau))\rho(\tau)^{m_t-1} \nonumber\\
&+  \left\lfloor \sfrac{t}{\tau}\right\rfloor \rho(\tau)^{\left\lfloor \sfrac{t}{\tau}\right\rfloor-1} \nonumber \\
= & \sum_{m_t=1}^{\left\lfloor \sfrac{t}{\tau}\right\rfloor}\rho(\tau)^{m_t-1}\nonumber\\
= &\begin{cases}
\frac{1-\rho(\tau)^{\left\lfloor \sfrac{t}{\tau}\right\rfloor}}{1-\rho(\tau)}, & \tau > \tau_0 \\
\left\lfloor \sfrac{t}{\tau_0}\right\rfloor, &  \tau = \tau_0, \label{eq:mt-sum}
\end{cases}
\end{align}
the mean block size with cutoff is defined as
\begin{align}
\overline{m_t}:= & \int_{0}^{t} d\tau \sum_{m_t=1}^{\left\lfloor \sfrac{t}{\tau}\right\rfloor} m_t\cdot \widetilde{p}_{\textrm{lin}}\left(m_{t},\tau\right) \nonumber \\
= &\int_{0}^{t}q_{\beta,z,\omega_{0}}\left(\tau\right)\frac{1-\rho(\tau)^{\left\lfloor \sfrac{t}{\tau}\right\rfloor }}{1-\rho(\tau)}d\tau \nonumber \\
 & + \left(1-Q\left(\beta,z,\omega_{0}\right)\right)\left\lfloor \sfrac{t}{\tau_{0}}\right\rfloor .
\end{align}
If $\int_{0}^{\infty}q_{\beta,z,\omega_{0}}\left(\tau\right)\left(1-\beta ze^{-\omega_{0}\tau}\right)^{-1}d\tau<\infty$,
\begin{equation}
\overline{m_t}=\left(1-Q\left(\beta,z,\omega_{0}\right)\right)\left\lfloor \frac{t}{\tau_{0}}\right\rfloor +\mathcal{O}\left(1\right) \label{eq:mt-asymptotics}
\end{equation}
If $k\left(\tau\right)\simeq A\left(\tau-\tau_{0}\right)^{\alpha-1}$
($A$ is a positive constant and $\alpha>1$) near $\tau_{0}$,
we can show that the leading term of $\overline{m_t}$ is $\left(1-Q\left(\beta,z,\omega_{0}\right)\right)\left\lfloor \sfrac{t}{\tau_0}\right\rfloor $
even when $\int_{0}^{\infty}q_{\beta,z,\omega_{0}}\left(\tau\right)\left(1-\beta ze^{-\omega_{0}\tau}\right)^{-1}d\tau=\infty$.
Let $1<\alpha\le2$.
We show $\int_{0}^{t}q_{\beta,z,\omega_{0}}\left(\tau\right)\frac{1-\left(\beta ze^{-\omega_{0}\tau}\right)^{\left\lfloor \sfrac{t}{\tau}\right\rfloor }}{1-\beta ze^{-\omega_{0}\tau}}d\tau$
is bounded by $\mathcal{O}\left(t^{2-\alpha}\right)$. Let $\tau_{c}:=\tau_{0}+\sfrac{\tau_{0}}{\left(\omega_{0}t\right)}$.
Then
\begin{equation}
\frac{1-\left(\beta ze^{-\omega_{0}\tau}\right)^{\left\lfloor \sfrac{t}{\tau}\right\rfloor }}{1-\beta ze^{-\omega_{0}\tau}}\le\begin{cases}
\left\lfloor \frac{t}{\tau_{0}}\right\rfloor , & \tau_{0}\le\tau<\tau_{c}\\
\left(1-\beta ze^{-\omega_{0}\tau}\right)^{-1}, & \tau\ge\tau_{c}
\end{cases}
\end{equation}
holds. Thus
\begin{align}
\int_{0}^{t} &q_{\beta,z,\omega_{0}}\left(\tau\right) \frac{1-\left(\beta ze^{-\omega_{0}\tau}\right)^{\left\lfloor \sfrac{t}{\tau}\right\rfloor }}{1-\beta ze^{-\omega_{0}\tau}}d\tau \le \\ \nonumber
&\left\lfloor \frac{t}{\tau_{0}}\right\rfloor \int_{\tau_{0}}^{\tau_{c}}q_{\beta,z,\omega_{0}}\left(\tau\right)d\tau +\int_{\tau_{c}}^{\tau_{m}}\frac{q_{\beta,z,\omega_{0}}\left(\tau\right)}{1-\beta ze^{-\omega_{0}\tau}}d\tau \\ &+\int_{\tau_{m}}^{\infty}\frac{q_{\beta,z,\omega_{0}}\left(\tau\right)}{1-\beta ze^{-\omega_{0}\tau}}d\tau \ , \nonumber
\end{align}
where $\tau_{m}>\tau_{c}$ is a constant independent
of $t$. For large enough $t$, $\tau_{m}$ can be close enough
to $\tau_{0}$ such that $k\left(\tau_{m}\right)\simeq A\left(\tau_{m}-\tau_{0}\right)^{\alpha-1}$and
$1-\beta ze^{-\omega_{0}\tau_{m}}\simeq\omega_{0}\left(\tau_{m}-\tau_{0}\right)$.
Note that the third term of RHS is constant. The first and second
terms are evaluated as
\begin{equation}
\int_{\tau_{0}}^{\tau_{c}}q_{\beta,z,\omega_{0}}\left(\tau\right)d\tau=\mathcal{O}\left(\left(\frac{\tau_{0}}{t}\right)^{\alpha-1}\right)
\end{equation}
and
\begin{equation}
\int_{\tau_{c}}^{\tau_{m}}\frac{q_{\beta,z,\omega_{0}}\left(\tau\right)}{1-\beta ze^{-\omega_{0}\tau}}d\tau=\begin{cases}
\mathcal{O}\left(\left(\frac{t}{\tau_{0}}\right)^{2-\alpha}\right), & 1<\alpha<2\\
\mathcal{O}\left(\ln\left(\frac{t}{\tau_{0}}\right)\right), & \alpha=2
\end{cases}
\end{equation}
Therefore, $\int_{0}^{t}q_{\beta,z,\omega_{0}}\left(\tau\right)\frac{1-\left(\beta ze^{-\omega_{0}\tau}\right)^{\left\lfloor \sfrac{t}{\tau}\right\rfloor }}{1-\beta ze^{-\omega_{0}\tau}}d\tau$
is increasing slower than $\mathcal{O}\left(t\right)$. Thus, for $\beta>\beta_{c}$,
the leading term of $\overline{m_t}$ is $\left(1-Q\left(\beta,z,\omega_{0}\right)\right)\left\lfloor \sfrac{t}{\tau}\right\rfloor $
not only when $\alpha>2$ but also when $1<\alpha\le2$. 
\subsection{Lineage IDT distribution for $\beta > \beta_{c}$}
Using the same approach as in the calculations above, we can compute $p_{\mathrm{lin}}(\tau)$ for $\beta > \beta_{c}$ as the $t \rightarrow \infty$ limit of the finite time version of expression Eq. (11); 
that is, we define
\begin{equation}
\widetilde{p}_{\textrm{lin}}\left(\tau\right) := \frac{\sum_{m_t=1}^{\left\lfloor \sfrac{t}{\tau}\right\rfloor} m_t \cdot \widetilde{p}_{\textrm{lin}}\left(m_t,\tau\right)}{\overline{m}_{t}}\ .
\label{eq:plin-of-tau-finite-t}
\end{equation}
We integrate the above distribution over a small interval $(\tau, \tau+\delta t)$, and using Eqs. \ref{eq:ptilde-lin} and \ref{eq:mt-sum}, we find
\begin{equation}
\int_{\tau}^{\tau+\delta t} \widetilde{p}_{\textrm{lin}}\left(\tau' \right) d\tau' = 
\begin{cases}
\frac{q_{\beta, z, \omega_0}(\tau)}{\overline m_t} \cdot \frac{1-\rho(\tau)^{\left\lfloor \sfrac{t}{\tau}\right\rfloor}}{1-\rho(\tau)} \delta t, & \tau > \tau_0 \vspace{5pt} \\
\frac{1 - Q(\beta, z, \omega_0)}{\overline m_t} \left\lfloor \sfrac{t}{\tau_0}\right\rfloor + \mathcal{O}(\delta t), &  \tau = \tau_0 .
\end{cases}
\end{equation}
Using Eq. \ref{eq:mt-asymptotics}, we see that as $t \rightarrow \infty$, $ \widetilde{p}_{\textrm{lin}}\left(\tau \right) \rightarrow \delta(\tau - \tau_0)$.

\subsection{On-lineage mean IDT}

For $\beta<\beta_{c}$, $\overline{m}_{\beta,z}<\infty$ always holds, while for $\beta> \beta_{c}$, $\overline{m}_{\beta,z} = \infty$, but $\lim_{\beta\uparrow\beta_{c}}\overline{m}_{\beta,z}$ may diverge or converge depending on $k\left(\tau\right)$. The distribution $p_{\mathrm{lin}}\left(\tau\right)$ converges to a density as $\beta\uparrow\beta_{c}$ when $\lim_{\beta\uparrow\beta_{c}}\overline{m}_{\beta,z}<\infty$.
In contrast, when $\lim_{\beta\uparrow\beta_{c}}\overline{m}_{\beta,z}=\infty$,
$p_{\mathrm{lin}}\left(\tau\right)$ weakly converges to $\delta\left(\tau-\tau_{0}\right)$
as $\beta\uparrow\beta_{c}$ because $\lim_{\beta\uparrow\beta_{c}}\int_{\tau_{0}+\epsilon}^{\infty}p_{\mathrm{lin}}\left(\tau\right)d\tau=0$
for any $\epsilon>0$.
Thus $\lim_{\beta\uparrow\beta_{c}}\overline{m}_{\beta,z}^{-1}$
is an important quantity to distinguish continuous or discontinuous
transition; $\lim_{\beta\uparrow\beta_{c}}\overline{m}_{\beta,z}^{-1}>0$ indicates a discontinuous transition and $\lim_{\beta\uparrow\beta_{c}}\overline{m}_{\beta,z}^{-1}=0$ a continuous one.
Under the condition $\int_{\tau_{0}}^{\infty}d\tau\sfrac{k\left(\tau\right)}{(\tau-\tau_{0})}<\infty$, 
\begin{align}
 & \int_{\tau_{0}}^{\infty}\frac{\left(\tau-\tau_{0}\right)q_{\beta_{c},z,\omega_{0}}\left(\tau\right)}{1-\beta_{c}ze^{-\omega_{0}\tau}}d\tau\nonumber \\
= & \frac{1-\beta_{c}}{\beta_{c}}\int_{\tau_{0}}^{\infty}\frac{\left(\tau-\tau_{0}\right)\left(\beta_{c}z\right)^{1-\tau/\tau_{0}}k\left(\tau\right)}{\left(1-\left(\beta_{c}z\right)^{1-\tau/\tau_{0}}\right)^{2}}d\tau<\infty
\end{align}
 holds. Therefore, the on-lineage mean IDT, $\overline{\tau}_{\beta,z}=\int_{0}^{\infty}\tau p_{\mathrm{lin}}\left(\tau\right)d\tau$
shows continuous/discontinuous phase transition depending on convergence/divergence
of $\lim_{\beta\uparrow\beta_{c}}\overline{m}_{\beta,z}$: 
\begin{align}
\lim_{\beta\uparrow\beta_{c}}\overline{\tau}_{\beta,z}-\tau_{0}\nonumber 
&=  \frac{1}{\lim_{\beta\uparrow\beta_{c}}\overline{m}_{\beta,z}}\int_{\tau_{0}}^{\infty}\frac{\left(\tau-\tau_{0}\right)q_{\beta_{c},z,\omega_{0}}\left(\tau\right)}{1-\beta_{c}ze^{-\omega_{0}\tau}}d\tau\nonumber \\
 & \begin{cases}
>0, & \lim_{\beta\uparrow\beta_{c}}\overline{m}_{\beta,z}<\infty\\
=0, & \lim_{\beta\uparrow\beta_{c}}\overline{m}_{\beta,z}=\infty
\end{cases}
\end{align}
Applying this result to Eq. \eqref{eq:growthrate-logbeta-derivative},
$\partial_{\ln\beta}\Lambda\left(\beta,z\right)$ is discontinuous
at $\beta=\beta_{c}$ if $\lim_{\beta\uparrow\beta_{c}}\overline{m}_{\beta,z}<\infty$
and otherwise continuous. If $\lim_{\tau\downarrow\tau_{0}}\frac{k\left(\tau\right)}{\left(\tau-\tau_{0}\right)^{\alpha-1}}$
converges a positive finite constant where $\alpha>1$, 
\begin{equation}
\lim_{\beta\uparrow\beta_{c}}\overline{m}_{\beta,z}\begin{cases}
<\infty, & \alpha>2\\
=\infty,  & 1 <\alpha\le2
\end{cases}
\end{equation}
and therefore
\begin{equation}
\lim_{\beta\uparrow\beta_{c}}\overline{\tau}_{\beta,z}-\tau_{0}\begin{cases}
>0, & \alpha>2\\
=0, & 1<\alpha\le2
\end{cases}
\end{equation}

\subsection{Block size distribution \label{subapp:Cluster-size-dist}}

The lineage distribution of $m$ for $\beta < \beta_{c}$ is
\begin{align}
p_{\mathrm{lin}}^{\mathrm{M}}\left(m\right) &:=
 \int_{0}^{\infty} p_{\textrm{lin}}\left(m,\tau\right) d\tau \nonumber \\
&= \frac{1-\beta}{\beta}\left(\beta ze^{-\Lambda_{\beta,z}\tau_{0}}\right)^{m} 
\hspace{-3pt} \int_{0}^{\infty} \hspace{-3pt} k\left(\tau\right)e^{-m\Lambda_{\beta,z}\left(\tau-\tau_{0}\right)}d\tau
\end{align}
To obtain an analytical expression for the integral $\int_{0}^{\infty}k\left(\tau\right)e^{-m\Lambda_{\beta,z}\left(\tau-\tau_{0}\right)}d\tau$,
we use a gamma distribution $k\left(\tau\right)=\Gamma\left(\alpha\right)^{-1}\theta^{-\alpha}\left(\tau-\tau_{0}\right)^{\alpha-1}e^{-\left(\tau-\tau_{0}\right)/\theta}$.
Then $\int_{0}^{\infty}k\left(\tau\right)e^{-m\Lambda_{\beta,z}\left(\tau-\tau_{0}\right)}d\tau=\left(1+m\Lambda_{\beta,z}\theta\right)^{-\alpha}$.
Thus 
\begin{equation}
p_{\mathrm{lin}}^{\mathrm{M}}\left(m\right)=\frac{1-\beta}{\beta}\left(\beta ze^{-\Lambda_{\beta,z}\tau_{0}}\right)^{m}\left(1+m\Lambda_{\beta,z}\theta\right)^{-\alpha}.
\end{equation}
For $\beta<\beta_{c}$, $p_{\mathrm{lin}}^{\text{M}}\left(m\right)$
exponentially decays with typical length $\left(\Lambda_{\beta,z}\tau_{0} - \ln \beta z \right)^{-1}$.
As $\beta$ approaches $\beta_{c}$, $\beta ze^{-\Lambda_{\beta,z}\tau_{0}}$
approaches 1 and thereby $\left(1-\beta ze^{-\Lambda_{\beta,z}\tau_{0}}\right)^{-1}$
diverges, meaning that the typical block size becomes much larger
as $\beta$ is approaching $\beta_{c}$. In the limit $\beta\uparrow\beta_{c}$,
$p_{\mathrm{lin}}^{\text{M}}\left(m\right)$ decays with $m^{-\alpha}$
for large $m$. For general forms of $k\left(\tau\right)$, the integral
$\int_{\tau_{0}}^{\infty}k\left(\tau\right)e^{-m\Lambda_{\beta,z}\left(\tau-\tau_{0}\right)}d\tau$,
for large enough $m$, is contributed dominantly by the values of
$k\left(\tau\right)$ for $\tau$ near $\tau_{0}$. Thus if $k\left(\tau\right)\simeq A\left(\tau-\tau_{0}\right)^{\alpha-1}$
($A$ is positive constant and $\alpha>1$), a power law $p_{\mathrm{lin}}^{\mathrm{M}}\left(m\right)\sim m^{-\alpha}$
holds in the limit $\beta\uparrow\beta_{c}$.

\section{Ancestral distribution\label{app:Ancestral-distribution}}
Let $\hat{N}(\tau;t^{\prime})$ denote the expected
number of cells after time $t^{\prime}$ has passed since a cell completed a cell-cycle with IDT $\tau$. This satisfies
\begin{equation}
\hat{N}(\tau;t^{\prime})=z\int_{0}^{t^{\prime}}\hat{N}(\tau^{\prime};t^{\prime}-\tau^{\prime})K\left(\tau^{\prime},\tau\right)d\tau^{\prime}.\label{eq:gentimedist-backward-equation}
\end{equation}
In a stationary growing population, it is natural to assume $\hat{N}(\tau;t^{\prime})=\phi\left(\tau\right)e^{\Lambda_{\beta,z}t^{\prime}}$, where $\phi(\tau) \geq 0$.
Then Eq. \eqref{eq:gentimedist-backward-equation} is rewritten as
\begin{equation}
\phi(\tau)=z\int_{0}^{\infty}\phi(\tau^{\prime})e^{-\Lambda_{\beta,z}\tau^{\prime}}K\left(\tau^{\prime},\tau\right)d\tau^{\prime}.\label{eq:backward-eq-general}
\end{equation}
where the limit $t^{\prime}\rightarrow\infty$ is taken. Since $\int_{0}^{\infty}p_{\text{div}}\left(\tau\right)\hat{N}(\tau;t)d\tau=e^{\Lambda_{\beta,z}t}$
holds, we have 
\begin{equation}
\int_{0}^{\infty}p_{\text{div}}\left(\tau\right)\phi(\tau)d\tau=1 \ ,
\label{eq:phi_norm}
\end{equation}
where the quantity $p_{\mathrm{div}}\left(\tau\right)\phi(\tau)$
is the distribution of $\tau$ of the ancestor in the
infinite past in a steadily growing population, which is also called the ancestral distribution \cite{Hermisson2002}; this requires that $\phi(\tau)$ is an integrable function with respect to $p_{\mathrm{div}}(\tau)d \tau$ (i.e. $\phi \in L^1(p_{\mathrm{div}})$, where $p_{\mathrm{div}}$ is a measure on $[\tau_0, \infty)$).  For example, if $K\left(\tau,\tau^{\prime}\right)=k\left(\tau\right)$,
$p_{\mathrm{div}}\left(\tau\right)=ze^{-\Lambda_{\beta,z}\tau}k\left(\tau\right)$
and $\phi\left(\tau\right)=1$, therefore $p_{\mathrm{lin}}\left(\tau\right)=p_{\mathrm{div}}\left(\tau\right)$.
The empirical distribution of IDT on lineages in a population converges
to the ancestral distribution as $t\rightarrow\infty$ (Sec. \ref{app:relationship-to-thermodynamics}).
In the main text, we derived $p_{\mathrm{lin}}\left(\tau\right)$ as Eq. (12). 
Here, we directly compute $p_{\mathrm{lin}}\left(\tau\right)=p_{\text{div}}\left(\tau\right)\phi(\tau)$. 

In the Lebowitz-Rubinow model, Eq. \eqref{eq:backward-eq-general} yields
\begin{align}
\phi(\tau) & =\beta ze^{-\Lambda_{\beta,z}\tau}\phi\left(\tau\right)\nonumber \\
 & +\left(1-\beta\right)z\int_{0}^{\infty}\phi(\tau^{\prime})e^{-\Lambda_{\beta,z}\tau^{\prime}}k\left(\tau^{\prime}\right)d\tau^{\prime}.\label{eq:stationary-backward-LR}
\end{align}
This implies, there exists a constant $C_{2}\in\R$
\begin{equation}
\phi(\tau) = \frac{C_{2}}{1-\beta ze^{-\Lambda_{\beta,z}\tau}} \label{eq:stat-backward-LR-sol}
\end{equation}
for $\tau\in\left[\tau_{0},\tau_{1}\right]$ if $\beta<\beta_c$ and for $\tau\neq\tau_{*}$ otherwise.
Substituting Eq. \eqref{eq:stat-backward-LR-sol} to Eq. \eqref{eq:stationary-backward-LR}, we have
\begin{equation}
C_{2}\left(1 - Q\left(\beta,z,\Lambda_{\beta,z}\right)\right) = 0.
\end{equation}

If $\beta<\beta_{c}$, $Q\left(\beta,z,\Lambda_{\beta,z}\right)=1$ and we find $C_{2} = \overline{m}_{\beta,z}^{-1}$ by the normalization Eq. \eqref{eq:phi_norm}.

If $\beta=\beta_{c}$, $Q\left(\beta,z,\Lambda_{\beta,z}\right)=1$ holds too and thus $C_{2} = 1/\lim_{\beta\uparrow\beta_{c}}\overline{m}_{\beta,z}$
if $\lim_{\beta\uparrow\beta_{c}}\overline{m}_{\beta,z} < \infty$.
If $\lim_{\beta\uparrow\beta_{c}}\overline{m}_{\beta,z} = \infty$, no strictly positive $C_{2}$ satisfy the normalization Eq. \eqref{eq:phi_norm}.
Additionally, $C_{2} = 0$ leads $\int_{0}^{\infty}p_{\text{div}}\left(\tau\right)\phi(\tau)d\tau=0$.
Thus there is no $\phi$ satisfying the normalization Eq. \eqref{eq:phi_norm} if $\lim_{\beta\uparrow\beta_{c}}\overline{m}_{\beta,z} = \infty$.

If $\beta>\beta_{c}$, $Q\left(\beta,z,\Lambda_{\beta,z}\right)<1$ implies $C_{2}=0$; i.e. $\phi(\tau) = 0$ for $\tau\neq\tau_{*}$ ($\tau_{*}$ is defined byEq. \eqref{smeq:taustar}).
Noting for $\phi \in L^1(p_{\mathrm{div}})$ that $p_{\mathrm{div}}$ includes an atomic measure at $\tau_0$ (see Eq. \ref{eq:stationary_solution}), by the normalization condition \eqref{eq:phi_norm} we have $\phi\left(\tau_0\right)=C_{1}^{-1} = (1 - Q_0(\beta, z))^{-1}$ and $\phi(\tau > \tau_0) = 0$;
equivalently, $\phi\left(\tau\right) = C_{1}^{-1} \delta_{\tau_0, \tau}$, where we define the Kronecker delta function $\delta_{x, y} = 1$ for $x = y$, and $\delta_{x, y} = 0$ for $x \neq y$.

In summary,
\begin{equation}
\phi\left(\tau\right)=\begin{cases}
\begin{array}{c}
\overline{m}_{\beta,z}^{-1}\cdot\left(1-\beta ze^{-\Lambda_{\beta,z}\tau}\right)^{-1},\\
\left(1-Q\left(\beta,z,\Lambda_{\beta,z}\right)\right)^{-1}\cdot\delta_{\tau_{0},\tau},
\end{array} & \begin{array}{c}
0\le\beta\le\beta_{c}\\
\beta_{c}<\beta<1
\end{array}\end{cases}\label{eq:stationary-lefteigenvector}
\end{equation}
Multiplying $\phi\left(\tau\right)$ by $p_{\text{div}}\left(\tau\right)$,
we have
\begin{equation}
p_{\mathrm{lin}}\left(\tau\right)=\begin{cases}
\begin{array}{c}
\overline{m}_{\beta,z}^{-1}\cdot\frac{\left(1-\beta\right)k\left(\tau\right)ze^{-\Lambda_{\beta,z}\tau}}{\left(1-\beta ze^{-\Lambda_{\beta,z}\tau}\right)^{2}},\\
\delta\left(\tau-\tau_{0}\right),
\end{array} & \begin{array}{c}
0\le\beta\le\beta_{c}\\
\beta_{c}<\beta<1
\end{array}\end{cases}\label{eq:stationary-ancestral-dist}
\end{equation}
Note that $\phi$ and $p_{\textrm{lin}}$ exist at $\beta=\beta_{c}$ if $\lim_{\beta\uparrow\beta_{c}}\overline{m}_{\beta,z}<\infty$.

\section{Derivatives of growth rate function \label{app:Derivatives-of-growth}}

For $\beta<\beta_{c}$, $Q\left(\beta,z,\Lambda_{\beta,z}\right)=1$
determines $\Lambda_{\beta,z}$. Taking the first derivatives of both
side of $Q\left(\beta,z,\Lambda_{\beta,z}\right)=1$ with respect
to $\ln z$, we obtain

\begin{equation}
\frac{\partial\Lambda_{\beta,z}}{\partial\ln z}=\overline{\tau}_{\beta,z}^{-1}.\label{eq:growthrate-logz-derivative}
\end{equation}
Namely, the inverse of ancestral mean IDT $\overline{\tau}_{\beta,z}^{-1}$
is equal to the first derivative of $\Lambda_{\beta,z}$ with respect
to $\ln z$. For $\beta>\beta_{c}$, Eq. \eqref{eq:growthrate-logz-derivative}
is clear because $\overline{\tau}_{\beta,z}=\tau_{0}$. On the other
hand, the derivative of $\Lambda_{\beta,z}$ with respect to $\ln\beta$
is
\begin{equation}
\frac{\partial\Lambda_{\beta,z}}{\partial\ln\beta}=\frac{1}{\overline{\tau}_{\beta,z}}\left(1-\left(1-\beta\right)^{-1}\overline{m}_{\beta,z}^{-1}\right).\label{eq:growthrate-logbeta-derivative}
\end{equation}
Eq. \eqref{eq:growthrate-logbeta-derivative} is also derived by taking
the first derivatives of both side of $Q\left(\beta,z,\Lambda_{\beta,z}\right)=1$
and holds as well for $\beta>\beta_{c}$. $\frac{\partial\Lambda_{\beta,z}}{\partial\ln\beta}>0$
holds because for $\beta<\beta_{c}$, 
\begin{align}
 & \overline{m}_{\beta,z}-\left(1-\beta\right)^{-1}\nonumber \\
 & =\int_{\tau_{0}}^{\infty}\frac{\left(1-\beta\right)ze^{-\Lambda_{\beta,z}\tau}k\left(\tau\right)}{\left(1-\beta ze^{-\Lambda\tau}\right)^{2}}d\tau-\int_{\tau_{0}}^{\infty}\frac{k\left(\tau\right)}{1-\beta}d\tau\nonumber \\
 & =\int_{\tau_{0}}^{\infty}\left(\frac{\left(1-\beta\right)ze^{-\Lambda_{\beta,z}\tau}}{1-\beta ze^{-\Lambda_{\beta,z}\tau}}-1\right)\frac{1-\beta^{2}ze^{-\Lambda_{\beta,z}\tau}}{1-\beta ze^{-\Lambda_{\beta,z}\tau}}\frac{k\left(\tau\right)}{1-\beta}d\tau\nonumber \\
 & >\int_{\tau_{0}}^{\infty}\left(\frac{\left(1-\beta\right)ze^{-\Lambda_{\beta,z}\tau}}{1-\beta ze^{-\Lambda_{\beta,z}\tau}}-1\right)\frac{k\left(\tau\right)}{1-\beta}d\tau=0.
\end{align}
Together, $\Lambda_{\beta,z}$ increases as $\beta$ or $z$ increases.
Since $\overline{m}_{\beta,1}=\left(1-\beta\right)^{-1}$ holds, this
inequality implies that the mean cluster size on lineage is always
greater in growing population ($z>1$) rather than in isolation.

\section{Thermodynamic analogy for lineage structures in the Lebowitz-Rubinow model \label{app:relationship-to-thermodynamics}}
Here we show that the population structure of lineages in the Lebowitz-Rubinow model is mapped to a statistical mechanical ensemble, which yields the thermodynamic interpretation of population growth rate and other quantities. The results hold for general forms of the transition kernel $K\left(\tau,\tau^{\prime}\right)$. 

Let $\Omega:=$$\left\{ \left(a,\tau\right)\vert0\le a\le\tau;\tau_{0}\le\tau<\infty\right\} $
be the domain for age $a$ and IDT $\tau$. The initial
condition is given by $n\left(a,\tau;0\right)=N_{0}\, p_{0}\left(a,\tau\right)$
where $N_{0}$ is the initial population size and $p_{0}$ is a probability
distribution on $\Omega$. Let $n^{\left(D\right)}\left(a,\tau;t\right)da\, d\tau$
denote the expected number of offspring in the interval $(a, a + da) \times (\tau, \tau+d\tau)$ at time $t$ after $D$ divisions. Recursively using Eq. \eqref{eq:PDE-age-idt-model-solution} and Eq. \eqref{eq:PDE-age-idt-model-BC}, we have
\begin{equation}
n\left(a,\tau;t\right)=N_{0}\sum_{D=0}^{\infty}z^{D}P^{\left(D\right)}\left(a,\tau,t\right)
\end{equation}
where $P^{\left(D\right)}\left(a,\tau;t\right)$ denotes the probability
distribution of $\left(a,\tau\right)$ at time $t$ after $D$ divisions
in isolation and the expected population size at time $t$ is
\begin{equation}
N_{t}=\iint_{\Omega} n\left(a,\tau;t\right)da \, d\tau = N_{0}\sum_{D=0}^{\infty}z^{D}P_{\mathrm{iso}}\left(D;t\right)
\end{equation}
where
\begin{equation}
P_{\mathrm{iso}}\left(D;t\right):=\iint_{\Omega} P^{\left(D\right)}\left(a,\tau,t\right) da \, d\tau
\end{equation}
denotes the probability distribution of number of divisions along
lineages in isolation. The probability distribution of number of divisions
along lineage in population is
\begin{equation}
P_{\mathrm{pop}}\left(D;t\right):=\frac{\iint_{\Omega} n^{\left(D\right)}\left(a,\tau;t\right)da \, d\tau}{N_{t}}=\frac{N_{0}}{N_{t}}z^{D}P_{\mathrm{iso}}\left(D;t\right)
\end{equation}
$P_{\mathrm{iso}}$ and $P_{\mathrm{pop}}$ are referred to as ``chronological''
and ``retrospective'' distribution respectively in \cite{Nozoe2017}.

A lineage can be identified as the sequences $\left(m_{0},\cdots,m_{S}\right)$
and $\left(x_{0},\cdots,x_{S-1},\tau\right)$ where $m_{j}$ denotes the
$j$-th block size with IDT $x_{j}$ $\left(x_{S}=\tau\right)$.
$S$ denotes the number of switching to the other IDT and $D=m_{0}+\cdots+m_{S}-1$
denotes the number of divisions on the lineage. Then, the joint probability
distribution of number of divisions $D$ and number of switches
$S$ on a lineage in isolation can be defined as
\begin{equation}
P_{\mathrm{iso}}\left(D,S;t\right):=\iint_{\Omega} P^{\left(D,S\right)}\left(a,\tau,t\right) da \, d\tau
\end{equation}
where $P^{\left(D,S\right)}\left(a,\tau;t\right)$ denotes probability
density of $\left(a,\tau\right)$ at time $t$ after $D$ divisions
and $S$ switches in isolation. Therefore
\begin{equation}
\frac{N_{t}}{N_{0}}=\sum_{D=0}^{\infty}\sum_{S=0}^{D}z^{D}P_{\mathrm{iso}}\left(D,S;t\right)
\end{equation}
holds and it is reasonable to define 
\begin{equation}
P_{\mathrm{pop}}\left(D,S;t\right):=\frac{N_{0}}{N_{t}}z^{D}P_{\mathrm{iso}}\left(D,S;t\right)
\end{equation}
It is noticeable that $P_{\mathrm{iso}}\left(D,S;t\right)$ depends
on $\beta$ in the form of $\left(1-\beta\right)^{S}\beta^{D-S}$. 

In this lineage formulation, the long-term growth rate $\Lambda_{\beta,z}$
can be interpreted as the cumulant generating function scaled by time.
Let $D$ denote the number of divisions on lineage and $S$ denote
the number of switching to the other $\tau$ on lineage. Let $\Phi_{\beta}\left(\xi,\eta;t\right)$
be the cumulant generating function of $D$ and $S$: 
\begin{equation}
\Phi_{\beta}\left(\xi,\eta;t\right):=\ln\sum_{D,S}P_{\mathrm{iso}}\left(D,S;t\right)e^{\xi D+\eta S}.
\end{equation}
$\Phi_{\beta}\left(\xi,\eta\right)$ depends on $\beta$ via $\left(1-\beta\right)^{S}\beta^{D-S}$
in $P_{\mathrm{iso}}\left(D,S;t\right)$. The biased probability distribution
of $\left(D,S\right)$ is defined by
\begin{equation}
P_{\xi,\eta}\left(D,S;t\right):=P_{\mathrm{iso}}\left(D,S;t\right)e^{\xi D+\eta S}e^{-\Phi_{\beta}\left(\xi,\eta;t\right)},
\end{equation}
Note that $\Phi_{\beta}\left(\ln z,0\right)=\ln\sfrac{N_{t}}{N_{0}}$
and $P_{\ln z,0}\left(D,S;t\right)=P_{\mathrm{pop}}\left(D,S;t\right)$.
In particular, $\Phi_{\beta}\left(\ln z,0\right)=t\Lambda_{\beta,z}$
holds at steady state. Below we assume steady state, that is, the
initial distribution $p_{0}$ the stationary solution. Let $\hat{D}_{t}$
and $\hat{S}_{t}$ denote the random variables taking values $D$
and $S$ along lineage with time length $t$ and let $\left\langle \cdot\right\rangle _{\mathrm{pop}}$
denote the expectation with respect to $P_{\mathrm{pop}}\left(D,S;t\right)$.
For any positive integer $j$, 
\begin{equation}
\frac{\partial^{j}\Lambda_{\beta,z}}{\left(\partial\ln z\right)^{j}}=\left.\frac{1}{t}\frac{\partial^{j}\Phi_{\beta}\left(\xi,0\right)}{\partial\xi^{j}}\right|_{\xi=\ln z}
\end{equation}
holds. In particular, for $j=1$, 
\begin{equation}
\frac{\partial\Lambda_{\beta,z}}{\partial\ln z}=\left.\frac{1}{t}\frac{\partial\Phi_{\beta}\left(\xi,0\right)}{\partial\xi}\right|_{\xi=\ln z}=\left\langle \frac{\hat{D}_{t}}{t}\right\rangle _{\text{pop}}
\end{equation}
Together with Eq. \eqref{eq:growthrate-logz-derivative}, we have
\begin{equation}
\left\langle \frac{\hat{D}_{t}}{t}\right\rangle _{\text{pop}}=\left\langle \tau\right\rangle _{\mathrm{anc}}^{-1}
\end{equation}
For $j=2$, 
\begin{equation}
\frac{\partial^{2}\Lambda\left(\beta,z\right)}{\left(\partial\ln z\right)^{2}}=\left.\frac{1}{t}\frac{\partial^{2}\Phi_{\beta}\left(\xi,0\right)}{\partial\xi^{2}}\right|_{\xi=\ln z}=\frac{1}{t}\mathrm{Var}\left[\hat{D}_{t}\right]_{\mathrm{pop}}
\end{equation}
holds, where $\mathrm{Var}\left[\hat{D}_{t}\right]_{\mathrm{pop}}=\left\langle \hat{D}_{t}^{2}\right\rangle _{\mathrm{pop}}-\left\langle \hat{D}_{t}\right\rangle _{\mathrm{pop}}^{2}$.
If $\beta>\beta_{c}$, $\Lambda_{\beta,z}=\tau_{0}^{-1}\ln\left(\beta z\right)$
implies $t^{-1}\mathrm{Var}\left[\hat{D}_{t}\right]_{\mathrm{pop}}=0$
holds. In terms of law of large number, a random variable $\sfrac{\hat{D}_{t}}{t}$
converges to $\left\langle \tau\right\rangle _{\mathrm{anc}}^{-1}$
as $t\rightarrow\infty$ with respect to $P_{\mathrm{pop}}\left(D,S;t\right)$.
We mean this convergence in probability $P_{\mathrm{pop}}$ by $\mathrm{plim}_{t\rightarrow\infty}\sfrac{\hat{D}_{t}}{t}=\left\langle \tau\right\rangle _{\mathrm{anc}}^{-1}$
(``plim'' for probability limit). In addition, the derivative of
$\Lambda_{\beta,z}$ with respect to $\ln\beta$ can be expressed
by derivatives of $\Phi_{\beta}\left(\xi,\eta\right)$ with respect
to $\xi$ and $\eta$. For a small $\epsilon>0$, 
\begin{equation}
\left(1-\beta e^{\epsilon}\right)^{S}\left(\beta e^{\epsilon}\right)^{D-S}=\left(1-\beta\right)^{S}\beta^{D-S}e^{\epsilon D-\frac{\epsilon S}{1-\beta}}+\mathcal{O}\left(\epsilon^{2}\right).
\end{equation}
That is, $\Phi_{\beta e^{\epsilon}}\left(\xi,\eta\right)=\Phi_{\beta}\left(\xi+\epsilon,\eta-\frac{\epsilon}{1-\beta}\right)+\mathcal{O}\left(\epsilon^{2}\right)$
and thereby
\begin{equation}
\frac{\partial\Phi_{\beta}\left(\xi,\eta\right)}{\partial\ln\beta}=\left(\frac{\partial}{\partial\xi}-\frac{1}{1-\beta}\frac{\partial}{\partial\eta}\right)\Phi_{\beta}\left(\xi,\eta\right)
\end{equation}
holds. Therefore, using the relation
\begin{equation}
\left.\frac{\partial\Phi_{\beta}\left(\ln z,\eta\right)}{\partial\eta}\right|_{\eta=0}=\left\langle \frac{\hat{S}_{t}}{t}\right\rangle _{\text{pop}}
\end{equation}
we have
\begin{align}
\frac{\partial\Lambda\left(\beta,z\right)}{\partial\ln\beta} & =\left.\frac{\partial\Phi_{\beta}\left(\xi,0\right)}{\partial\xi}\right|_{\xi=\ln z}-\frac{1}{1-\beta}\left.\frac{\partial\Phi_{\beta}\left(\ln z,\eta\right)}{\partial\eta}\right|_{\eta=0}\nonumber \\
 & =\left\langle \frac{\hat{D}_{t}}{t}\right\rangle _{\text{pop}}-\frac{1}{1-\beta}\left\langle \frac{\hat{S}_{t}}{t}\right\rangle _{\text{pop}}\label{eq:growthrate-logbeta-derivative-2}
\end{align}
Comparing it to Eq. \eqref{eq:growthrate-logbeta-derivative}, we have 
\begin{equation}
\left\langle \frac{\hat{S}_{t}}{t}\right\rangle _{\text{pop}}=\overline{m}_{\beta,z}^{-1}\overline{\tau}_{\beta,z}^{-1}
\end{equation}
In terms of law of large number,
\begin{align}
\left\langle \frac{\hat{S}_{t}}{t}\right\rangle _{\text{pop}} & =\mathrm{plim}_{t\rightarrow\infty}\frac{\hat{S}_{t}}{t}\nonumber \\
 & =\mathrm{plim}_{t\rightarrow\infty}\frac{\hat{S}_{t}}{\hat{D}_{t}}\cdot\mathrm{plim}_{t\rightarrow\infty}\frac{\hat{D}_{t}}{t}\nonumber \\
 & =\overline{m}_{\beta,z}^{-1}\cdot\overline{\tau}_{\beta,z}^{-1}
\end{align}
implies $\mathrm{plim}_{t\rightarrow\infty}\sfrac{\hat{S}_{t}}{\hat{D}_{t}}=\overline{m}_{\beta,z}^{-1}$,
which is reasonable in that $\overline{m}_{\beta,z}$ represents the
mean cluster size on lineage in population. 

The cumulant generating function $t\cdot\Phi_{\beta}\left(\xi,\eta\right)$
plays the role of free energy in thermodynamics. A lineage is analogous to a specific configuration of a macroscopic system. The total time $t$, denoting the lineage length, corresponds to the physical size of the system, i.e. its volume. $\left\langle \hat{D}_{t}\right\rangle _{\text{pop}}$
and $\left\langle \hat{S}_{t}\right\rangle _{\text{pop}}$ are extensive
variables, like energy and entropy in thermodynamics, as they are typically proportional to $t$. $\xi\left(=\ln z\right)$ and $\eta$ are the conjugate variables to $\left\langle \hat{D}_{t}\right\rangle _{\text{pop}}$
and $\left\langle \hat{S}_{t}\right\rangle _{\text{pop}}$; and as in thermodynamics, the first derivative of free energy with respect to an intensive variable becomes its conjugate extensive variable. In this point of view, a discontinuous transition of $\overline{\tau}_{\beta,z}^{-1}$ at $\beta=\beta_{c}$ is a first-order phase transition. 

\section{Numerical simulations \label{app:Numerical-simulations}}

\subsection{IDT distribution in isolation}

In simulations, $k\left(\tau\right)$ is chosen as
\begin{equation}
k\left(\tau\right)=\begin{cases}
\frac{\left(\tau-\tau_{0}\right)^{\alpha-1}}{\Gamma\left(\alpha\right)\theta^{\alpha}}e^{-\sfrac{\left(\tau-\tau_{0}\right)}{\theta}}, & \tau\ge\tau_{0}\\
0, & otherwise
\end{cases}
\end{equation}
To fix division time scale, we always set the mean inter-division
time in isolation to 1: $\int_{0}^{\infty}\tau k\left(\tau\right)d\tau=\tau_{0}+\alpha\theta=1$. 

\subsection{Growth rate }

To sample $\left(a,\tau\right)$ for initial conditions, we need to
compute $\Lambda_{\beta,z}$. For $\beta=0$, $\Lambda_{0,z}$ is
computed using Lambert W function $W\left(x\right)$: $W\left(x\right)$
satisfies $W\left(x\right)e^{W\left(x\right)}=x$ and $W\left(x\right)\ge-1$
with its domain $\left[-e^{-1},\infty\right)$. Since $\Lambda_{0,z}$
satisfies
\begin{align}
1 & =\int_{0}^{\infty}ze^{-\lambda\tau}k\left(\tau\right)d\tau=\frac{ze^{-\tau_{0}\Lambda_{0,z}}}{\left(1+\theta\Lambda_{0,z}\right)^{\alpha}}\nonumber \\
 & =\left(\frac{\frac{\tau_{0}}{\alpha\theta}e^{\sfrac{\tau_{0}}{\left(\alpha\theta\right)}}z^{\sfrac{1}{\alpha}}}{\frac{\tau_{0}}{\alpha}\left(\Lambda_{0,z}+\theta^{-1}\right)\exp\left(\frac{\tau_{0}}{\alpha}\left(\Lambda_{0,z}+\theta^{-1}\right)\right)}\right)^{\alpha}
\end{align}
Thus we get 
\begin{equation}
\Lambda_{0,z}=\frac{\alpha}{\tau_{0}}W\left(\frac{\tau_{0}}{\alpha\theta}e^{\sfrac{\tau_{0}}{\left(\alpha\theta\right)}}z^{\sfrac{1}{\alpha}}\right)-\frac{1}{\theta}
\end{equation}
 When $\beta>0$, we numerically solve the integral equation
\begin{equation}
\int_{0}^{\infty}\frac{\left(1-\beta\right)ze^{-\lambda\tau}k\left(\tau\right)}{1-\beta ze^{-\lambda\tau}}d\tau=1
\end{equation}
in the range of $\lambda>\max\left(0,\tau_{0}^{-1}\ln\left(\beta z\right)\right)$
to get $\Lambda_{\beta,z}$, and if there is no solution in that range,
we set $\Lambda_{\beta,z}=\tau_{0}^{-1}\ln\left(\beta z\right)$. 

\subsection{Initial conditions}

Let $\Omega:=$$\left\{ \left(a,\tau\right)\vert0\le a\le\tau;\tau_{0}\le\tau<\infty\right\} $
be the domain for age $a$ and IDT $\tau$. The stationary
joint probability distribution of $\left(a,\tau\right)$ is given
by Eq (\ref{eq:pst_pbirth_relation}). For LR model, this becomes
\begin{align}
\varphi_{\beta,z}\left(a,\tau\right) & :=\frac{z}{z-1}\Lambda_{\beta,z}e^{-\Lambda_{\beta,z}a}\nonumber \\
 & \times\left(\left(1-G_{\beta,z}\right)\delta\left(\tau-\tau_{0}\right)+g_{\beta,z}\left(\tau\right)\right)
\end{align}
where
\begin{equation}
g_{\beta,z}\left(\tau\right)=\frac{\left(1-\beta\right)k\left(\tau\right)}{1-\beta ze^{-\Lambda_{\beta,z}\tau}}
\end{equation}
and 
\begin{equation}
G_{\beta,z}=\int_{0}^{\infty}g_{\beta,z}\left(\tau\right)d\tau
\end{equation}
Noting that $\varphi_{\beta,z}\left(a,\tau\right)$ is proportional
to the product of $\Lambda_{\beta,z}e^{-\Lambda_{\beta,z}a}$ and
of $\left(1-G_{\beta,z}\right)\delta\left(\tau-\tau_{0}\right)+g_{\beta,z}\left(\tau\right)$,
each of which is a univariate probability distribution, sampling $\left(a,\tau\right)$
from $\varphi_{\beta,z}$ is done as follows. First, generate a
random number $\hat{a}$, which is exponentially distributed with
mean $\Lambda_{0,z}^{-1}$. Next generate a random number $\tau$,
which follows $\left(1-G_{\beta,z}\right)\delta\left(\tau-\tau_{0}\right)+g_{\beta,z}\left(\tau\right)$.
To do so, set $\tau=\tau_{0}$ with probability $1-G_{\beta,z}$.
Otherwise, generate uniformly distributed $\left(x,y\right)$ in sufficiently
large rectangular region in $\R^{2}$ and set $\tau=x$
if $g_{\beta,z}\left(x\right)>y$. 

In the numerical simulations described in the main text, we set 2 different
initial conditions.
We mean by \emph{delocalized state} that $\left(a,\tau\right)$ is sampled from $\varphi_{0,z}\left(a,\tau\right)$
at time 0. Each $\left(a,\tau\right)$ among $N$ cells is independent
of each other. The other condition is \emph{predicted stationary}, for which
$\left(a,\tau\right)$ is sampled from $\varphi_{\beta,z}\left(a,\tau\right)$
at time 0, where $\beta$ is the same number as the one used in
time-evolution.

\subsection{Simulation}

Fix an integer $N>0$ for population size. Fix a real number $T>0$
for the simulation time. The offspring number per cell is set to $z=2$.
Let $\hat{t}_{i}$ denote the time of $i$-th division in the population
and let $B_{i}$ denote the number of divisions at time $\hat{t}_{i}$.
To keep the population size constant, $B_{i}$ cells among $N+B_{i}$
are randomly excluded. Let $\tau_{ij}$ denote the IDT of the $j$-th dividing cell among $B_{i}$ cells. Let $\varDelta t$
be the length of the time window within which the growth rate is averaged
and the locally minimum generation time is determined. Let $t_{n}:=n\varDelta t$,
$n=0,1,\dots$. Then the empirical growth rate
\begin{equation}
\Lambda^{\left(n\right)}:=\frac{1}{\varDelta t}\sum_{i:t_{n-1}\le\hat{t}_{i}<t_{n}}\ln\left(1+\frac{B_{i}}{N}\right)
\end{equation}
and the empirical minimum IDT
\begin{equation}
\hat{\tau}_{0}^{\left(n\right)}:=\min_{i:t_{n-1}\le\hat{t}_{i}<t_{n}}\min_{1\le j\le B_{i}}\tau_{ij}
\end{equation}
are computed. We also record all the lineages that remain by
the end of the simulation. We validated that $N$ individuals at time
$T$ coalesce to a single ancestor. In other words, the simulation
time $T$ is taken much longer than the coalescence time for $N$
individuals.

\section{$N$-dependence of population growth and definition of $\beta_{c,N}$
\label{app:N-dependence-of-growth-and-betac}}

To observe the population size dependence of population growth and lineage quantities,
we computed empirical counterparts for $\Lambda_{\beta,z}$, $\overline{\tau}_{\beta,z}^{-1}$ and $\overline{m}_{\beta,z}^{-1}$ with varied population size $N$ and with fixed simulation time $T$.
As the counterpart of  $\Lambda_{\beta,z}$ in a finite population, time-average of $\Lambda^{\left(n\right)}$ is taken, denoted by $\Lambda_{\beta,z,N}$.
To compute the counterparts of $\overline{\tau}_{\beta,z}^{-1}$ and $\overline{m}_{\beta,z}$, a single-cell lineage is randomly chosen and we let $D$ and $S$ be the number of divisions and number of switching events to a different IDT, respectively.
As the counterpart of  $\overline{\tau}_{\beta,z}^{-1}$ in finite population, $D/T$ is taken, denoted by $\overline{\tau}_{\beta,z,N}^{-1}$.
As the counterpart of  $\overline{m}_{\beta,z}^{-1}$ in finite population, $S/D$ is taken, denoted by $\overline{m}_{\beta,z,N}^{-1}$.
For each empirical evaluation, the mean and standard deviation are calculated over 10 simulations.

\begin{figure}
\begin{centering}
\includegraphics[scale=0.55]{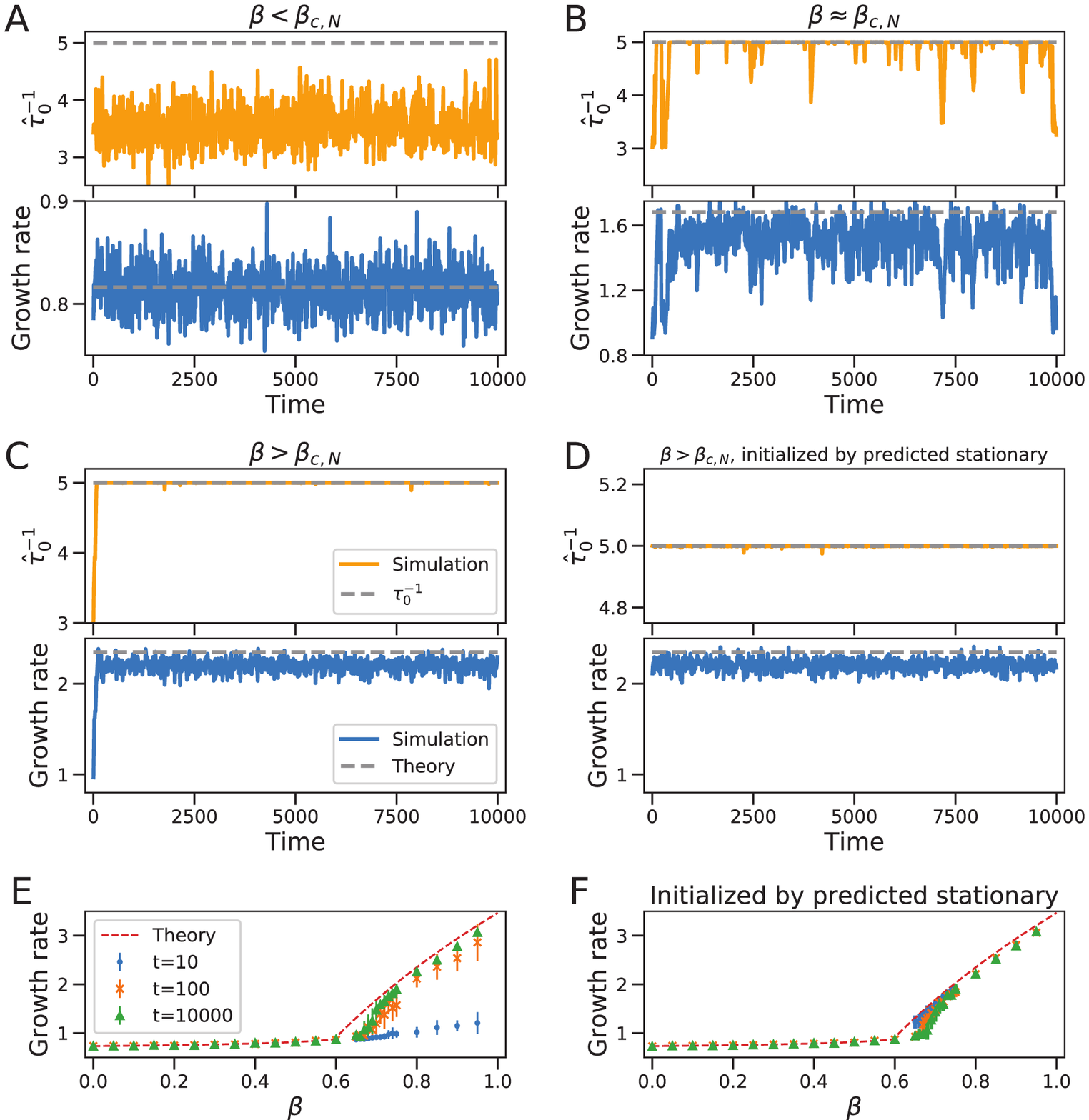}
\par\end{centering}
\caption[Disappearing aging dynamics with 1\% inheritance noise]{Disappearing aging dynamics with 1\% inheritance noise ($\sigma=0.01\tau_{0}$
in Eq. (13)).
The parameters are as in Fig. 1 using $\alpha = 4$, which exhibits a first-order localization transition. 
See the caption of Fig. 2 for the description of figures.
\label{appfig:dyn_1pc_noise}}
\end{figure}

\begin{figure}
\begin{centering}
\includegraphics[scale=0.55]{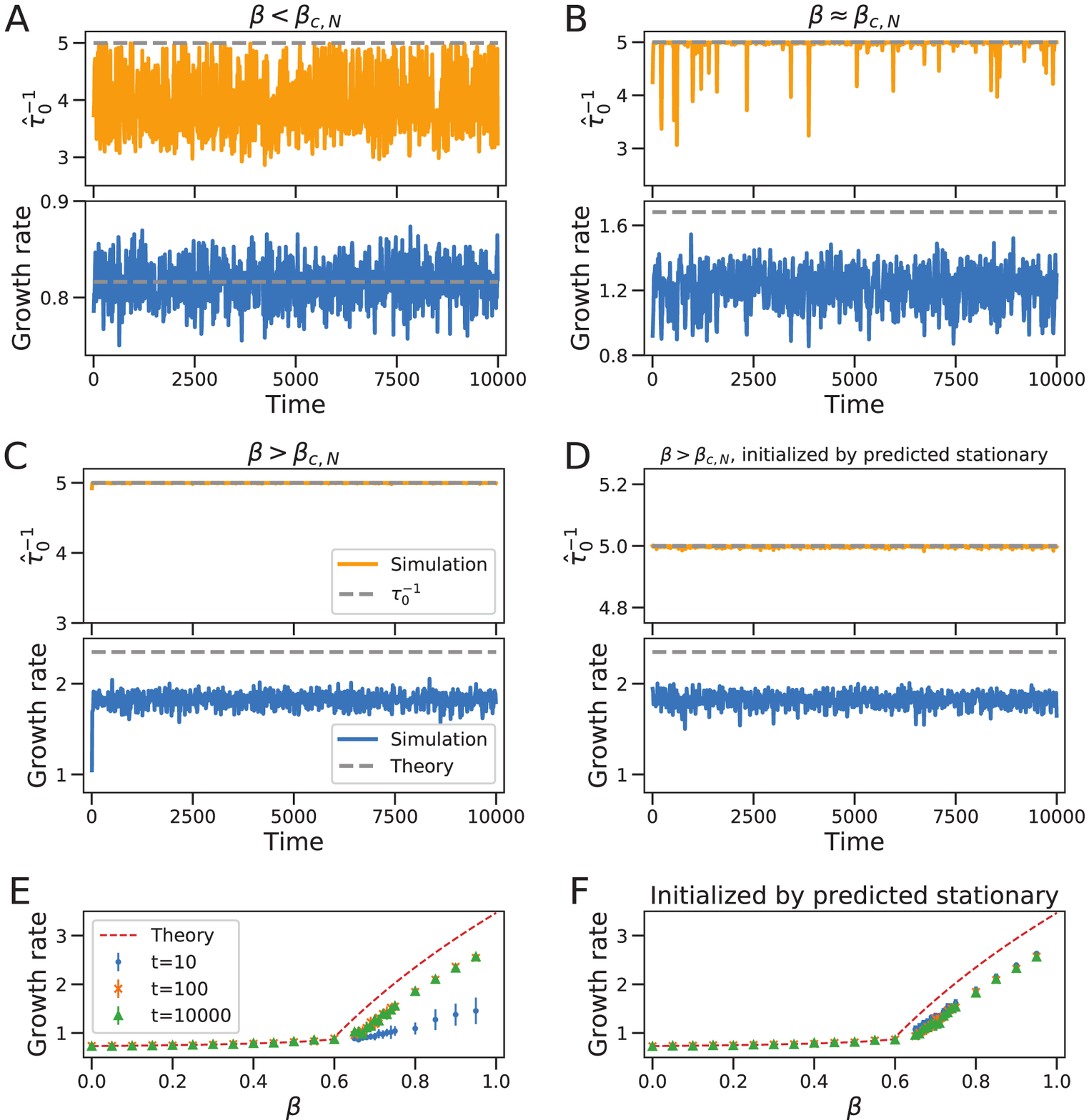}
\par\end{centering}
\caption[Disappearing aging dynamics with 10\% inheritance noise]{Disappearing aging dynamics with 10\% inheritance noise ($\sigma=0.1\tau_{0}$
in Eq. (13)).
The parameters are as in Fig. 1 using $\alpha = 4$, which exhibits a first-order localization transition. 
See the caption of Fig. 2 for the description of figures.
\label{appfig:dyn_10pc_noise}}
\end{figure}

We denote by $\beta_{c,N}$ the critical heritability of IDT at finite population size $N$.
This is defined as $\beta$ which maximizes
\begin{equation}
\overline{\tau}_{\beta,z,N}^{-1} \ln z - \Lambda_{\beta,z,N} \ .
\end{equation}
If $N=\infty$, this is discontinuous at $\beta_{c}$ and
\begin{equation}
\lim_{\beta\downarrow\beta_{c}}\left(\tau_{\beta,z}^{-1}\ln z-\Lambda_{\beta,z}\right)=-\tau_{0}^{-1}\ln\beta_{c}
\end{equation}
is the supreme of $\tau_{\beta,z}^{-1}\ln z-\Lambda_{\beta,z}$ for $\beta\in[0,1)$.
For finite $N$, this function becomes smoother
around $\beta_{c}$ and have a peak at $\beta_{c,N}>\beta_{c}$. Still,
the result can depend on initial state due to aging dynamics. Thus
introducing small noise in inheritance will allow convergence over a feasible
time-scale (Figs \ref{appfig:dyn_1pc_noise} and \ref{appfig:dyn_10pc_noise}).

\bibliography{ref_supplement}

\clearpage
\clearpage
\onecolumngrid

\begin{table} 
\begin{centering}
\caption[List of mother-daughter correlation coefficients of IDTs]
{List of mother-daughter correlation coefficients of IDTs.
For the same cell type in the same reference, brief summary
of the acquisition conditions of the data is described in the ``Data
set'' column. ``Correlation type'' indicates Spearman rank correlation
or Pearson correlation. ``{*}'' denotes the definition of correlation
coefficient is not explicitly described in the reference. ``N''
denotes the sample size. ``M-D corr.'' denotes the mother-daughter
correlation of IDTs. \label{tab:List-of-mdcorr}}
\begin{ruledtabular}
\begin{tabular}{|c|>{\centering}p{4cm}|>{\centering}m{3cm}|c|c|c|>{\centering}p{2cm}|c|}
 & Cell type & Data set & Correlation type & $N$ & M-D corr. & Error range & Ref\\
\hline 
\hline 
\multirow{17}{*}{Bacteria} & Cornebacterium glutamicum & n/a & Spearman & 51 & -0.21 & -0.21$\pm$0.12 & \cite{Mosheiff2018}\\
 & E.coli B Rel606 & n/a & Spearman & 60 & -0.06 & -0.06$\pm$0.25 & \cite{Mosheiff2018}\\
 & E.coli B/r & n/a & Pearson & 160 & 0.11 & n/a & \cite{Kubitschek1962}\\
 & E.coli B/r $\Delta$fimA$\Delta$flu rpsL-gfp & M9 glucose, 37C & Pearson & n/a & -0.18 & -0.18$\pm$0.03 & \cite{Hashimoto2016apdx}\\
 & '' & M9 glycerol, 37C & Pearson & n/a & 0.02 & 0.02$\pm$0.02 & \cite{Hashimoto2016apdx}\\
 & E.coli W3110 $\Delta$fliC$\Delta$fimA$\Delta$flu LVS & M9 glucose, 37C & Pearson & n/a & 0.11 & 0.11$\pm$0.03 & \cite{Hashimoto2016apdx}\\
 & E.coli W3110 $\Delta$fliC$\Delta$fimA$\Delta$flu rpsL-gfp & M9 cas. Acids, 37C & Pearson & n/a & -0.06 & -0.06$\pm$0.03 & \cite{Hashimoto2016apdx}\\
 & '' & M9 glucose, 30C & Pearson & n/a & 0.13 & 0.13$\pm$0.03 & \cite{Hashimoto2016apdx}\\
 & '' & M9 glucose, 37C & Pearson & n/a & 0.26 & 0.26$\pm$0.02 & \cite{Hashimoto2016apdx}\\
 & '' & M9 glycerol, 30C & Pearson & n/a & -0.08 & -0.08$\pm$0.03 & \cite{Hashimoto2016apdx}\\
 & '' & M9 glycerol, 37C & Pearson & n/a & 0.08 & 0.08$\pm$0.03 & \cite{Hashimoto2016apdx}\\
 & '' & M9 LB, 37C & Pearson & n/a & -0.06 & -0.06$\pm$0.04 & \cite{Hashimoto2016apdx}\\
 & E.coli W3110 $\Delta$fliC$\Delta$fimA$\Delta$flu T7-venus & M9 glucose, 30C & Pearson & n/a & 0.02 & 0.02$\pm$0.03 & \cite{Hashimoto2016apdx}\\
 & '' & M9 glucose, 37C & Pearson & n/a & 0.08 & 0.08$\pm$0.02 & \cite{Hashimoto2016apdx}\\
 & Synechococcus elongatus WT & n/a & Spearman & 65 & -0.16 & -0.16$\pm$0.08 & \cite{Mosheiff2018}\\
 & Synechococcus elongatus $\Delta$kaiBC & n/a & Spearman & 74 & -0.17 & -0.17$\pm$0.08 & \cite{Mosheiff2018}\\
 & Synechocystis sp. PCC6803 & n/a & Pearson{*} & n/a & -0.1 & n/a & \cite{Yu2017b}\\
\hline 
\multirow{30}{*}{Mammalian} & B lymphocyte & Fam2 & Spearman & n/a & 0.66 & n/a & \citep{Hawkins2009}\\
 & '' & Fam3 & Spearman & n/a & 0.658 & n/a & \citep{Hawkins2009}\\
 & C3H & n/a & Pearson{*} & 144 & -0.09 & n/a & \citep{Sasaki1977}\\
 & EMT6 & 1st generation & Spearman & 66 & 0.46 & n/a & \citep{Staudte1984}\\
 & '' & 2nd generation & Spearman & 106 & 0.63 & n/a & \citep{Staudte1984}\\
 & '' & 3rd generation & Spearman & 131 & 0.64 & n/a & \citep{Staudte1984}\\
 & '' & 4th generation & Spearman & 37 & 0.14 & n/a & \citep{Staudte1984}\\
 & GPK & n/a & Pearson{*} & 40 & 0.27 & n/a & \citep{Hola1987}\\
 & HCT116 p53V-KI & n/a & Pearson & 71 & -0.03 & {[}\textminus 0.26, 0.16{]} (95\%CI) & \citep{Chakrabarti2018}\\
 & HeLa & series case I\hspace{-.1em}I, $\tau_{d}<15$ & Pearson{*} & n/a & 0.57 & n/a & \citep{Froese1964}\\
 & '' & series case I\hspace{-.1em}I, $15<\tau_{d}<25$ & Pearson{*} & n/a & -0.79 & n/a & \citep{Froese1964}\\
 & '' & series case I\hspace{-.1em}I, $\tau_{d}>25$ & Pearson{*} & n/a & 0.19 & n/a & \citep{Froese1964}\\
 & HelaS & n/a & Pearson{*} & n/a & 0.12 & n/a & \citep{Froese1964}\\
 & HTC & n/a & Pearson{*} & n/a & 0.55 & n/a & \citep{VanWijk1979}\\
 & human embryonic lung fibroblasts & n/a & Pearson{*} & 28 & -0.245 & n/a & \citep{Absher1983}\\
 & IMR-90 & young & Pearson{*} & 41 & 0.1 & n/a & \citep{Absher1984}\\
 & '' & aged & Pearson{*} & 14 & 0.31 & n/a & \citep{Absher1984}\\
 & L1210 & closed system & Spearman & 432 & 0.04 & 0.04$\pm$0.08 & \citep{Sandler2015}\\
 & '' & constant media flow & Spearman & 381 & 0.3 & 0.3$\pm$0.03 & \citep{Sandler2015}\\
 & mouse L-cells & n/a & Pearson{*} & 100 & 0.648 & n/a & \citep{Miyamoto1973}\\
 & mouse osteosarcoma & 1st generation & Pearson{*} & 133 & 0.33 & n/a & \citep{VanMeeteren1984}\\
 & '' & 2nd generation & Pearson{*} & 122 & 0.28 & n/a & \citep{VanMeeteren1984}\\
 & '' & n/a & Pearson{*} & 117 & 0.52 & n/a & \citep{Grasman1990}\\
 & neuroblastoma & n/a & Pearson{*} & 60 & 0.61 & n/a & \citep{Grasman1990}\\
 & NIH3T3 & n/a & Pearson{*} & n/a & 0.06 & n/a & \citep{Shields1977}\\
 & '' & rich medium & Pearson{*} & 141 & -0.418 & n/a & \citep{Staudte1997}\\
 & '' & poor medium & Pearson{*} & 166 & -0.241 & n/a & \citep{Staudte1997}\\
 & PCC3 & n/a & Pearson & 100 & 0.41 & n/a & \citep{Sennerstam1988}\\
 & PCC4 azal & n/a & Pearson{*} & 84 & 0.07 & n/a & \citep{Sennerstam1988}\\
 & rat S6 sarcoma & n/a & Pearson{*} & n/a & 0.194 & n/a & \citep{Dawson1965}\\ 
\end{tabular}
\end{ruledtabular}
\par\end{centering}
\end{table}
\clearpage

\begin{table}
\begin{centering}
\caption[Classification heritability threshold of LR model]
{Classification heritability threshold of LR model. First column denotes
the cases explained in Sec. \ref{subsec:Classification-of-stationary}.
Second column shows the type of heritability threshold. Third to fifth
columns show each limiting case. n/a implies it is the same as the
one in the second column. \label{tab:LR-stationary-class}}
\par\end{centering}
\centering{}
\begin{tabular}{|c|c|c|c|c|c|}
\hline 
 & $0<\tau_{0}<\tau_{1}<\infty$ & $\omega_0$ & $\tau_{0}\downarrow0$ & $\tau_{1}\uparrow\infty$ & $k\left(\tau\right)\rightarrow\delta\left(\tau-\bar{\tau}\right)$\tabularnewline
\hline 
\hline 
$z>1$ (case I) & Localize to $\tau_{0}$ above $\beta_{c}\left(>z^{-1}\right)$\footnote{$\beta_{c}<1$ when $\int_{\tau_{0}}^{\tau_{1}}d\tau\sfrac{k\left(\tau\right)}{\left(\tau-\tau_{0}\right)}<\infty$}
&$ \ln(\beta z) / \tau_0$ & $\beta_{c}\downarrow z^{-1}$ & n/a & $\beta_{c}=z^{\sfrac{\tau_{0}}{\bar{\tau}}-1}$\tabularnewline
\hline 
$z=1$ (case I\hspace{-.1em}I) & No threshold ($\beta_{c}=1$) & 0 & n/a & n/a & $\beta_{c}=1$\tabularnewline
\hline 
$0<z<1$ (case I\hspace{-.1em}I\hspace{-.1em}I) & Localize to $\tau_{1}$ above $\beta_{c}\left(>0\right)$ \footnote{$\beta_{c}<1$ when $\int_{\tau_{0}}^{\tau_{1}}d\tau\sfrac{k\left(\tau\right)}{\left(\tau_{1}-\tau\right)}<\infty$} & $\ln (\beta z) / \tau_1$ & n/a & $\beta_{c}\downarrow0$  & $\beta_{c}=z^{\sfrac{\tau_{1}}{\bar{\tau}}-1}$\tabularnewline
\hline 
\end{tabular}
\end{table}

\clearpage

\begin{figure}
\begin{centering}
\includegraphics[scale=0.55]{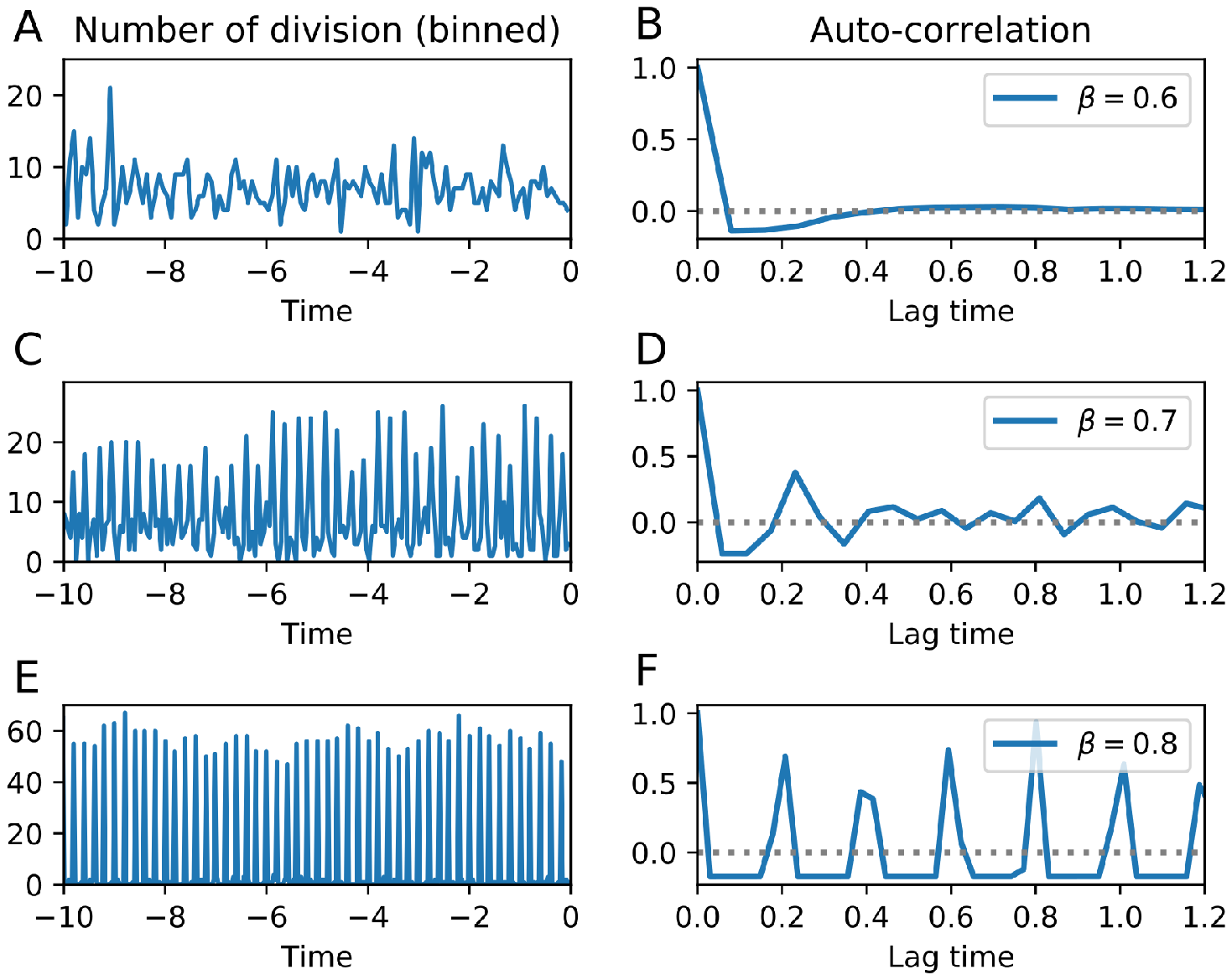}
\par\end{centering}
\caption[Synchronization of cell-cycles without inheritance noise]{Synchronization of cell-cycles without inheritance noise ($\sigma=0$
in Eq. (13)), starting from
predicted stationary.
Except for the initial condition, parameters used for the simulation are the same as Fig. 3 in the main text.
Captions for each figure are the same with those in Fig. 3.
\label{appfig:divsynch_N100_sig0_init2}}
\end{figure}

\begin{figure}
\begin{centering}
\includegraphics[scale=0.55]{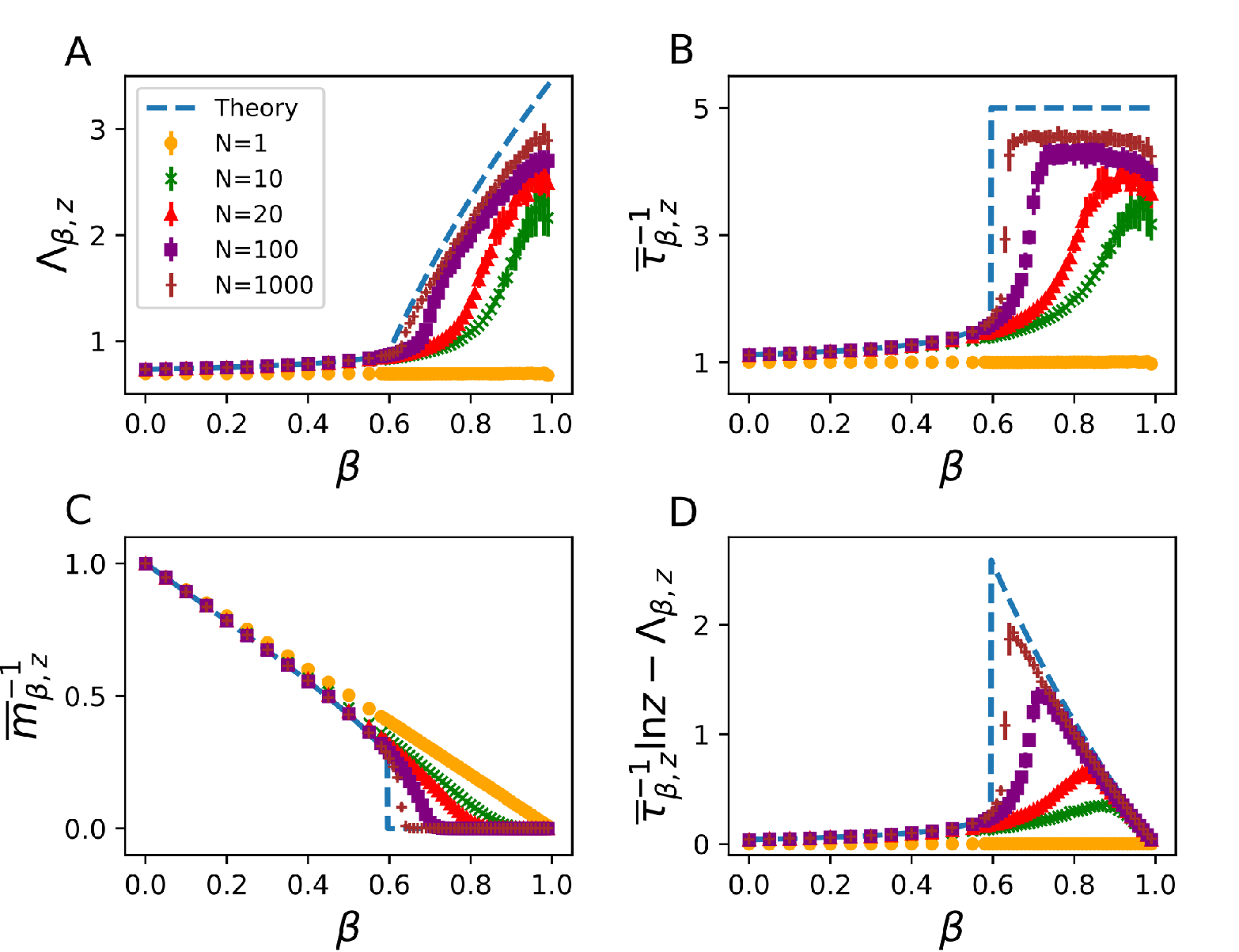}
\par\end{centering}
\caption[Finite $N$ effects, initial delocalized state]{The effect of finite population size $N$ on population growth rate
and lineage quantities, starting from delocalized. Simulation
time $t=10000$. The parameters are the same with $\alpha=4$ in Fig.
1. (A) Population growth rate. (B) The frequency of
divisions per unit time on lineage. (C) The frequency of switching
to the other $\tau$ per generation on lineage. (D) Results in (B)
multiplied by $\ln2$ subtracted by results in (A). \label{appfig:N-dependence-lineage-init0}}
\end{figure}

\begin{figure}
\begin{centering}
\includegraphics[scale=0.55]{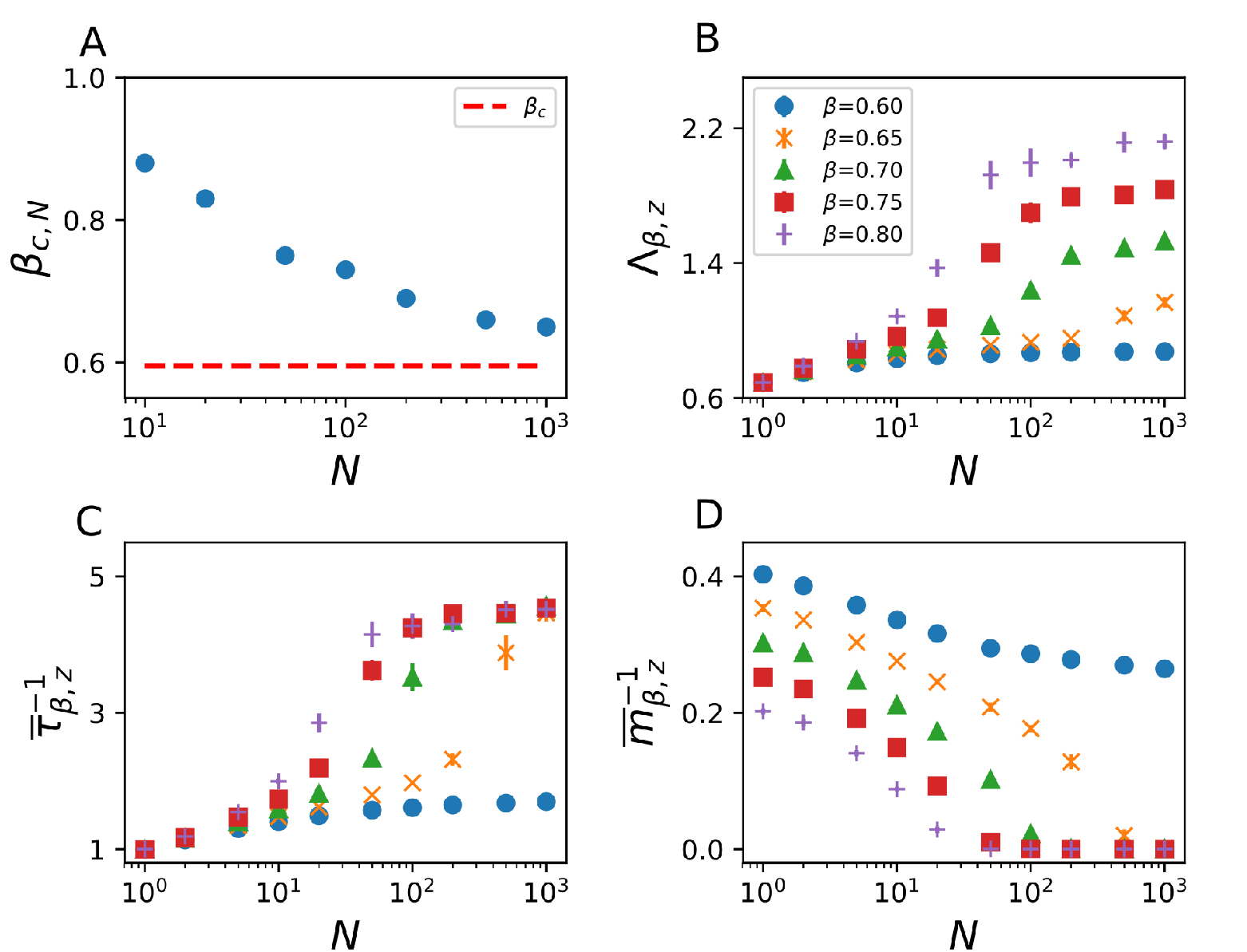}
\par\end{centering}
\caption[Phase transition at various $N$, initial delocalized state]{Phase transition across population size $N$, starting from delocalized state.
The same results with Fig. \ref{appfig:N-dependence-lineage-init0}
are used for these figures. (A) Threshold of $\beta$ in finite population
of fixed size, computed as the position of peak in Fig \ref{appfig:N-dependence-lineage-init0}.
(B-D) Rearrangements of Fig \ref{appfig:N-dependence-lineage-init0}A-C
against $N$ with the certain range of $\beta$ around transition
point. \label{appfig:N-dependence-PT-init0}}
\end{figure}

\begin{figure}
\centering{}\includegraphics[scale=0.55]{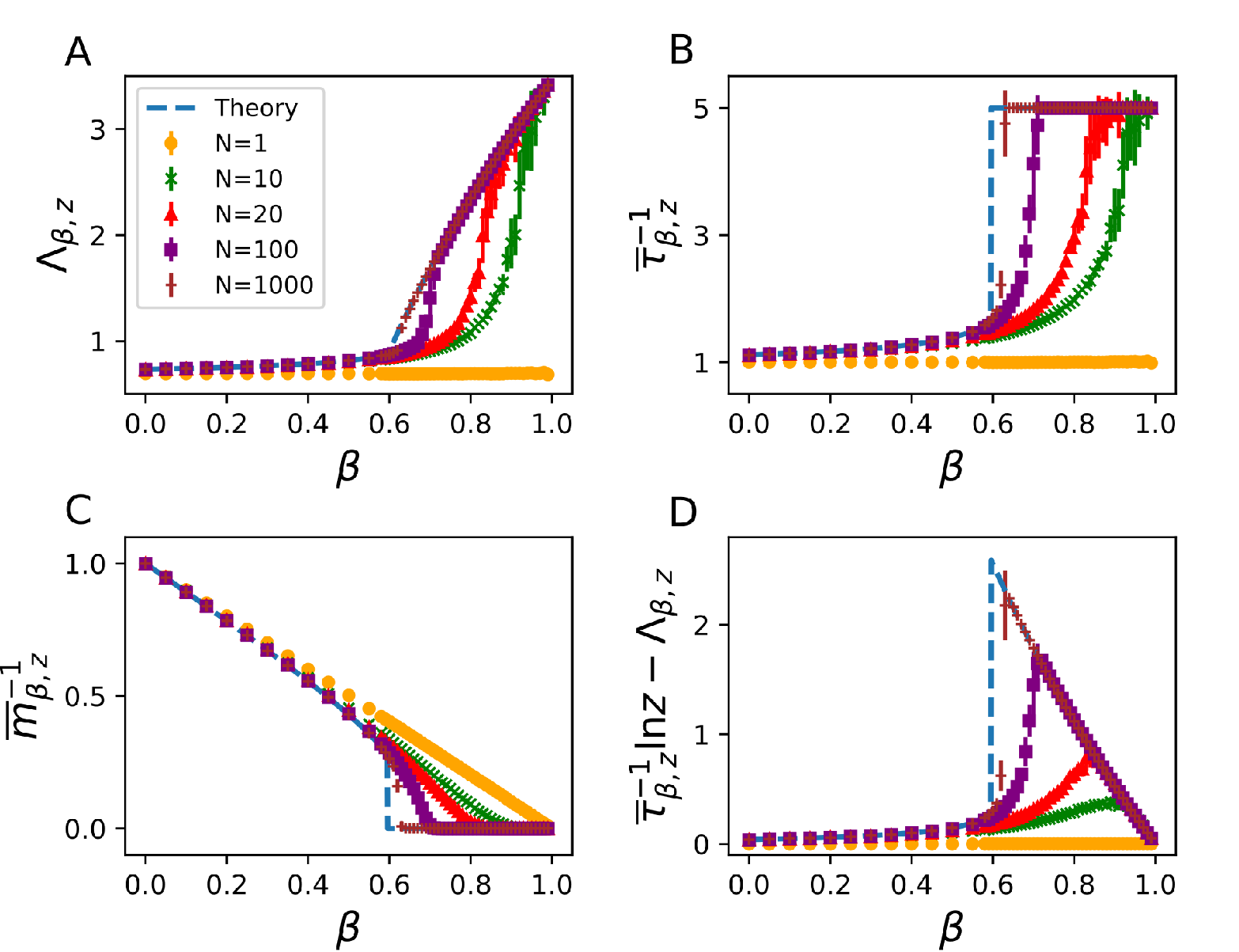}
\caption[Finite $N$ effects, initial predicted stationary state]{The effect of finite population size $N$ on population growth rate
and lineage quantities, starting from predicted stationary. Captions
for each figure are the same with those in Fig. \ref{appfig:N-dependence-lineage-init0}.
\label{appfig:N-dependence-lineage-init2}}
\end{figure}

\begin{figure}
\centering{}\includegraphics[scale=0.55]{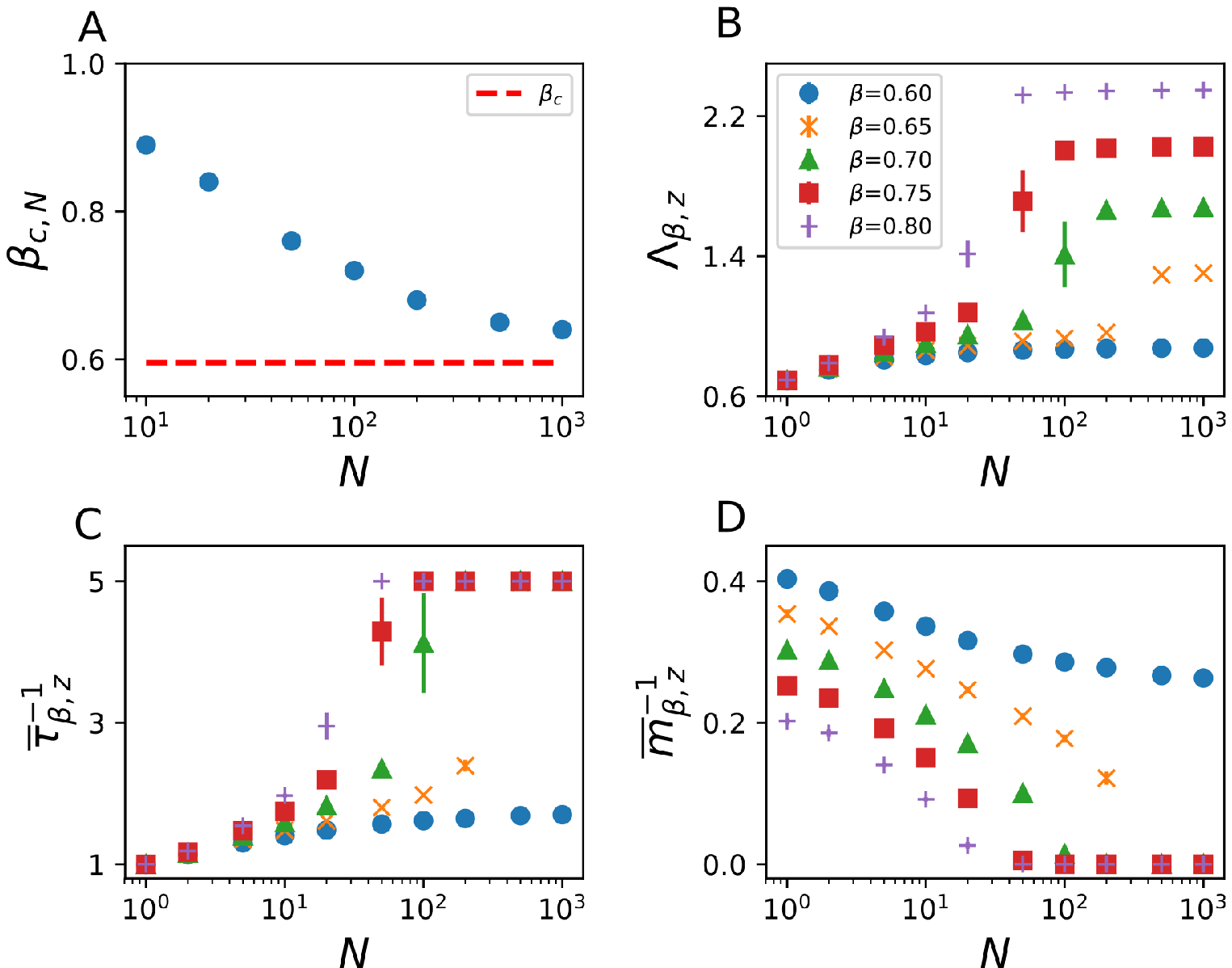}\caption[Phase transition at various $N$, initial predicted stationary state]{Phase transition across population size $N$,
starting from predicted stationary. The same results with Fig. \ref{appfig:N-dependence-lineage-init2}
are used for these figures. Captions for each figure are the same
with those in Fig. \ref{appfig:N-dependence-PT-init0}. \label{appfig:N-dependence-PT-init2}}
\end{figure}

\begin{figure}
\centering{}\includegraphics[scale=0.55]{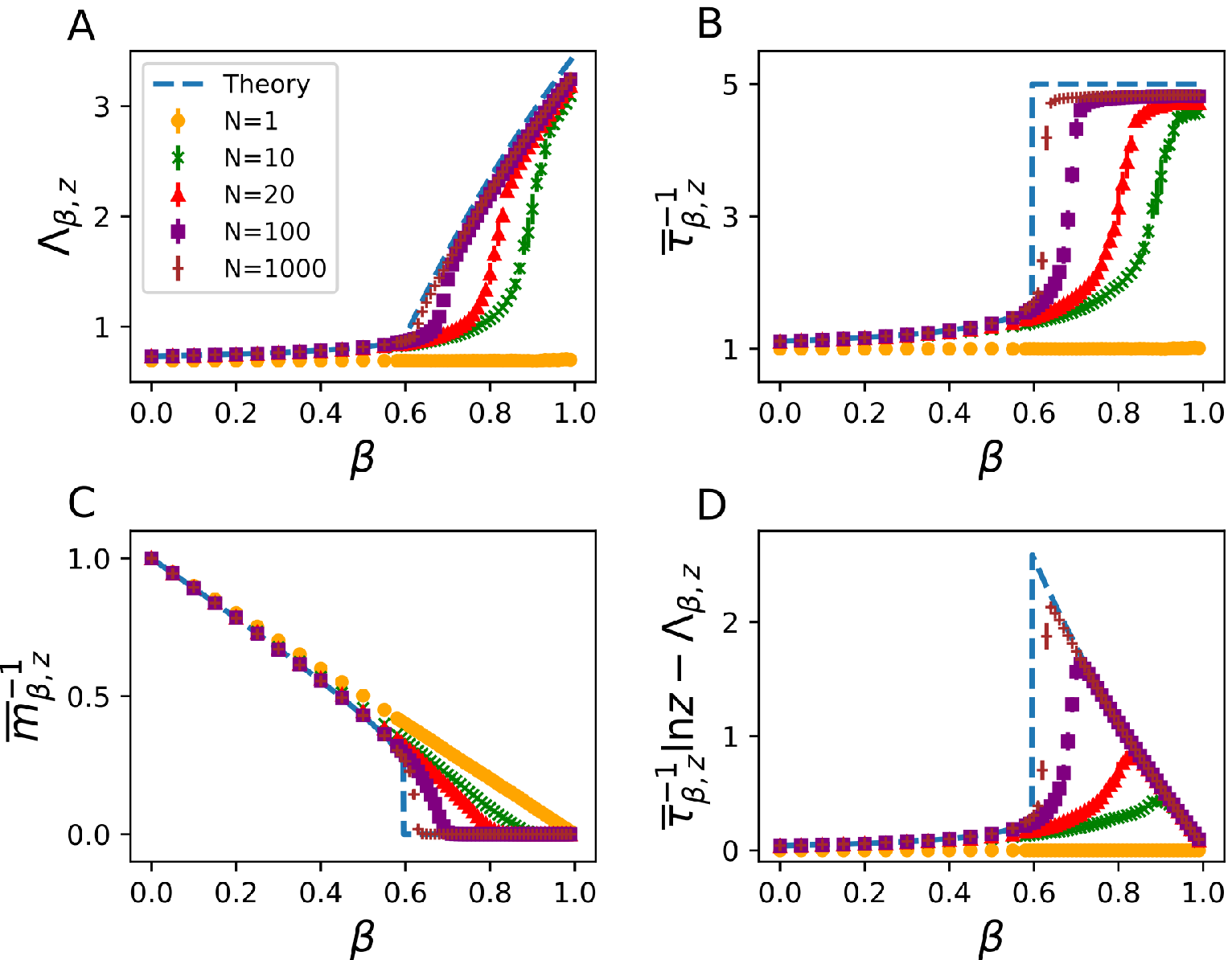}\caption[Finite $N$ effects with 1\% inheritance noise, initial delocalized state]{The effect of finite population size $N$ on population growth rate
and lineage quantities, with 1\% inheritance noise ($\sigma=0.01\tau_{0}$
in Eq. (13)), starting from
delocalized. Simulation time $t=10000$. The parameters
are the same with $\alpha=4$ in Fig. 1. Captions for
each figure are the same with those in Fig. \ref{appfig:N-dependence-lineage-init0}.
\label{appfig:N-dependence-lineage-init0-1pc-noise}}
\end{figure}

\begin{figure}
\centering{}\includegraphics[scale=0.55]{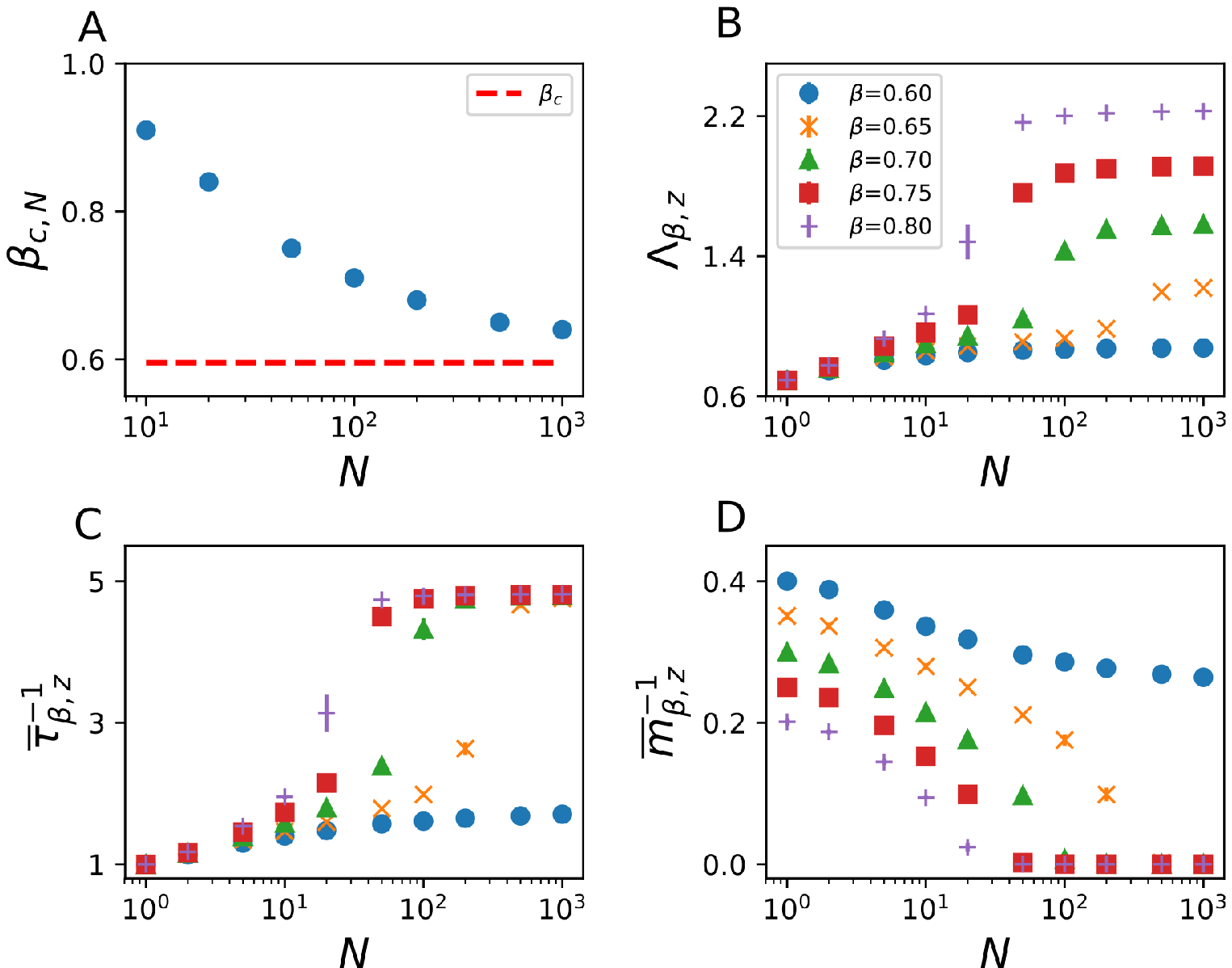}\caption[Phase transition at various $N$ with 1\% inheritance
noise, initial delocalized state]{Phase transition across population size $N$, with 1\% inheritance
noise, starting from delocalized. The same results with
Fig. \ref{appfig:N-dependence-lineage-init0-1pc-noise} are used for
these figures. Captions for each figure are the same with those in
Fig. \ref{appfig:N-dependence-PT-init0}. \label{appfig:N-dependence-PT-init0-1pc-noise}}
\end{figure}

\begin{figure}
\centering{}\includegraphics[scale=0.55]{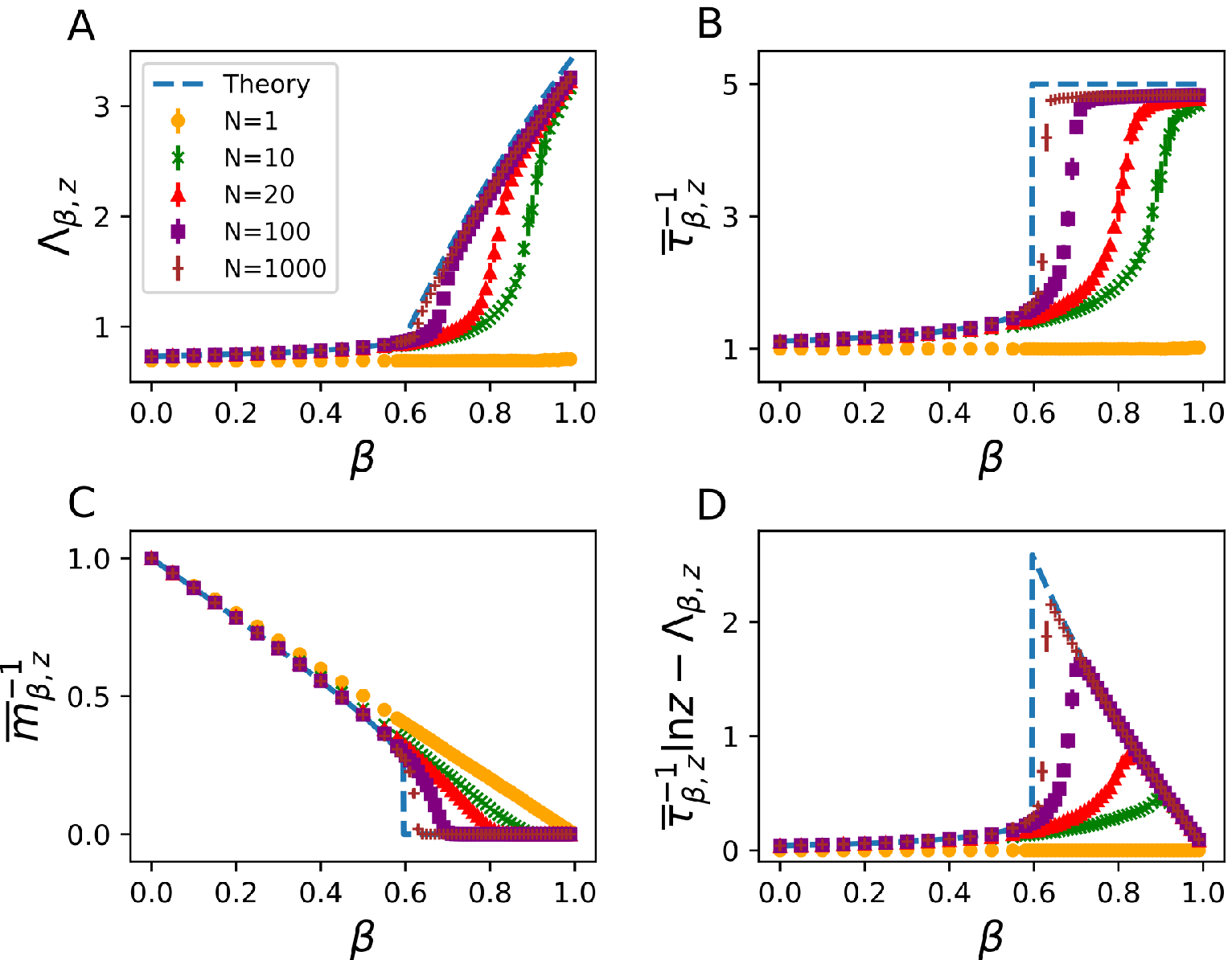}\caption[Finite $N$ effects with 1\% inheritance noise, initial predicted stationary state]{The effect of finite population size $N$ on population growth rate
and lineage quantities, with 1\% inheritance noise ($\sigma=0.01\tau_{0}$
in Eq. (13)), starting from
predicted stationary. Simulation time $t=10000$. The parameters
are the same with $\alpha=4$ in Fig. 1. Captions for
each figure are the same with those in Fig. \ref{appfig:N-dependence-lineage-init0}.
\label{appfig:N-dependence-lineage-init2-1pc-noise}}
\end{figure}

\begin{figure}
\centering{}\includegraphics[scale=0.55]{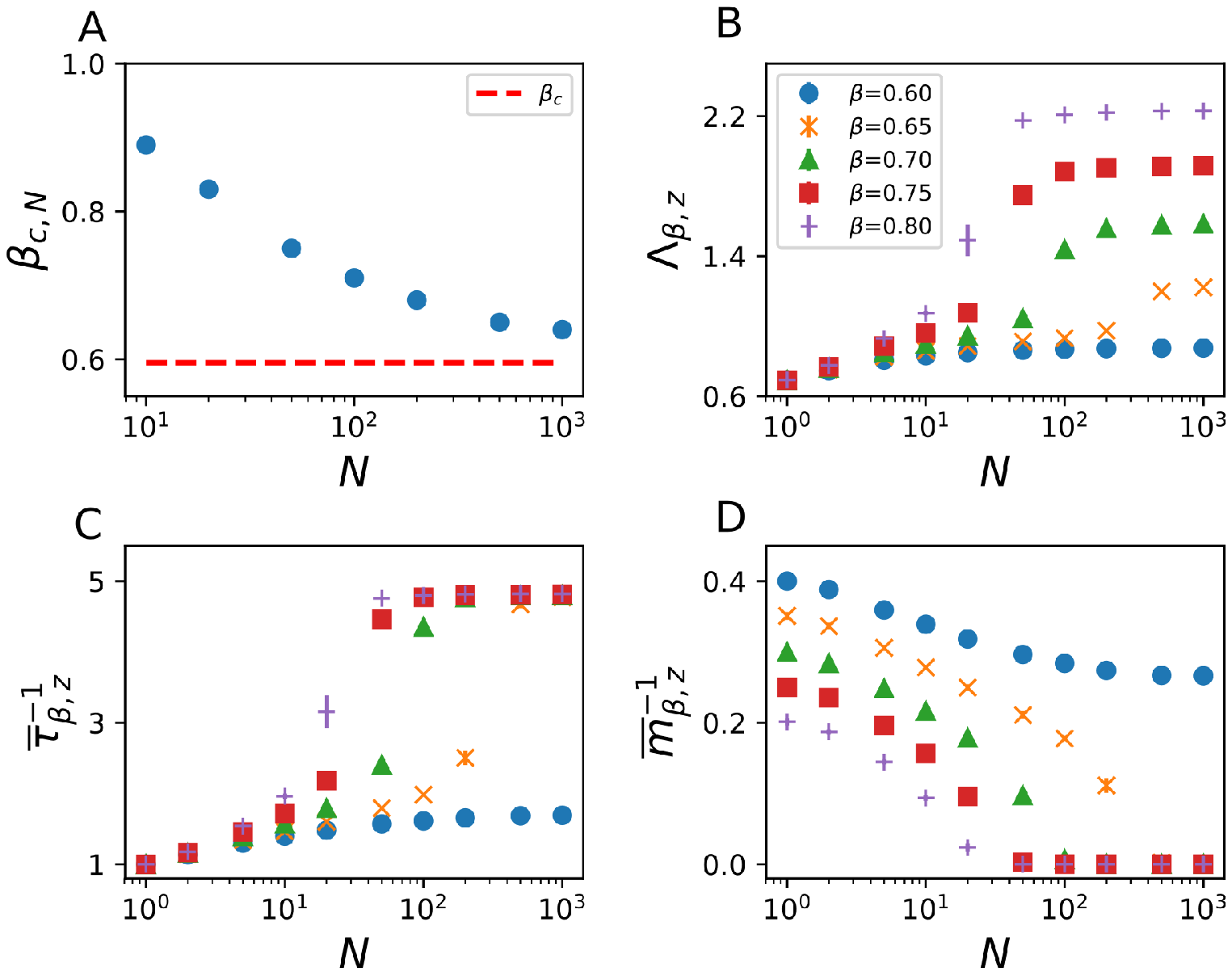}\caption[Phase transition at various $N$ with 1\% inheritance
noise, initial predicted stationary state]{Phase transition across population size $N$, with 1\% inheritance
noise, starting from predicted stationary. The same results with
Fig. \ref{appfig:N-dependence-lineage-init2-1pc-noise} are used for
these figures. Captions for each figure are the same with those in
Fig. \ref{appfig:N-dependence-PT-init0}. \label{appfig:N-dependence-PT-init2-1pc-noise}}
\end{figure}

\begin{figure}
\centering{}\includegraphics[scale=0.55]{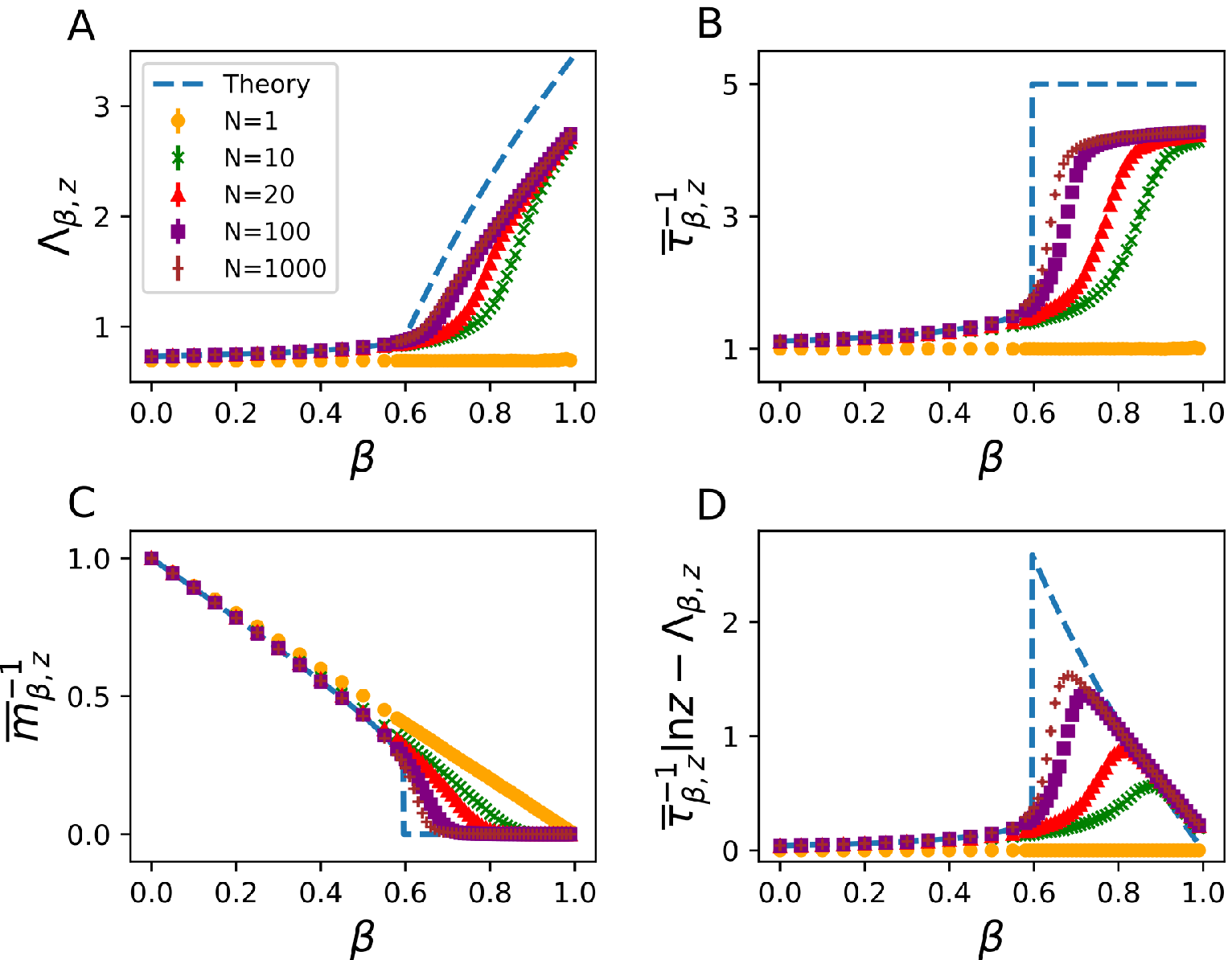}\caption[Finite $N$ effects with 10\% inheritance noise, initial delocalized state]{The effect of finite population size $N$ on population growth rate
and lineage quantities, with 10\% inheritance noise ($\sigma=0.1\tau_{0}$
in Eq. (13)), starting from
delocalized. Simulation time $t=10000$. The parameters
are the same with $\alpha=4$ in Fig. 1. Captions for
each figure are the same with those in Fig. \ref{appfig:N-dependence-lineage-init0}.
\label{appfig:N-dependence-lineage-init0-10pc-noise}}
\end{figure}

\begin{figure}
\centering{}\includegraphics[scale=0.55]{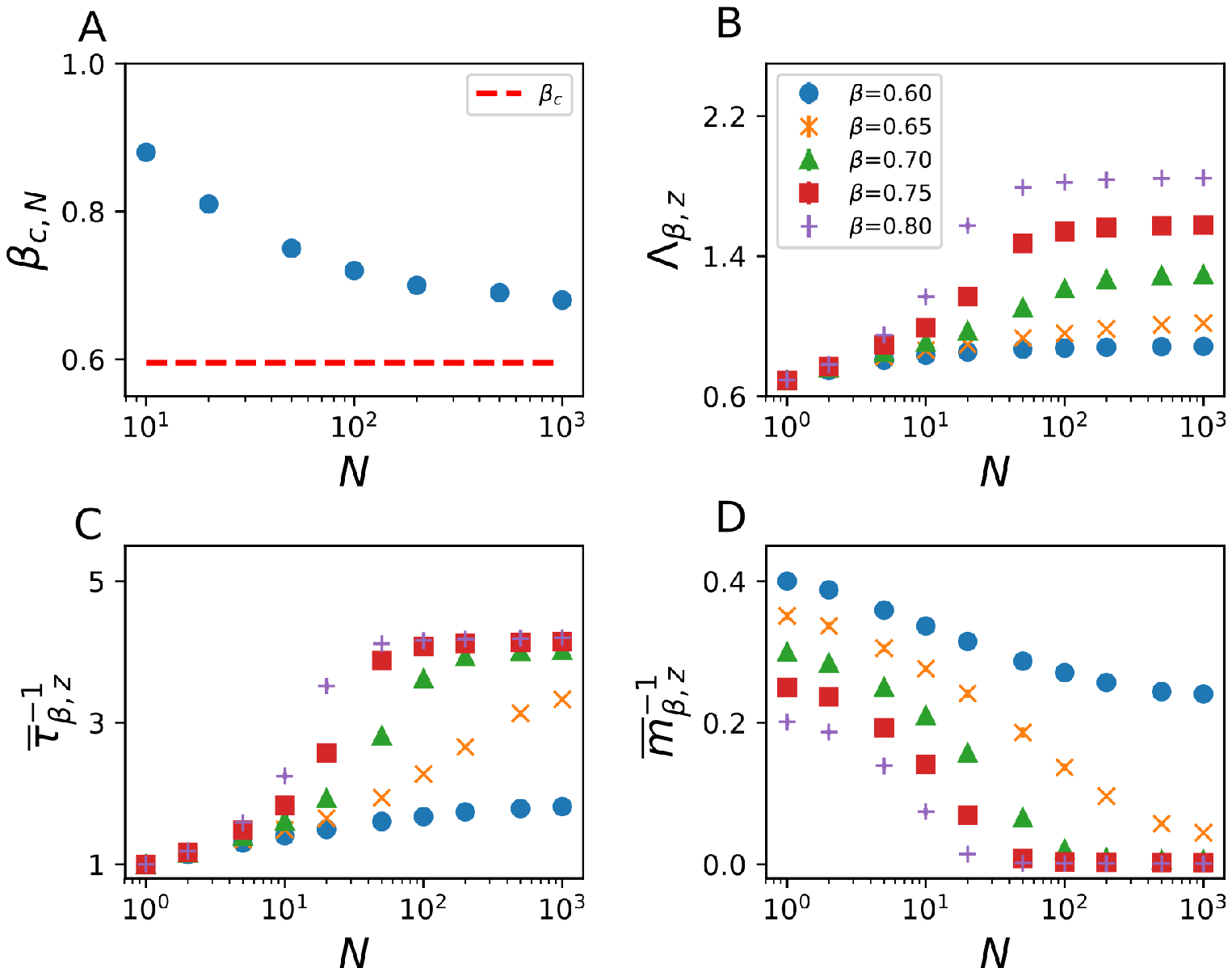}\caption[Phase transition at various $N$ with 10\% inheritance
noise, initial delocalized state]{Phase transition across population size $N$, with 10\% inheritance
noise, starting from delocalized. The same results with
Fig. \ref{appfig:N-dependence-lineage-init0-10pc-noise} are used
for these figures. Captions for each figure are the same with those
in Fig. \ref{appfig:N-dependence-PT-init0}. \label{appfig:N-dependence-PT-init0-10pc-noise}}
\end{figure}

\begin{figure}
\centering{}\includegraphics[scale=0.55]{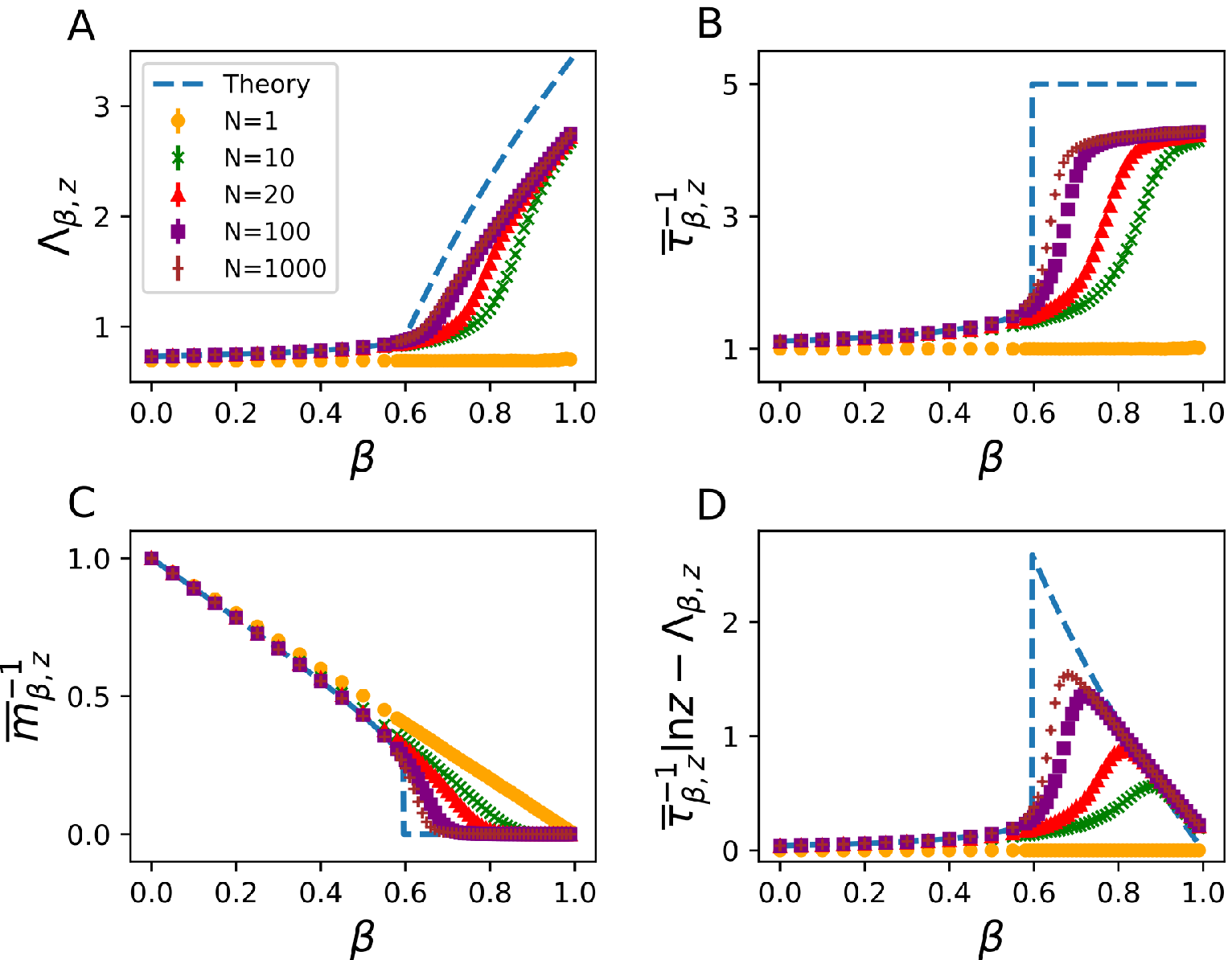}\caption[Finite $N$ effects with 10\% inheritance noise, initial predicted stationary state]{The effect of finite population size $N$ on population growth rate
and lineage quantities, with 10\% inheritance noise ($\sigma=0.1\tau_{0}$
in Eq. (13)), starting from
predicted stationary. Simulation time $t=10000$. The parameters
are the same with $\alpha=4$ in Fig. 1. Captions for
each figure are the same with those in Fig. \ref{appfig:N-dependence-lineage-init0}.
\label{appfig:N-dependence-lineage-init2-10pc-noise}}
\end{figure}

\begin{figure}
\centering{}\includegraphics[scale=0.55]{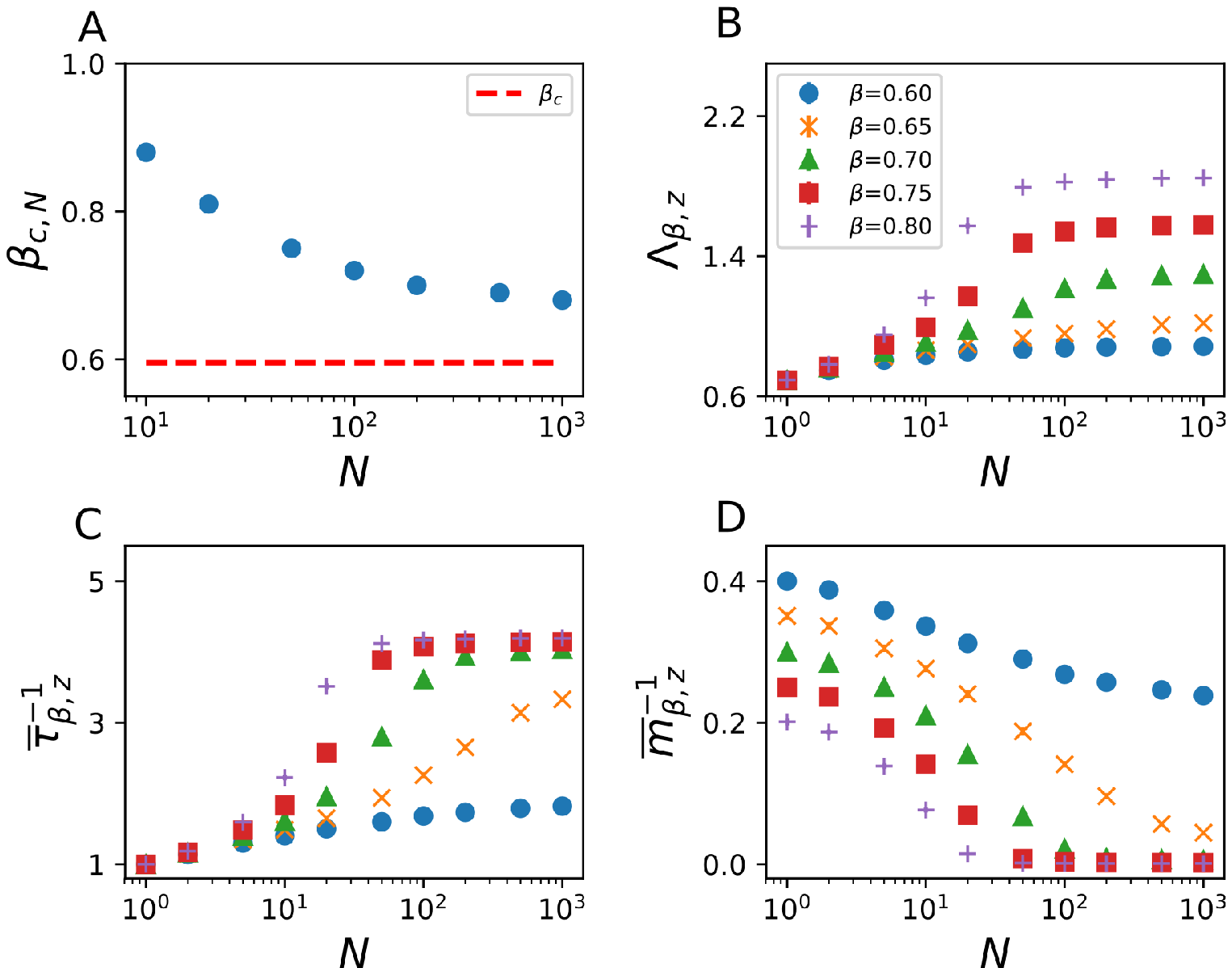}\caption[Phase transition at various $N$ with 10\% inheritance
noise, initial predicted stationary state]{Phase transition across population size $N$, with 10\% inheritance
noise, starting from predicted stationary. The same results with
Fig. \ref{appfig:N-dependence-lineage-init2-10pc-noise} are used
for these figures. Captions for each figure are the same with those
in Fig. \ref{appfig:N-dependence-PT-init0}. \label{appfig:N-dependence-PT-init2-10pc-noise}}
\end{figure}

\begin{figure}
\centering{}\includegraphics[scale=0.55]{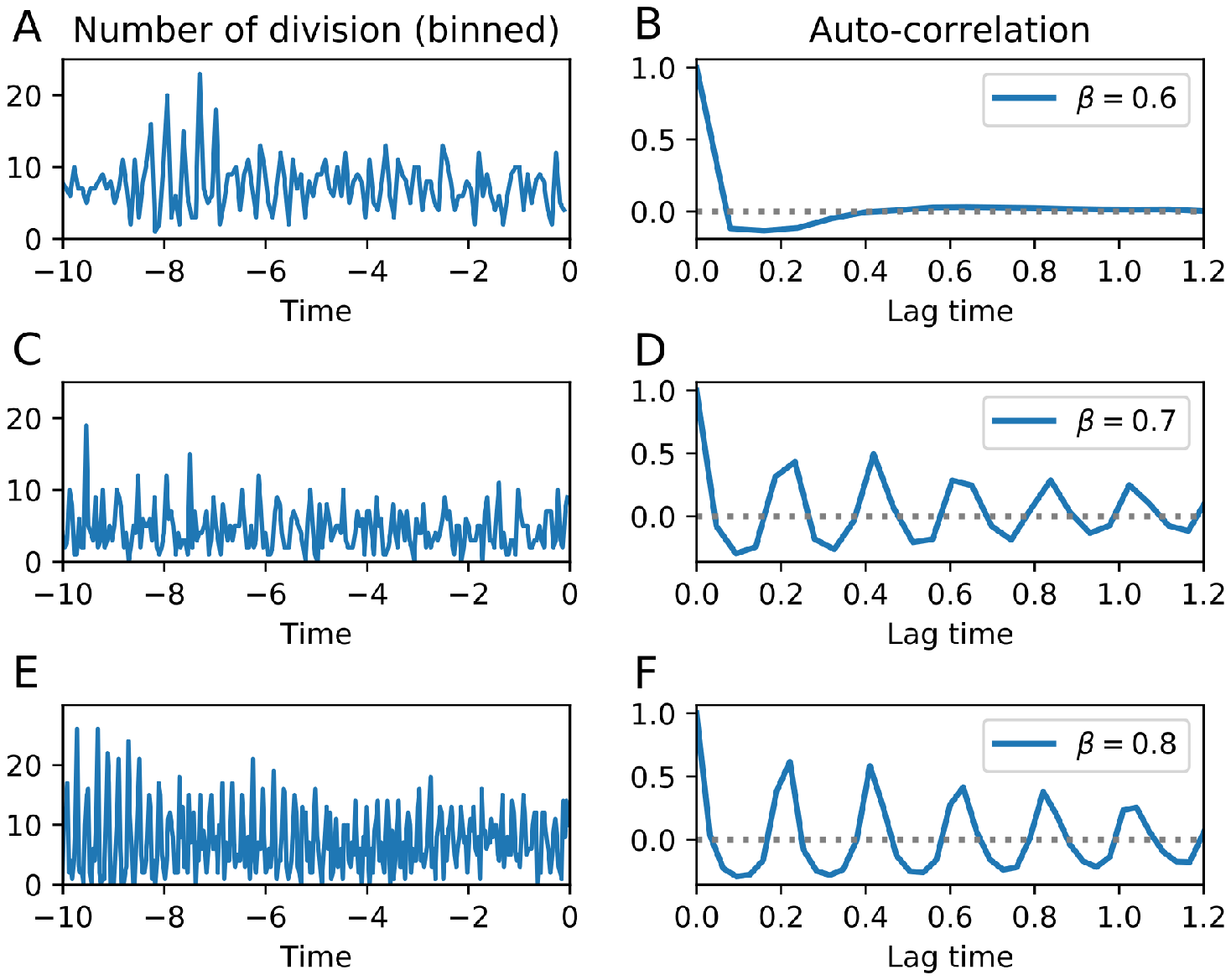}\caption[Synchronization of cell-cycles with 1\% inheritance noise]{Synchronization of cell-cycles with 1\% inheritance noise ($\sigma=0.01\tau_{0}$
in Eq. (13)), starting from delocalized.
The parameters are the same with $\alpha=4$ in Fig. 1.
$N=100$. Simulation time $t=10000$.
Captions for each figure are the same with those in Fig. 3.
\label{appfig:divsynch_N100_sig0.01_init0}}
\end{figure}

\begin{figure}
\centering{}\includegraphics[scale=0.55]{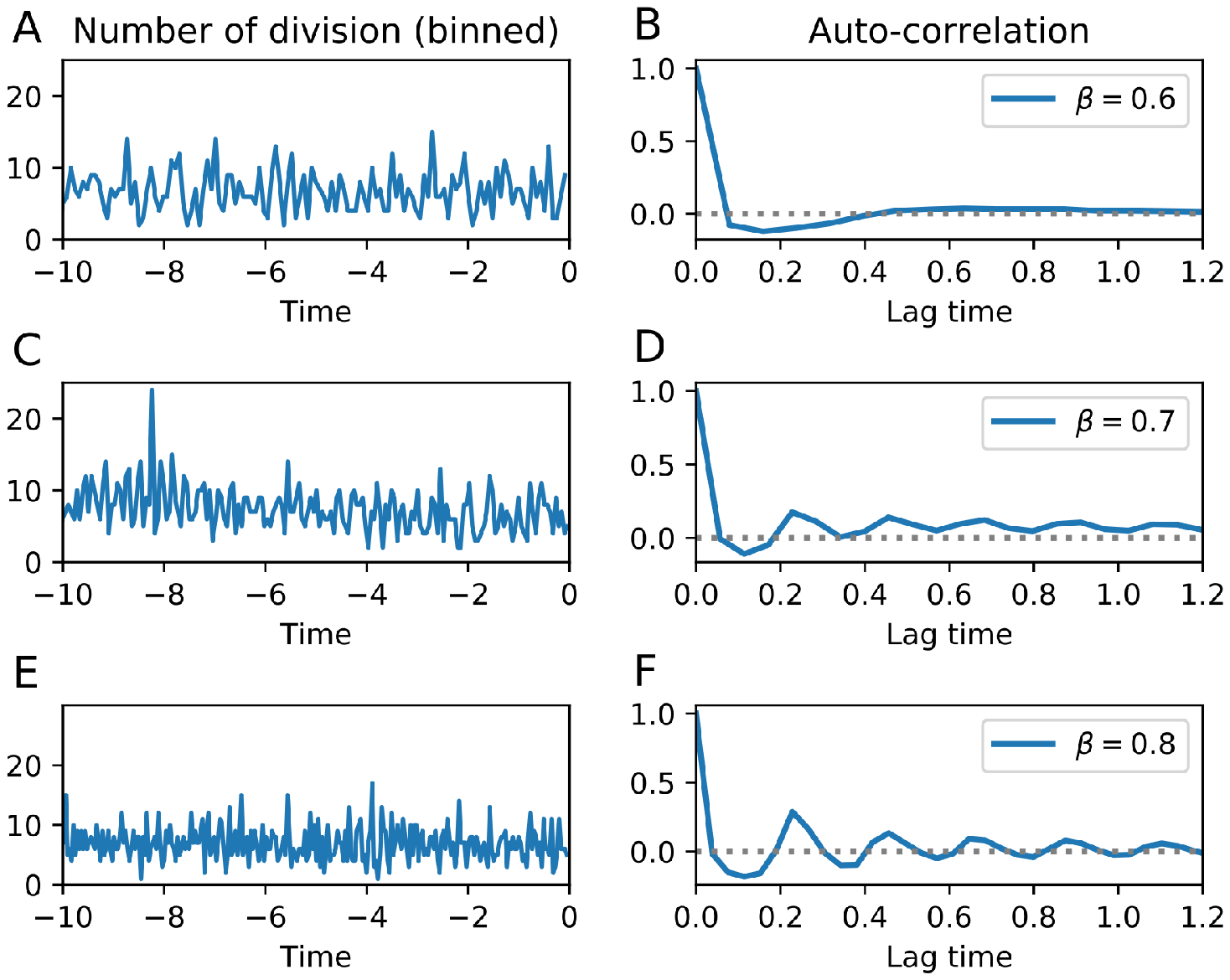}\caption[Synchronization of cell-cycles with 10\% inheritance noise]{Synchronization of cell-cycles with 10\% inheritance noise ($\sigma=0.1\tau_{0}$
in Eq. (13)), starting from delocalized.
The parameters are the same with $\alpha=4$ in Fig. 1.
$N=100$. Simulation time $t=10000$. Captions for each figure are
the same with those in Fig. 3. 
\label{appfig:divsynch_N100_sig0.1_init0}}
\end{figure}